\documentclass[usenatbib]{mn2e}
\usepackage{multirow}
\usepackage{rotating}
\usepackage{colortbl}
\usepackage{color}
\usepackage[fleqn]{amsmath}
\usepackage{amssymb}
\usepackage{amsfonts}
\usepackage{verbatim}
\usepackage{scalefnt}
\usepackage[percent]{overpic}

\newcommand{\araa}{ARA\&A}
\newcommand{\apj}{ApJ}
\newcommand{\apjl}{ApJ}
\newcommand{\aap}{A\&A}
\newcommand{\mnras}{MNRAS}
\newcommand{\nar}{New A Rev.}

\newcommand{\msun}{\,{\rm M_\odot}}

\newcommand{\sfrn}{{\rm SFR}_{100{\rm pc}}}

\newcommand{\sfrg}{{\rm SFR}_{5{\rm kpc}}}
\newcommand{\beq}{\begin{equation}}
\newcommand{\eeq}{\end{equation}}
\newcommand{\ba}{\begin{eqnarray}}
\newcommand{\ea}{\end{eqnarray}}
\def\spose#1{\hbox to 0pt{#1\hss}}
\newcommand{\lta}{\mathrel{\spose{\lower 3pt\hbox{$\mathchar'218$}}
      \raise 2.0pt\hbox{$\mathchar"13C$}}}
\newcommand{\gta}{\mathrel{\spose{\lower 3pt\hbox{$\mathchar"218$}}
      \raise 2.0pt\hbox{$\mathchar"13E$}}}

\newcommand\T{\rule{0pt}{2.6ex}}       
\newcommand\B{\rule[-1.2ex]{0pt}{0pt}} 

\newcommand{\comments}[1]{} 

\title[BH accretion and star formation rate]{Growing black holes and galaxies: black hole accretion versus star formation rate}

\author[Volonteri et al.]{Marta Volonteri,$^{1}$\thanks{E-mail: martav@iap.fr}  Pedro R. Capelo$^{2}$, Hagai Netzer$^{3}$,\newauthor Jillian Bellovary$^{4}$, Massimo Dotti$^{5}$, Fabio Governato$^6$\\
$^1$Institut d'Astrophysique de Paris, 98bis Boulevard Arago, F-75014 Paris, France\\
$^2$Department of Astronomy, University of Michigan, Ann Arbor, MI 48109, USA\\
$^3$School of Physics and Astronomy, The Sackler Faculty of Exact Sciences, Tel-Aviv University, Tel-Aviv 69978, Israel\\
$^4$Department of Physics and Astronomy, Vanderbilt University, Nashville, TN 37235, USA\\
$^5$Dipartimento di Fisica G. Occhialini, Universit$\grave{a}$ degli Studi di Milano Bicocca, Piazza della Scienza 3, I-20126 Milano, Italy\\
$^6$Department of Astronomy, University of Washington, Box 351580, Seattle, WA 98195, USA}

\begin{document}

\maketitle

\begin{abstract}
We present a new suite of hydrodynamical simulations and use it to study, in detail, black hole and galaxy properties. 
The high time, spatial and mass resolution, and realistic orbits and mass ratios, down to 1:6 and 1:10, enable us to meaningfully compare star formation rate 
(SFR) and BH accretion rate (BHAR) timescales, temporal behaviour and relative magnitude.  
We find that (i) BHAR and galaxy-wide SFR are  typically temporally uncorrelated, and have different variability timescales, except during the merger proper, lasting $\sim 0.2-0.3$~Gyr. BHAR and nuclear ($<100$ pc) SFR are better correlated,  and their variability are similar.
 Averaging over time, the merger phase leads typically to an increase by a factor of a few in the BHAR/SFR ratio. 
(ii) BHAR and nuclear SFR are intrinsically proportional, but the correlation lessens if the long-term SFR is measured. 
(iii) Galaxies in the remnant phase are the ones most likely to be selected as systems dominated by an active galactic nucleus (AGN),
because of the long time spent in this phase. 
(iv) The timescale over which a given diagnostic probes the SFR has a profound impact on the recovered correlations with BHAR, and on the interpretation of 
observational data. 

\end{abstract}


\section{Introduction}\label{sec:Introduction}
Several known scaling relationships between supermassive black holes (BHs) and large-scale properties of their host galaxies, such as mass, luminosity, and velocity dispersion, primarily of the bulge component, suggest a joint galaxy and BH cosmic evolution 
 \citep{Magorrian1998,fm00,MarconiHunt2003,Haring2004,Gultekin2009,2013ARA&A..51..511K}. 
In particular, the 
almost linear correlation between BH mass and bulge mass suggests parallel growth. More specifically, 
`for every $\sim$ 1000 units of star formation (SF) there is $\sim$ 1--2 units of BH accretion'' \citep{2012NewAR..56...93A}. 
Several observational studies attempted to compare BH accretion rate (BHAR) and  SF rates (SFRs) on galactic scale
\citep[e.g.,][]{2007ApJ...666..806N,2007MNRAS.381..543W,2008ApJ...684..853L,2009MNRAS.399.1907N,2010MNRAS.405..933W,2012A&A...545A..45R} 
and sub-galactic scales \citep[$<$1~kpc,][]{2012ApJ...746..168D}. In general, such a comparison shows a large scatter which is somewhat reduced when the SFR is 
measured over $<$1~kpc scales, the region more easily influenced by the BH, and more directly identified with the bulge (although bulges can be significantly larger).  

Taking a statistical approach, \cite{2004ApJ...613..109H,Merlonietal2004,2008ApJ...679..118S,2009ApJ...696..396S} argued that the volume averaged ratio of BHAR to SFR 
is about constant up to $z\sim 3$. \cite{2012ApJ...753L..30M,2013ApJ...773....3C} further suggested that the measured BHAR/SFR ratio 
may vary wildly, mostly because the time-variability of BH accretion is much faster than that of SF \citep[see also][]{Aird2012,Hickox2014}. In this view, BHAR and SFR 
may appear uncorrelated in sources taken one by one, but once a large sample is averaged the underlying correlation emerges.

Theoretical models have investigated AGN activity and SF on different levels \citep{Kauffmann2000,Hopkins2006,2011MNRAS.412.2154B,Hayward_et_al_2014}. 
\cite{DiMatteo2005} and \cite{Springel2005Red} 
suggested that galaxy mergers enhance both BH activity and SFR. \cite{2014MNRAS.443.1125T} performed a series of simulations of equal-mass galaxy mergers 
to study the SFR-BHAR correlation. They found that the evolution of BHAR and SFR in a single merger is highly complex and that  the volume averaged  
 correlation is only approximate. 
\cite{Silk2013}  develops a feedback model that 
couples SFR and BHAR via outflow-induced pressure-enhanced SF. This model predicts that, on average, BHAR $ \sim 10^{-3}$ SFR, 
modulated by the radiative  and mechanical efficiencies. \cite{Gabor2013} focus, instead, on isolated high-redshift gas-rich galaxies. They found that  
a wide range of SFRs is possible if the BHAR is low, because such low rates are characterized by high variability  
driven by the structure of the interstellar medium and by AGN feedback. 
\cite{2014MNRAS.437.3373N} developed a semi-analytic model where BHs grow only during star-formation bursts caused by galaxy mergers. 
They naturally explained the 
lack of correlation between SFR and BHAR at low AGN luminosities as the measured SFR in such phases is being polluted by secular SF that occurred before the
 burst and hence is
unrelated to the AGN activity. High luminosity AGN, on the other hand, are observed at times that are close to the peak of the BH accretion event. In this case the 
measured SFR traces the merger-driven burst, concurrent with the merger-driven AGN-activity. 

Previous calculations of galaxy mergers with different mass ratios do not have the required spatial and time resolution to follow 
nuclear inflows and resolve the different timescales involving BH accretion and SF. Improving these resolutions would allow us to  
address, at the same time, the question of whether the time-dependent BHAR washes out an underlying correlation with SFR, and whether  
merging galaxies behave differently from quiescent ones. 

The calculations presented in this paper focus on 1) the temporal correlation between SFR and BHAR, 2) the time variability of SFR and BHAR,
3) the relative growth of stellar mass and BH mass before, during and after a merger;
and 4) the relative magnitude of SFR and BHAR through all the phases of the merger event.
The purpose is to address the assembly of stellar and BH mass, and the establishment of scaling relations. 
We take both the theorists' view, asking  if an underlying correlation between SFR and BHAR exists, and the observers' view, asking  
if a putative underlying correlation between SFR and BHAR can be measured.  
The simulations represent a major improvement in this direction. Our new suite of hydrodynamical simulations provides
very high spatial and temporal resolution (gas mass of $\sim 5 \times 10^3$~M$_{\odot}$, softening length of 20 pc for gas and 5~pc for the BHs, 
BH properties output every 0.1~Myr), a large range of initial mass ratios (1:1 to 1:10), several orbital configurations, and various gas fractions. 
In particular, we keep our time, masses and spatial resolution very high throughout the entire merger process
and are able to evolve the galaxies for a long time before and after the merger proper. This means that we are capturing the properties of galaxies in quiescence 
(hereafter `stochastic') phases and between the `merger' and the re-establishment of quiescence (hereafter `remnant' phase).
The main limitation of our suite is that it does not allow to simulate large galaxies. Each of our mergers requires $\sim 10^7$ particles 
and the entire suite required  $\sim 10^8$ particles. The total equivalent simulated time amounted to $\sim$30 Gyr of evolution.  

The structure of the paper is as follows:
In \S2 we present the numerical setup. In \S3 we describe the general behaviour of a typical merger and in \S4 we discuss the temporal correlation between SFR and BHAR, while  in \S5 we compare the time variability of SFR and BHAR. 
In \S6 we study the relative growth of stellar mass and BH mass; and in \S7 we explain the various BH and SF relationships extracted from the simulations. In \S8 we compare the relationship between BH and SF to observations.

\section{Numerical Setup}\label{sec:Numerical_setup}

The numerical setup includes a suite of hydrodynamical simulations applied to mergers of disc galaxies with mass ratios of 1:1, 1:2, 1:4, 1:6, and 1:10.
The chosen redshift, $z=3$, corresponds to the peak of the cosmic merger rate. The calculations and main results are presented below and the
Appendix adds the necessary information about the dependences on the numerical resolution and the assumed strength of AGN feedback.

\begin{table} \centering
\vspace{5pt}
{\small
\begin{tabular*}{0.48\textwidth}{ccccc}
\hline
Name & Mass ratio ($q$) & $\theta_1$ & $\theta_2$  & gas fraction \\
\hline
m1.gf0.3.pro & 1:1 & 0 & 0 & 0.3 \\
m2.gf0.3.pro & 1:2 & 0 & 0 & 0.3 \\
m2.gf0.3.incl & 1:2 & $\pi/4$ & 0  & 0.3 \\
m2.gf0.3.retprim & 1:2 & $\pi$ & 0 & 0.3 \\
m2.gf0.3.retsec & 1:2 & 0 & $\pi$ & 0.3 \\
m2.gf0.6.pro & 1:2 & 0 & 0 & 0.6 \\
m4.gf0.3.pro & 1:4 & 0 & 0 & 0.3 \\
m4.gf0.3.incl &1:4 & $\pi$/4 & 0 & 0.3 \\
m6.gf0.3.pro &1:6 & 0 & 0 & 0.3 \\
m10.gf0.3.pro &1:10 & 0 & 0 & 0.3 \\
\hline
\end{tabular*}
}
\caption[Simulation Parameters]{Parameters for our simulations. $\theta_1$ and $\theta_2$ are the angles between the spin axis and the total orbital angular momentum axis for each galaxy. $q$ is the initial mass ratio between the merging galaxies. 
\label{tab:params}}
\end{table}

\begin{table*} \centering
\vspace{-3.5pt}
\caption[Galactic simulation parameters]{Main galactic parameters at the beginning of the simulation. (1) Galaxy (primary -- G1 or secondary -- G2) and merger. (2)~Virial mass. (3)~Stellar bulge mass. (4)~Stellar disc mass. (5)~Gas disc mass. (6)~Disc scale radius. (7)~BH mass. (8)~dark matter  particle mass. (9)~dark matter  particle softening length. The disc mass is the sum of the stellar disc mass and the gas disc mass. The stellar bulge scale radius and the disc scale height are always equal to $0.2\,r_{\rm disc}$ and $0.1\,r_{\rm disc}$, respectively. All other parameters are the same for all galaxies and all mergers: gas and stellar particle mass ($4.6 \times 10^{3}$ and $3.3 \times 10^{3}\; {\rm M}_{\odot}$, respectively) and softening (20 and 10~pc, respectively); BH softening (5~pc); dark matter  halo spin and concentration parameters ($\lambda=0.04$ and $c=3$, respectively); and redshift ($z=3$).
\label{agn2014:tab:galactic_params}}
\vspace{5pt}
{\small
\begin{tabular*}{0.96\textwidth}{cccccccccc}
\hline
Galaxy & $M_{\rm vir}$ & $M_{\rm stell.\,bulge}$ & $M_{\rm stell.\,disc}$ & $M_{\rm gas.\,disc}$ & $r_{\rm disc}$ & $M_{\rm BH}$ & $M_{\rm DM\,part.}$ & $\epsilon_{\rm DM\,part.}$ \T \B \\
{[}Merger] & [$10^{11} \hbox{M}_{\odot}$] &  [$10^{9} \hbox{M}_{\odot}$] &  [$10^{9} \hbox{M}_{\odot}$] &  [$10^{9} \hbox{M}_{\odot}$] & [kpc] & [$10^{6} \hbox{M}_{\odot}$] &  [$10^{5} \hbox{M}_{\odot}$] & [pc] \B \\
\hline
G1 [1:1, 1:2, 1:4 low-gas-frac] & $2.21$ & $1.77$ & $6.19$ & $2.65$ & 1.13 & $3.53$ & $1.1$ & 30 & \T \B \\
G1 [1:2 high-gas-frac]              & $2.21$ & $1.77$ & $3.54$ & $5.30$ & 1.13 & $3.53$ & $1.1$ & 30 & \B \\
G1 [1:6]                                    & $2.21$ & $1.77$ & $6.19$ & $2.65$ & 1.13 & $3.53$ & $0.8$ & 27 & \B \\
G1 [1:10]                                  & $2.21$ & $1.77$ & $6.19$ & $2.65$ & 1.13 & $3.53$ & $0.5$ & 23 & \B \\
\hline
G2 [1:2 low-gas-frac]               & $1.11$ & $0.88$ & $3.09$ & $1.33$ & 0.90 & $1.77$ & $1.1$ & 30 & \T \B \\
G2 [1:2 high-gas-frac]              & $1.11$ & $0.88$ & $1.77$ & $2.65$ & 0.90 & $1.77$ & $1.1$ & 30 & \B \\
G2 [1:4]                                    & $0.55$ & $0.44$ & $1.55$ & $0.66$ & 0.71 & $0.88$ & $1.1$ & 30 & \B \\
G2 [1:6]                                    & $0.37$ & $0.30$ & $1.03$ & $0.44$ & 0.62 & $0.59$ & $0.8$ & 27 & \B \\
G2 [1:10]                                  & $0.22$ & $0.18$ & $0.62$ & $0.27$ & 0.52 & $0.35$ & $0.5$ & 23 & \B \\
\hline
\end{tabular*}
\vspace{5pt}
}
\end{table*}

\subsection{Orbital configuration}\label{sec:Orbital_parameters}

We chose an orbital configuration that matches those of the most common halo mergers in cosmological simulations of galaxy formation \citep{Benson05}, where almost half of all mergers have an eccentricity $e$ between 0.9 and 1.1. \citet{Khochfar2006} find that 85 percent of merging halo orbits have initial pericentre distances in excess of 10 percent of the virial radius of $G_1$ ($G_1$ and $G_2$ are the larger and smaller galaxies, respectively). Most simulations of galaxy mergers consider smaller pericentre distances, to save computational time, producing more direct collisions. 
Instead, we set the initial pericentre distance near 20 percent of the virial radius of $G_1$, in order to be consistent with cosmological orbits. The initial separation between the galaxies is set near the sum of the two virial radii. We summarize the orbital configuration for each simulation in Table \ref{tab:params}.

We vary the angle between each galaxy's angular momentum axis and the overall orbital angular momentum vector, given by $\theta$ in Table \ref{tab:params}. We consider coplanar, prograde-prograde mergers, in which $\theta_1$ and $\theta_2$, the angles for $G_1$ and $G_2$, respectively, are both zero. In our inclined mergers, we set $\theta_1=\pi/4$ and $\theta_2=0$. Lastly, we consider coplanar, retrograde mergers, in which one of the galaxies is anti-aligned with the overall orbital angular momentum axis. In the coplanar, retrograde-prograde merger, $\theta_1 = \pi$ and $\theta_2 = 0$.  In the coplanar, prograde-retrograde merger, $\theta_1=0$ and $\theta_2 = \pi$.

\subsection{Galaxies}\label{sec:Galaxies}

All galaxies are composite systems of dark matter, gas, stars, and a central BH (described in the next section). See Table~\ref{agn2014:tab:galactic_params} for a complete list. Most of this description follows \citet{Springel_White_1999} and \citet{springel2005b}.  Most values in this section were chosen for consistency with previous work \citep{Callegari2009,Callegari2011,VW2012,VW2014} and in Table~\ref{agn2014:tab:galactic_params} we report the complete list of their properties and those of their central BHs (described in the next section). The dark matter halo is described by a spherical Navarro-Frenk-White profile \citep{NFW1997} with spin parameter $\lambda=0.04$. The dark matter halo concentration parameter is initialized to $c=3$. The disc has an exponential density profile with total mass equal to 4 percent of the virial mass of the galaxy. The disc scale radius $r_{\rm disc}$ is then determined by imposing conservation of specific angular momentum of the material that forms the disc, whereas the disc scale height $z_{\rm disc}$ is set to be 10 per cent of $r_{\rm disc}$. The gas in the disc has a mass fraction $f_{\rm gas}=0.3$ or $f_{\rm gas}=0.6$.  The stellar bulge is described by a spherical \citet{Hernquist1990} density profile with total mass equal to 0.8 percent of the virial mass of the galaxy. In each merger, $G_1$ has a virial mass of $2.24 \times 10^{11}$~M$_{\odot}$ (consistent with \citealt{adelberger2005b}), and, consequently, a bulge mass of $1.77 \times 10^{9}\; {\rm M}_{\odot}$, a disc mass of $8.84 \times 10^{9}\; {\rm M}_{\odot}$, and a disc scale radius of 1.13~kpc. The mass and all the other properties of $G_2$ scale according to the mass ratio. 

Stellar and gas particles initially have the same particle mass ($3.3 \times 10^{3}$ and $4.6 \times 10^{3}\; {\rm M}_{\odot}$, respectively) and softening length (10 and 20~pc, respectively) in all the ten mergers of the suite. In order to limit excursions of BHs from the centre of each galaxy, we impose the dark matter particles to have a mass smaller than 15 per cent of that of the smaller BH in each merger. For this reason, the mass and softening length of dark matter  particles in the 1:1, 1:2, and 1:4 mergers were set to $1.1 \times 10^{5}\; {\rm M}_{\odot}$ and 30~pc, respectively. In the other mergers, on the other hand, because of the much lower mass of the secondary BH, dark matter  particle masses and softening lengths were lowered accordingly (1:6 merger: $8 \times 10^{4}\; {\rm M}_{\odot}$ and 27~pc; 1:10 merger: $5 \times 10^{4}\; {\rm M}_{\odot}$ and 23~pc). 

Each galaxy is initialized with solar metallicity and a uniform stellar population with an age of 2~Gyr to reflect the young age of the Universe at $z=3$. 
Without any feedback to heat the gas at the beginning of the simulation, much of the gas initially cools and forms stars. To avoid an unphysical burst of supernovae 
at the beginning of our merger simulations, we evolve the galaxies in isolation over $\sim$100~Myr (relaxation period), during which the  SF efficiency is gradually increased, by 50 per cent every $3\times 10^4$~yr, up to the value $c^*=0.015$, in order to obtain galaxies that start the main part of the simulation from the $z=3$ sequence of star-forming galaxies \citep{elbaz2007} and that obey the Kennicutt-Schmidt relation. Data on SFR is extracted every 1~Myr.

We performed all our simulations using the N-body SPH code {\scshape gasoline} \citep{wadsley04}, an extension of the pure gravity tree code {\scshape pkdgrav} \citep{stadel01}. {\scshape gasoline} includes explicit line cooling for atomic hydrogen, helium and metals, as well as a physically motivated prescription for SF, supernova feedback and stellar winds \citep{Stinson2006}. In particular, stars are allowed to form if the parent gas particle is colder than 6000~K and denser than 100~cm$^{-3}$, and supernovae release $10^{51}$~erg into the surrounding gas, according to the blast wave formalism of \citet{Stinson2006}.

\subsection{Black holes}\label{sec:Black_holes}

A recent implementation in the {\scshape gasoline} code has been the inclusion of a recipe for BH physics \citep{Bellovary10}, in which BHs are implemented as sink particles that accrete from nearby gas particles according to an Eddington-limited Bondi--Hoyle--Littleton accretion formula.  In order to realistically model accretion from an inhomogeneous mix of hot and cold gas particles around the BH, the accretion rate is computed as the sum of the Bondi accretion rate of each individual gas particle near the BH, including the relative velocity with respect to the BH, rather than simply averaging the gas quantities over all the neighboring particles. This method allows the accretion rate to be weighted more heavily by nearby, cold, dense gas particles (and less by more distant, hot ones, or particles moving fast with respect to the BH) rather than treating them all equally. Additional information is provided in Capelo et al. (2014).

BH accretion gives rise to feedback, implemented as thermal energy injected into the nearest gas particle according to $\dot{E} = \epsilon_{\rm f} \epsilon_{\rm r} \dot{M}_{\rm BH} c^2$, where $c$ is the speed of light in vacuum, $\epsilon_{\rm r}=0.1$ is the radiative efficiency and $\epsilon_{\rm f}$ is the AGN feedback efficiency, chosen to be equal to 0.001, which is lower than other numerical implementations \citep[see ][for a review]{2014MNRAS.443.1125T} to match the local M$_{\rm BH}$-M$_{\rm bulge}$ relation over the galaxy evolution  (see the Appendix for a discussion of how the results depend on the feedback strength).

We place a single BH at the centre of each galaxy, after the galaxy has been initialized. Its mass, $M_{\rm BH}=2 \times 10^{-3} M_{\rm Bulge}$,  is set according to the local M$_{\rm BH}$-M$_{\rm bulge}$ relation \citep{MarconiHunt2003}. The mass of the primary BH ($BH_1$) in each simulation is initially set to $3.53 \times 10^6$~M$_{\odot}$, whereas $BH_2$ has a mass proportional to the mass ratio between the galaxies, producing a minimum initial mass of $3.53 \times 10^5$~M$_{\odot}$ in the 1:10 merger. The softening length of all BHs is set to 5~pc, regardless of their mass. Data on BHAR is extracted every 0.1~Myr.   The distance between a BH and the local centre of mass remains small throughout the simulation. The mean distance between the local centre of mass and the BH itself is of the same order as the gravitational softening of the stellar particles, 10 pc.

\begin{figure}
\centering
\vspace{5pt}
\includegraphics[width=1.0\columnwidth,angle=0]{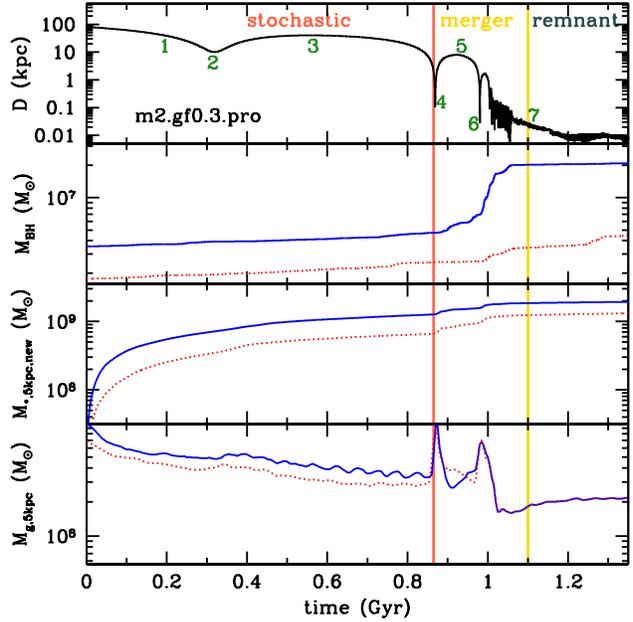}
\vspace{5pt}
\caption{Merger properties as a function of time for a 1:2 coplanar, prograde-prograde merger (m2.gf0.3.pro). 
The entire process is divided into three phases: stochastic, merger and remnant (see text for details).
First panel: BH separation. 
 Second panel: masses of the two BHs. $BH_1$ (blue solid line) and $BH_2$ (red dotted line).  
Third panel:  cumulative new stellar mass in the central 5~kpc of $G_1$ (blue solid line) and $G_2$ (red, dotted line). 
Fourth panel: gas mass in the central $5$~kpc of $G_1$ (blue, solid line) and $G_2$ (red dotted line). 
The first seven snapshots of the simulation of Fig.~\ref{fig:stellar_and_gas_density_snapshots_m2} are marked in green.
The comparison between BH growth and SFR is shown in Fig. ~\ref{fig:sfravebharall}.}
\label{fig:4panelpro}
\end{figure}

\section{General merger behaviour}\label{sec:General}
We present the behaviour of one of our mergers in
Fig.~\ref{fig:4panelpro}.  The reference case is m2.gf0.3.pro, a 1:2 merger, and the 
differences with other mass ratios, orbital configurations and gas content are discussed at the end of this section. 

\begin{figure*}
\centering
\begin{overpic}[width=0.3\textwidth,angle=0]{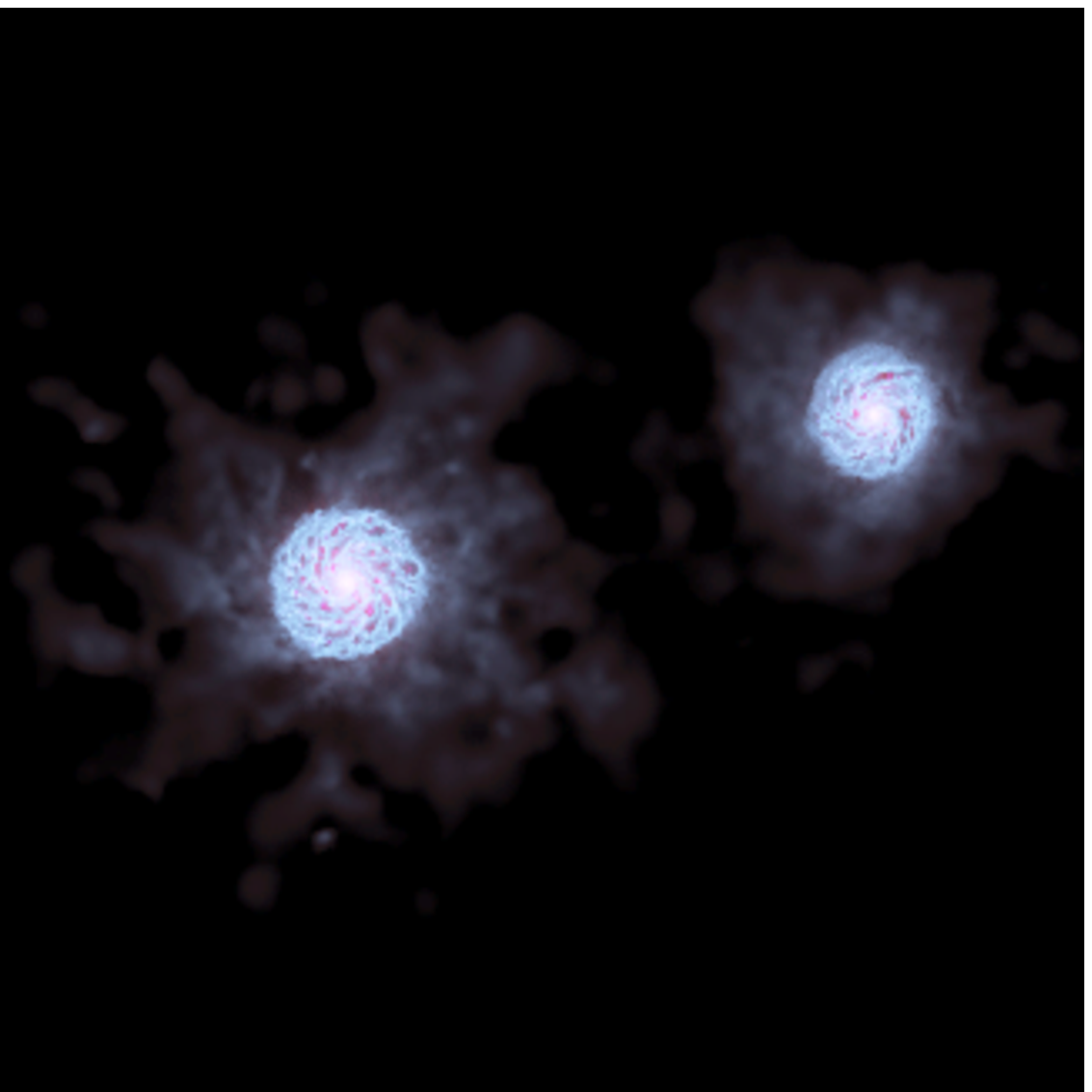}
\put (2,8) {\textcolor{white}{$1$}}
\put (20,8) {\textcolor{white}{First approach}}
\end{overpic}
\hskip -1mm
\begin{overpic}[width=0.30\textwidth,angle=0]{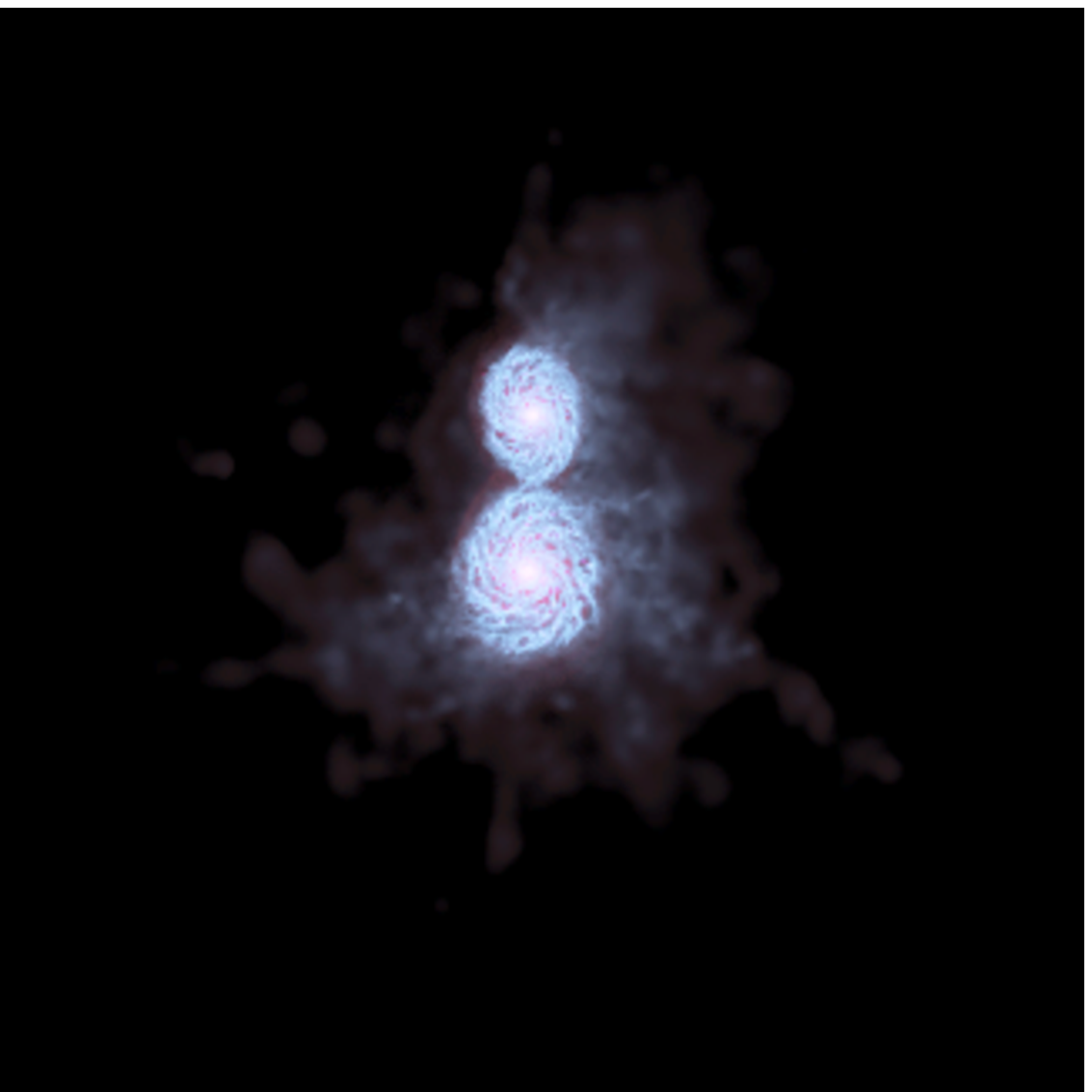}
\put (2,8) {\textcolor{white}{$2$}}
\put (20,8) {\textcolor{white}{First pericentre}}
\end{overpic}
\hskip -1mm
\begin{overpic}[width=0.3\textwidth,angle=0]{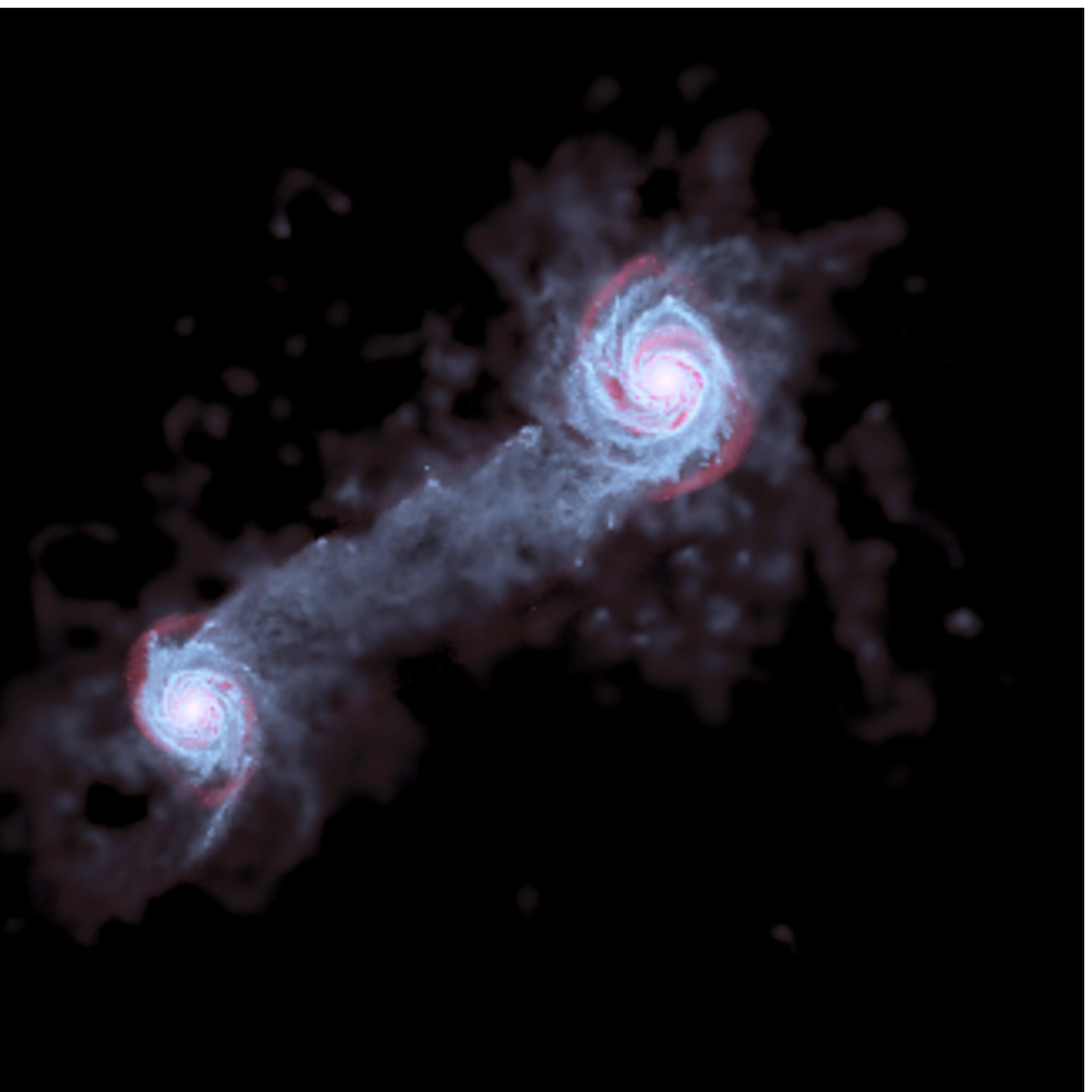}
\put (2,8) {\textcolor{white}{$3$}}
\put (26,8) {\textcolor{white}{First apocentre}}
\end{overpic}
\vskip -1.0mm
\begin{overpic}[width=0.3\textwidth,angle=0]{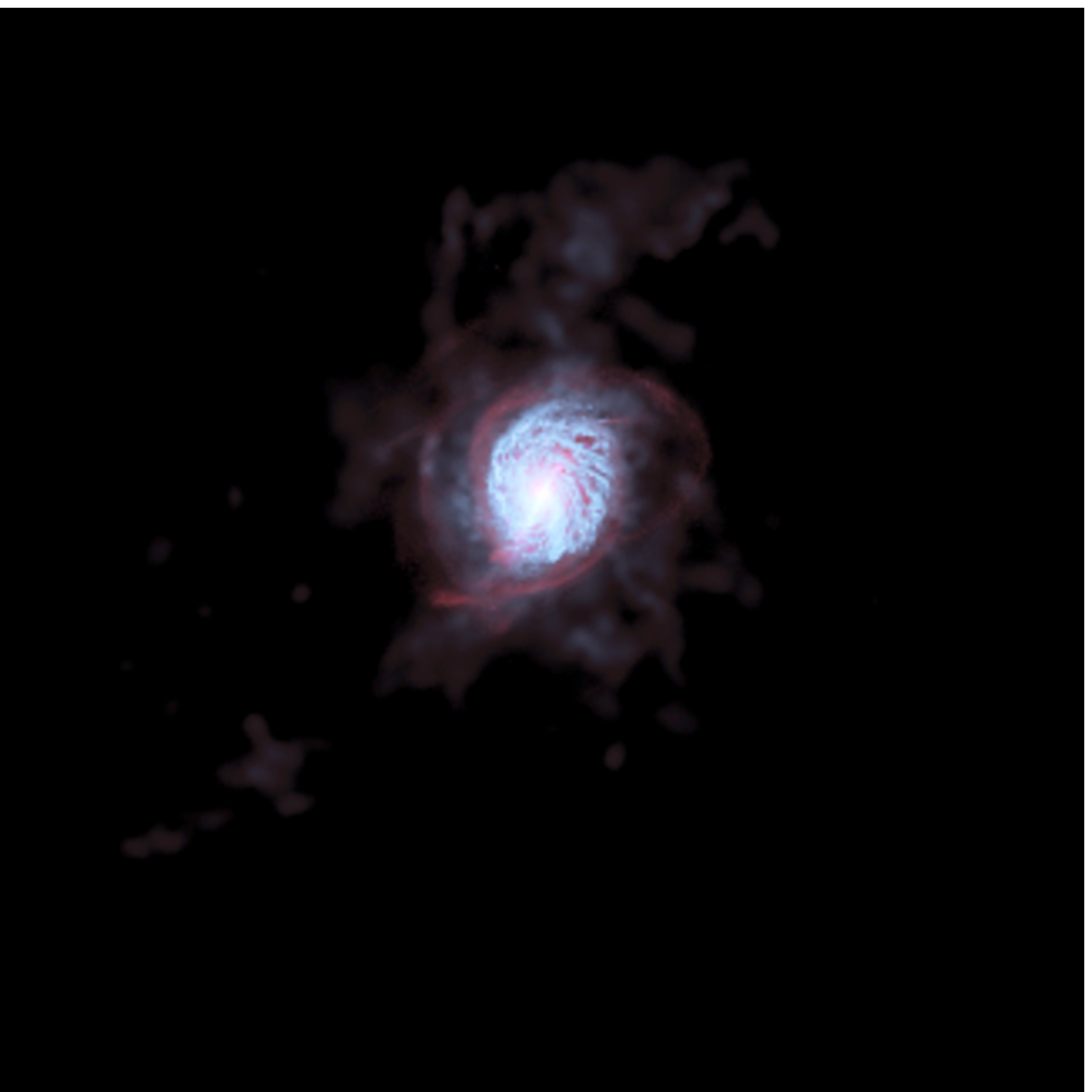}
\put (2,9) {\textcolor{white}{$4$}}
\put (8,9) {\textcolor{white}{End of the stochastic stage}}
\end{overpic}
\hskip -1mm
\begin{overpic}[width=0.3\textwidth,angle=0]{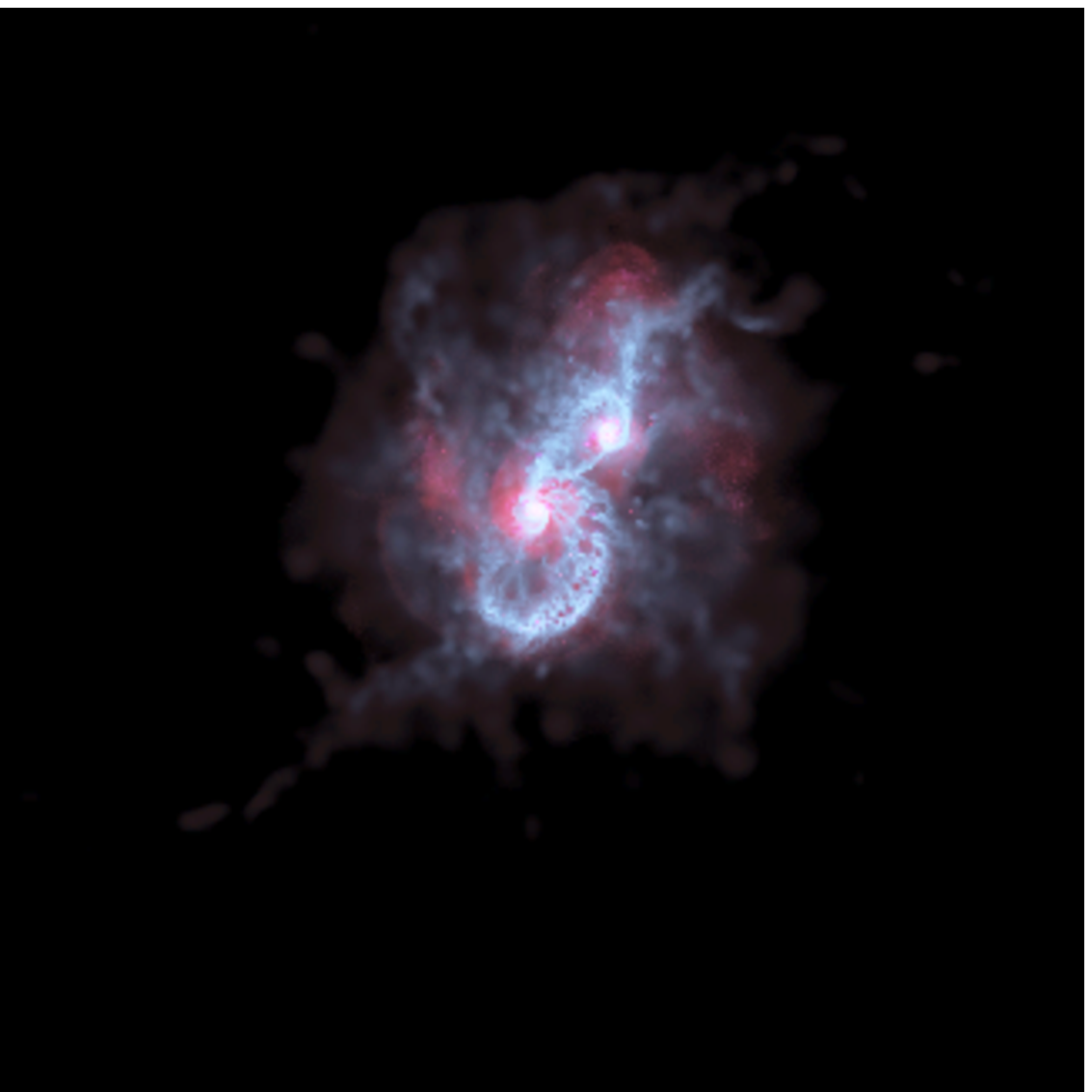}
\put (2,9) {\textcolor{white}{$5$}}
\put (24,9) {\textcolor{white}{Second apocentre}}
\end{overpic}
\hskip -1mm
\begin{overpic}[width=0.3\textwidth,angle=0]{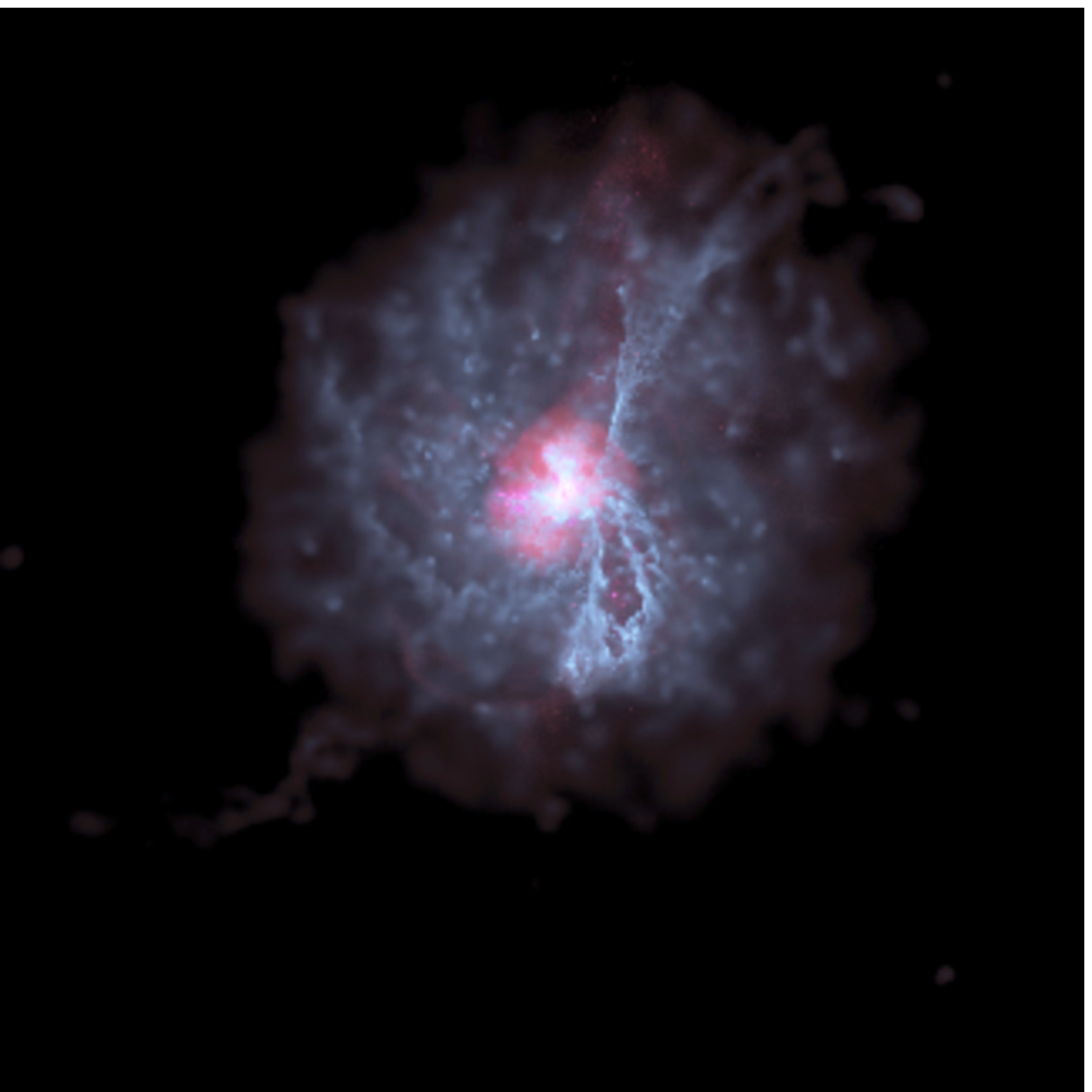}
\put (2,9) {\textcolor{white}{$6$}}
\put (26,9) {\textcolor{white}{Third pericentre}}
\end{overpic}
\vskip -1.0mm

\begin{overpic}[width=0.3\textwidth,angle=0]{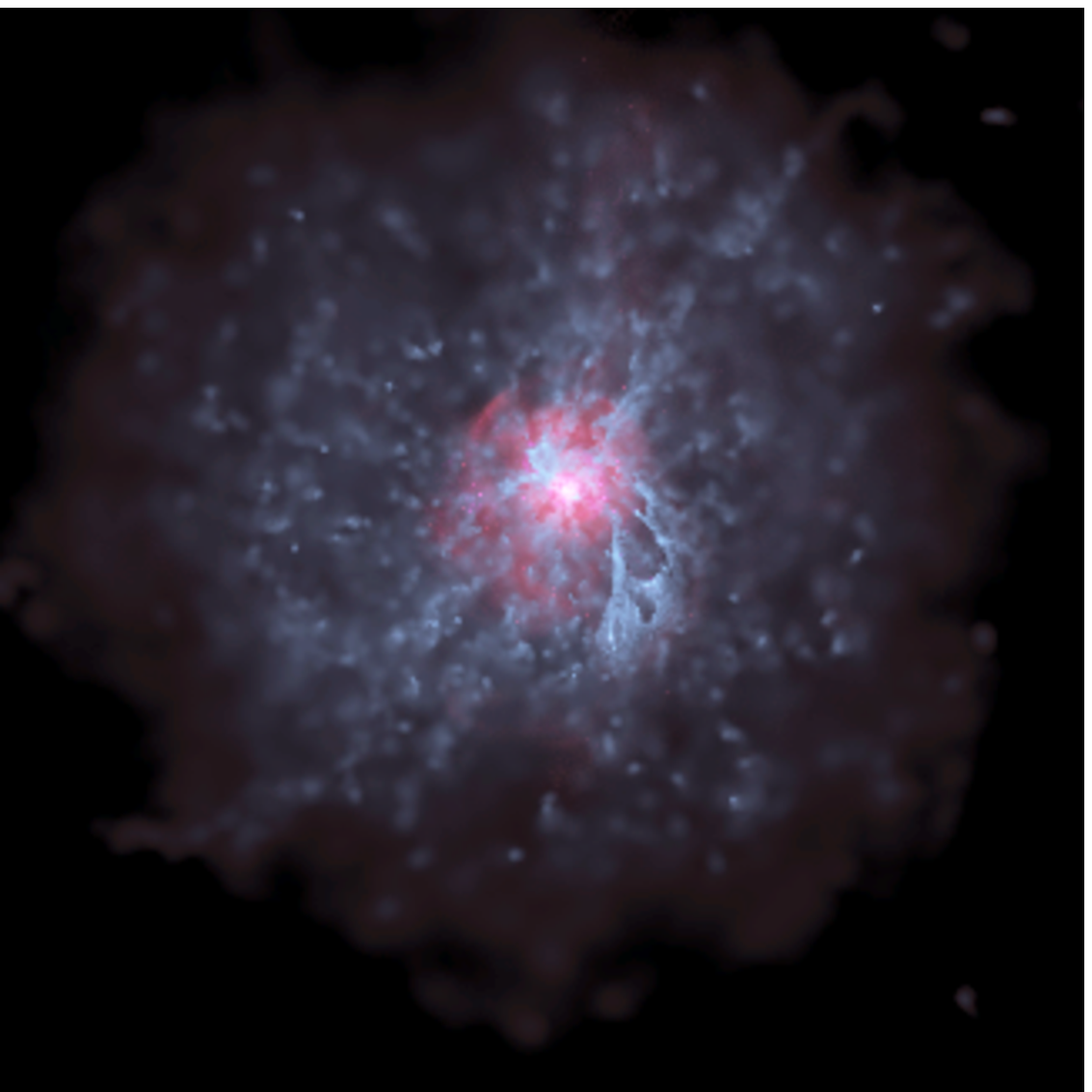}
\put (2,8) {\textcolor{white}{$7$}}
\put (26,8) {\textcolor{white}{Fifth apocentre}}
\end{overpic}
\hskip -0.8mm
\begin{overpic}[width=0.301\textwidth,angle=0]{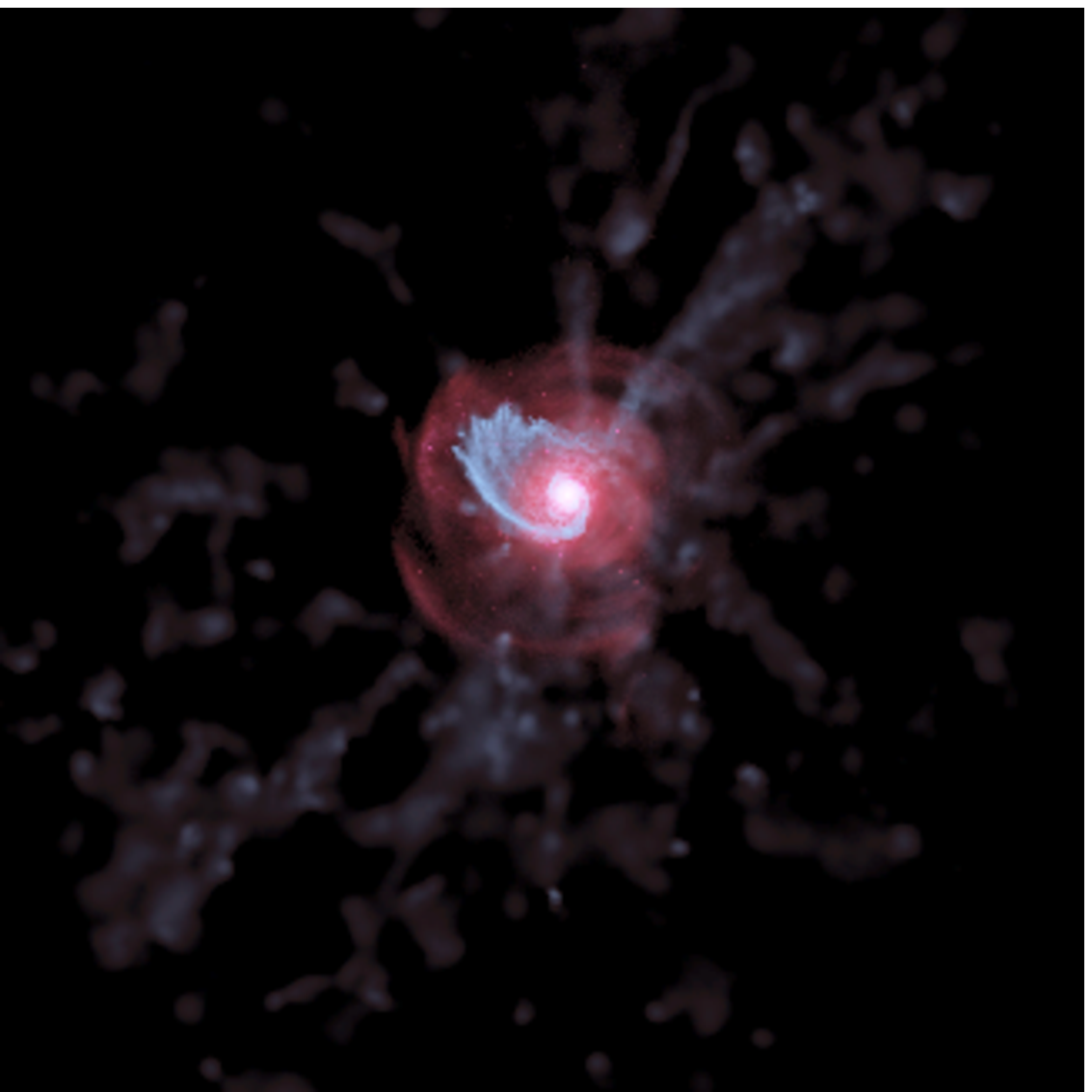}
\put (2,8) {\textcolor{white}{$8$}}
\put (8,8) {\textcolor{white}{End of the merger stage}}
\end{overpic}
\hskip -1.1mm
\begin{overpic}[width=0.299\textwidth,angle=0]{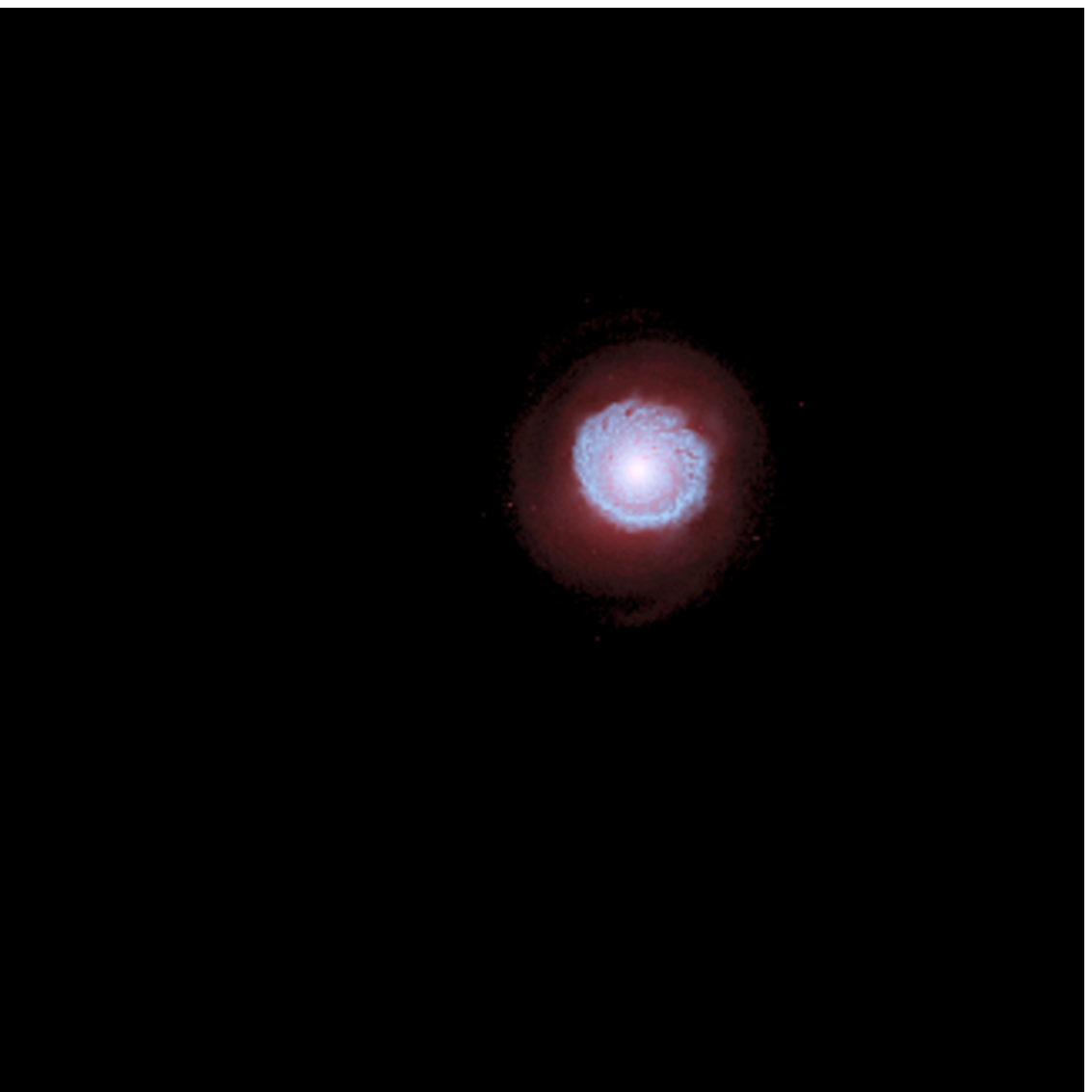}
\put (2,8) {\textcolor{white}{$9$}}
\put (10.5,8) {\textcolor{white}{End of the remnant stage}}
\end{overpic}
\caption[Stellar (red) and gas density snapshots]{Face-on stellar (red) and gas (blue) density snapshots at representative times, in Gyr, of the 1:2 coplanar, 
prograde--prograde merger: (1) 0.20, (2) 0.32 (first pericentric passage), (3) 0.55 (first apocentric passage), (5) 0.85 (second pericentric passage -- end of the 
stochastic stage), (6) 0.92 (second apocentric passage), (7) 0.98 (fifth apocentric passage), (8) 1.1 (end of the merger stage), (9) 2 (end of the remnant stage), 
respectively. The image size is 75x75 kpc. The gas density is over-emphasized with respect to stellar density in order to make the gas more visible.}
\label{fig:stellar_and_gas_density_snapshots_m2}
\end{figure*}

We divide each merger into three
phases, that we dub `stochastic', `merger' proper, and
`remnant'. The definition we adopt is based on the behaviour of the
specific angular momentum in shells within 1~kpc from the galaxy
centre (see Capelo et al. 2014 for details). The stochastic phase
lasts until the second pericentric passage, when the galaxies
enter in close contact. During this phase the galaxies behave as they do in
isolation. This phase is characterised by a non-evolving specific
angular momentum. 
The merger phase starts at the
second pericentre, when the specific angular momentum drops abruptly,
because of strong dynamical torques. 
This phase ends when the specific angular momentum
returns to be constant in time, specifically, as the first time after
the second pericentric passage when the relative change of specific
angular momentum over time increments of 0.05~Gyr is less than 0.3, as
in Capelo et
al. 2014. 
 The remnant phase lasts from this moment until the end of the simulation. 
We stop when the remnant phase has reached the same duration as 
the stochastic phase.
Fig.~\ref{fig:stellar_and_gas_density_snapshots_m2} shows
snapshots of the galaxies at different times. The
three bottom panels of Fig.~\ref{fig:4panelpro} highlight the differences 
in the evolution of gas, SF and BH evolution in the three phases.

In the stochastic phase, which, as discussed above, represents also isolated galaxies not involved in mergers, the gas content within 5 kpc steadily decreases because of its consumption by SF. 
Concurrently the mass-growth of the BHs is smooth and limited, albeit non-zero: the primary BH grows by $\sim 1.2\times 10^6 \msun$ in 0.85 Gyr, the secondary by $\sim 6.3\times 10^5 \msun$.  

When the merger phase starts,  strong gas inflows reach the centres of both galaxies (see Van Wassenhove et al. 2014, and Capelo et al. 2014 for details), enhancing SFR and BHAR. The highest peaks of gas inflows, SFR and BHAR coincide with the second and third pericentres. 

After the merger proper ends,  the behaviour in the remnant phase is initially erratic, partly because of feedback effects, and partly because the galaxies are still disturbed. 
At later times the conditions return to be similar to what they were in the stochastic phase, although with a somewhat higher BHAR. 

\begin{figure}
\centering
\includegraphics[width=1.0\columnwidth,angle=0]{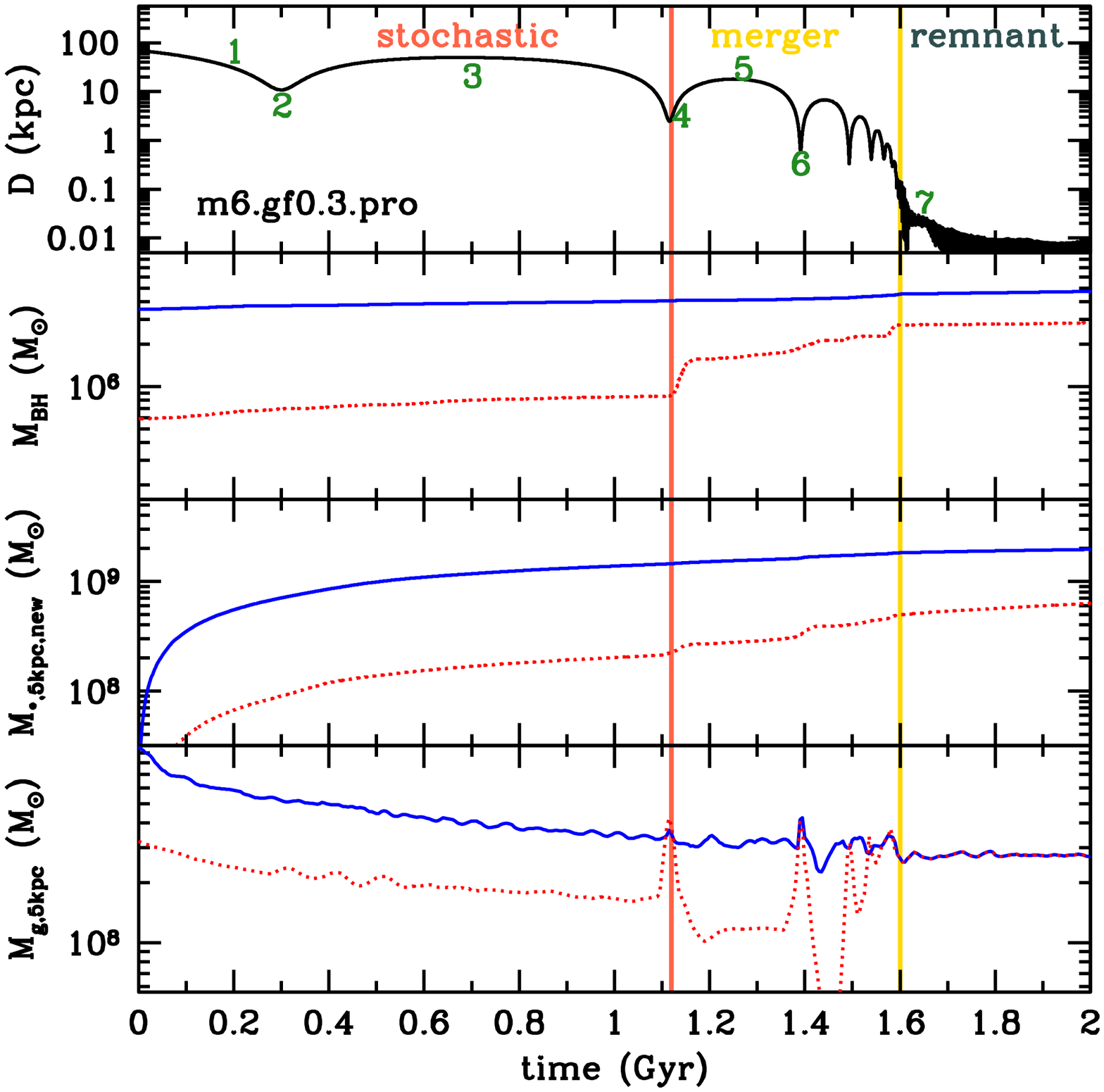}
\caption{As figure \ref{fig:4panelpro} but for the 1:6 coplanar, prograde-prograde merger (m6.gf0.3.pro).  The first seven snapshots of the simulation of Fig.~\ref{fig:stellar_and_gas_density_snapshots_m6} are marked in green.
 The comparisons between BH growth and SFR are shown in Fig. ~\ref{fig:sfravebharmergerm6}.}
\label{fig:4panelpro_m6}
\end{figure}

\begin{figure*}
\centering
\begin{overpic}[width=0.3\textwidth,angle=0]{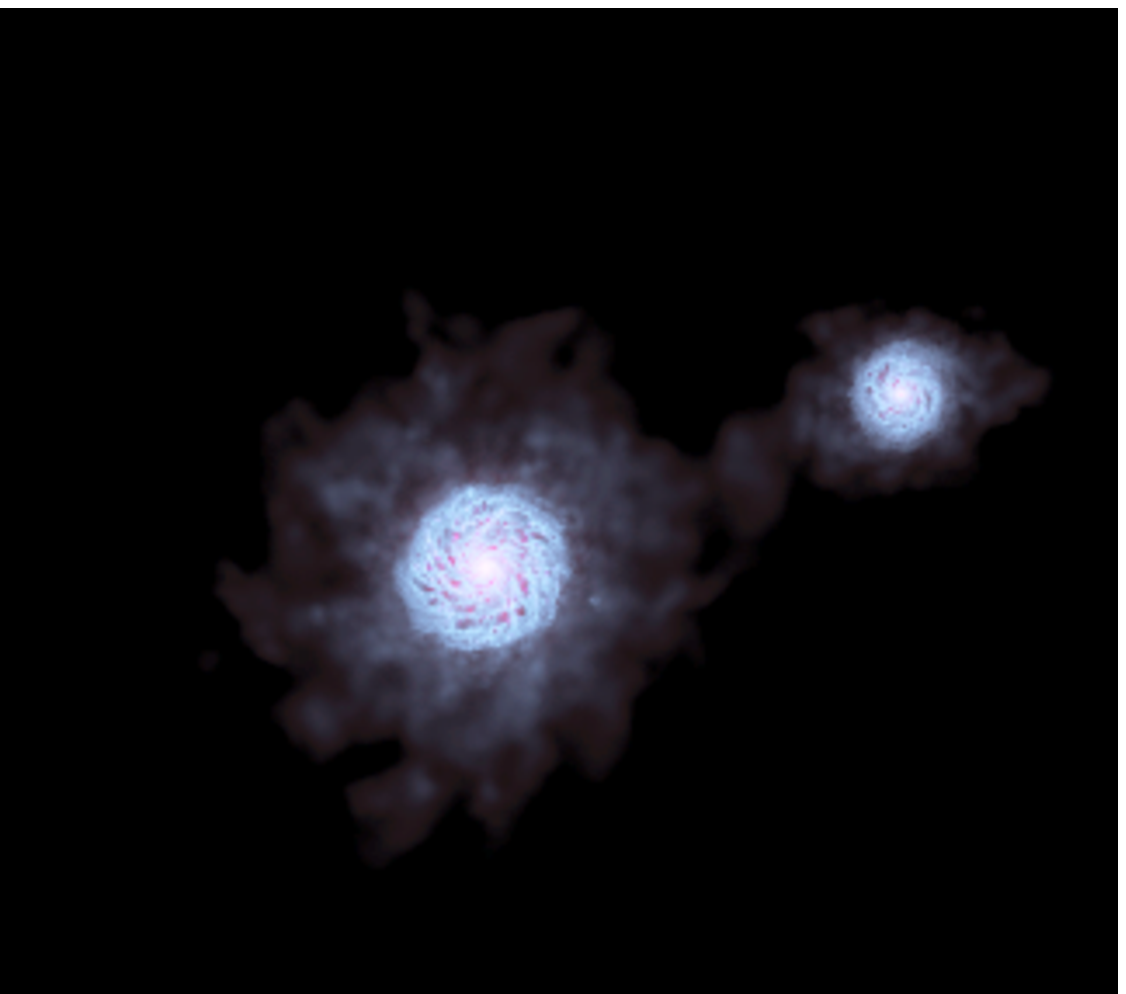}
\put (2,80) {\textcolor{white}{$1$}}
\put (20,80) {\textcolor{white}{First approach}}
\end{overpic}
\hskip -1mm
\begin{overpic}[width=0.301\textwidth,angle=0]{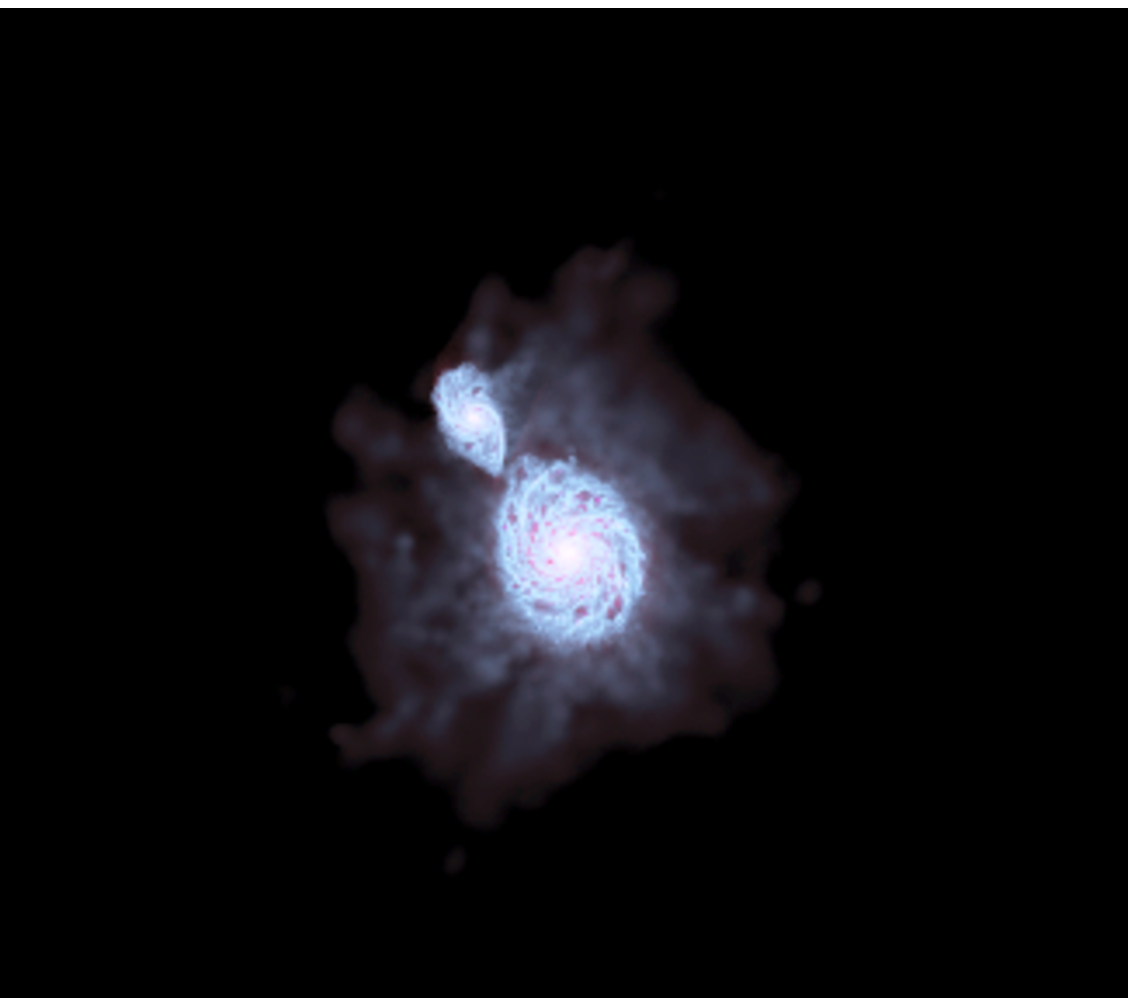}
\put (2,80) {\textcolor{white}{$2$}}
\put (20,80) {\textcolor{white}{First pericentre}}
\end{overpic}
\hskip -1mm
\begin{overpic}[width=0.299\textwidth,angle=0]{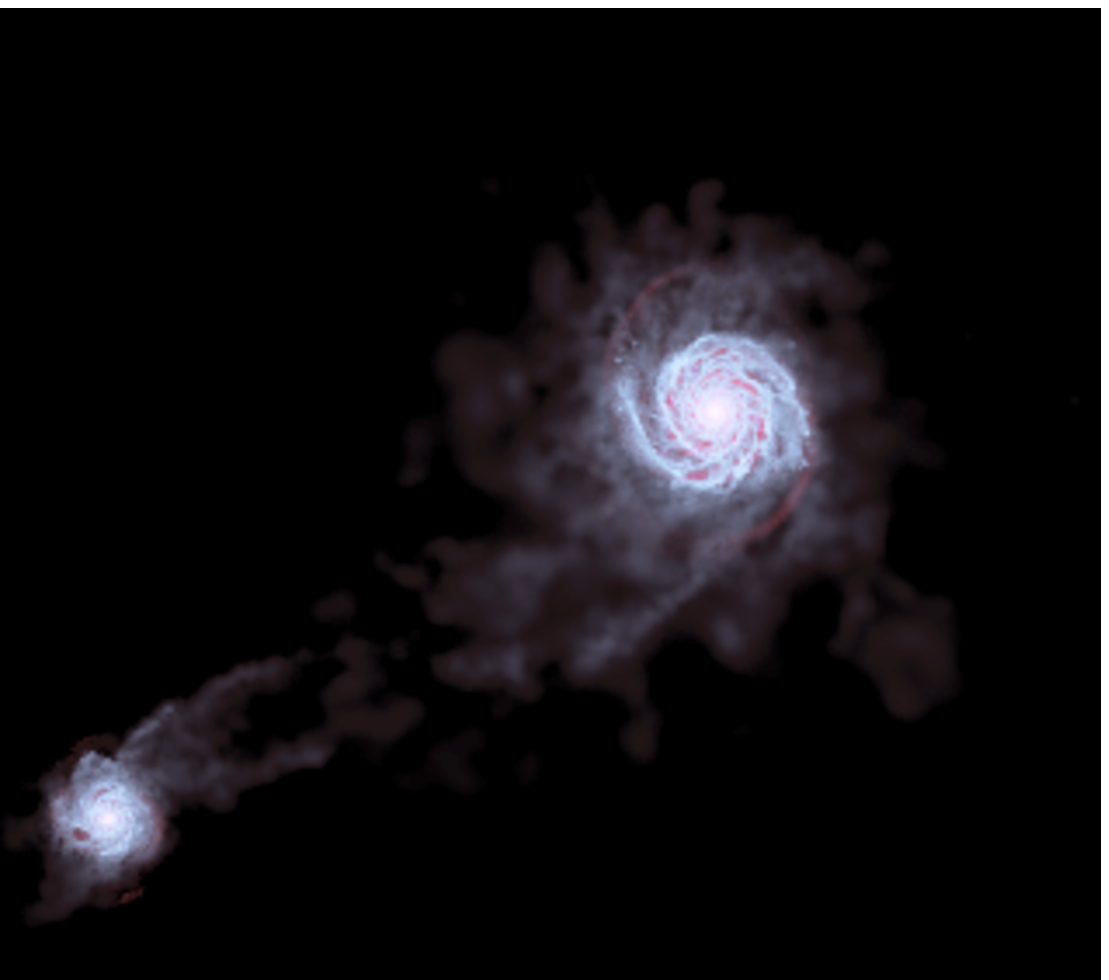}
\put (2,80) {\textcolor{white}{$3$}}
\put (26,80) {\textcolor{white}{First apocentre}}
\end{overpic}
\vskip -1.0mm
\begin{overpic}[width=0.3\textwidth,angle=0]{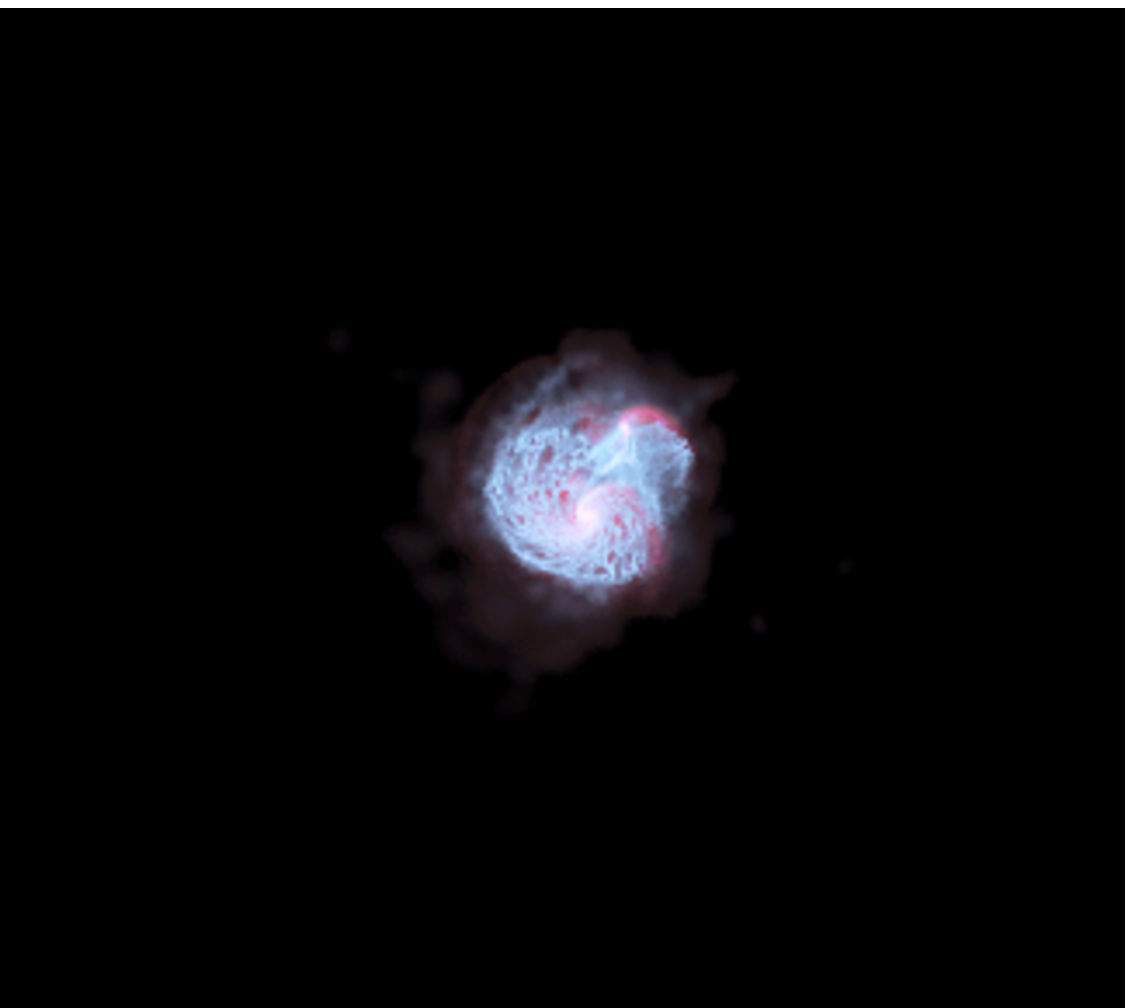}
\put (2,80) {\textcolor{white}{$4$}}
\put (8,80) {\textcolor{white}{End of the stochastic stage}}
\end{overpic}
\hskip -1mm
\begin{overpic}[width=0.3\textwidth,angle=0]{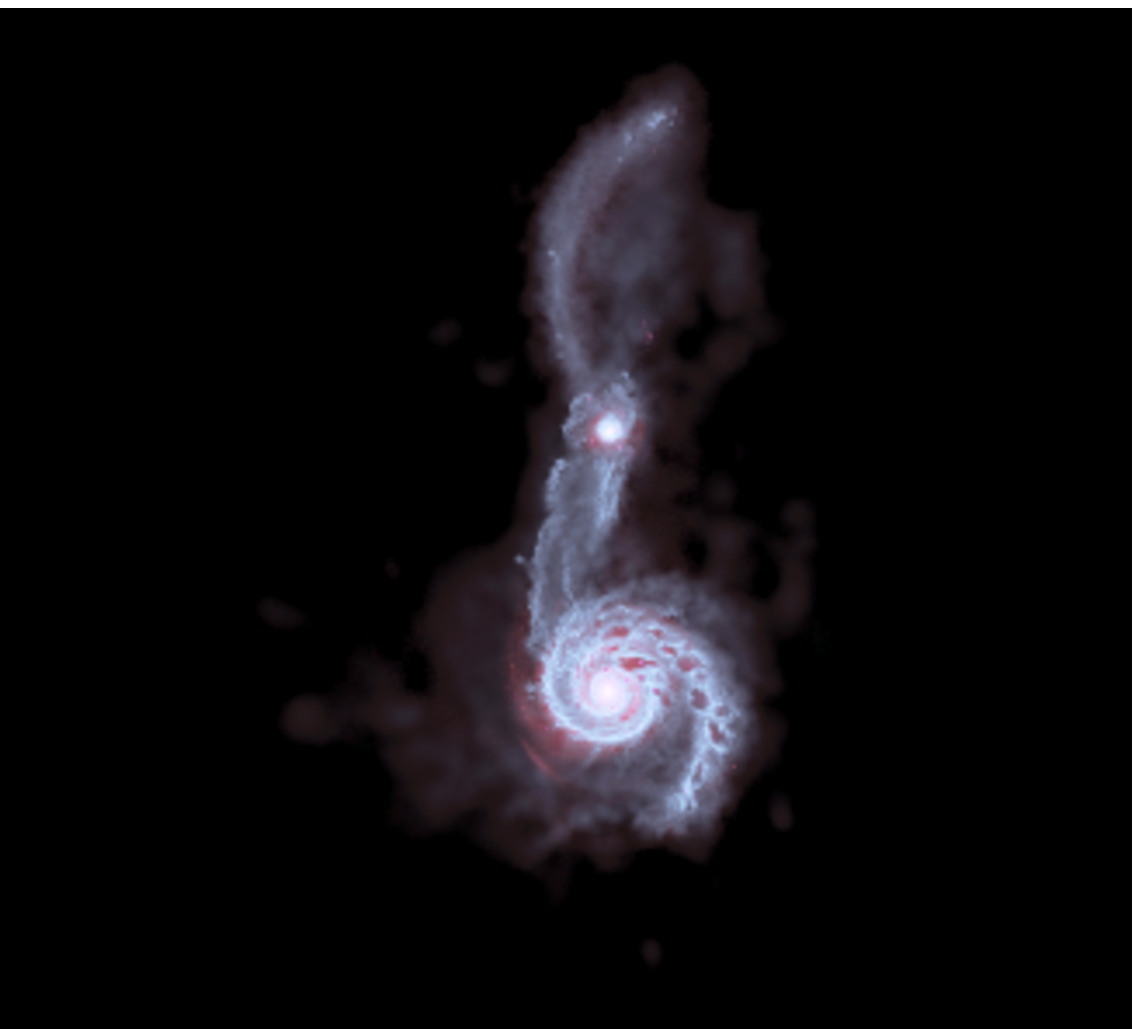}
\put (2,80) {\textcolor{white}{$5$}}
\put (24,80) {\textcolor{white}{Second apocentre}}
\end{overpic}
\hskip -1mm
\begin{overpic}[width=0.3\textwidth,angle=0]{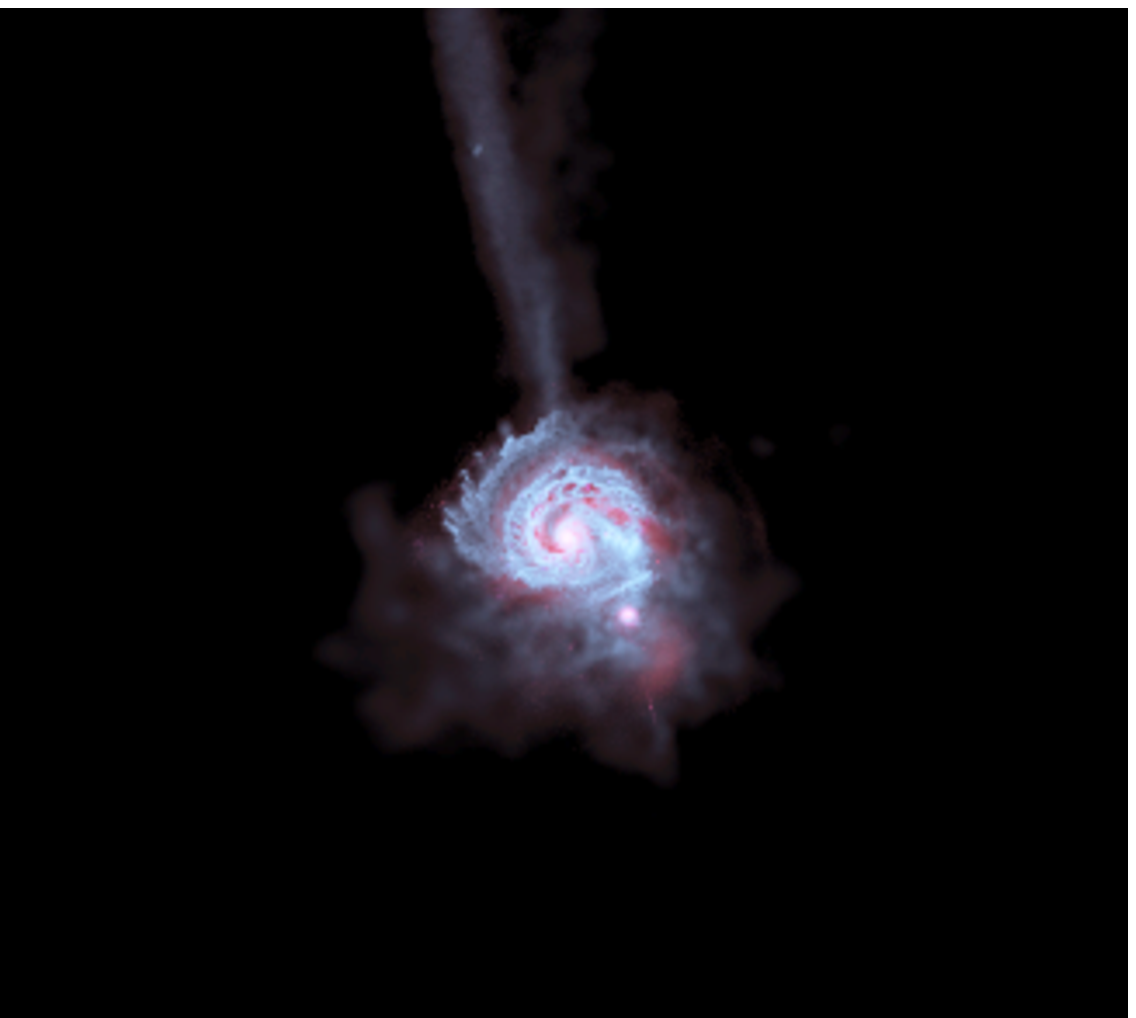}
\put (2,80) {\textcolor{white}{$6$}}
\put (26,80) {\textcolor{white}{Third pericentre}}
\end{overpic}
\vskip -1.0mm
\hskip 1mm
\begin{overpic}[width=0.3\textwidth,angle=0]{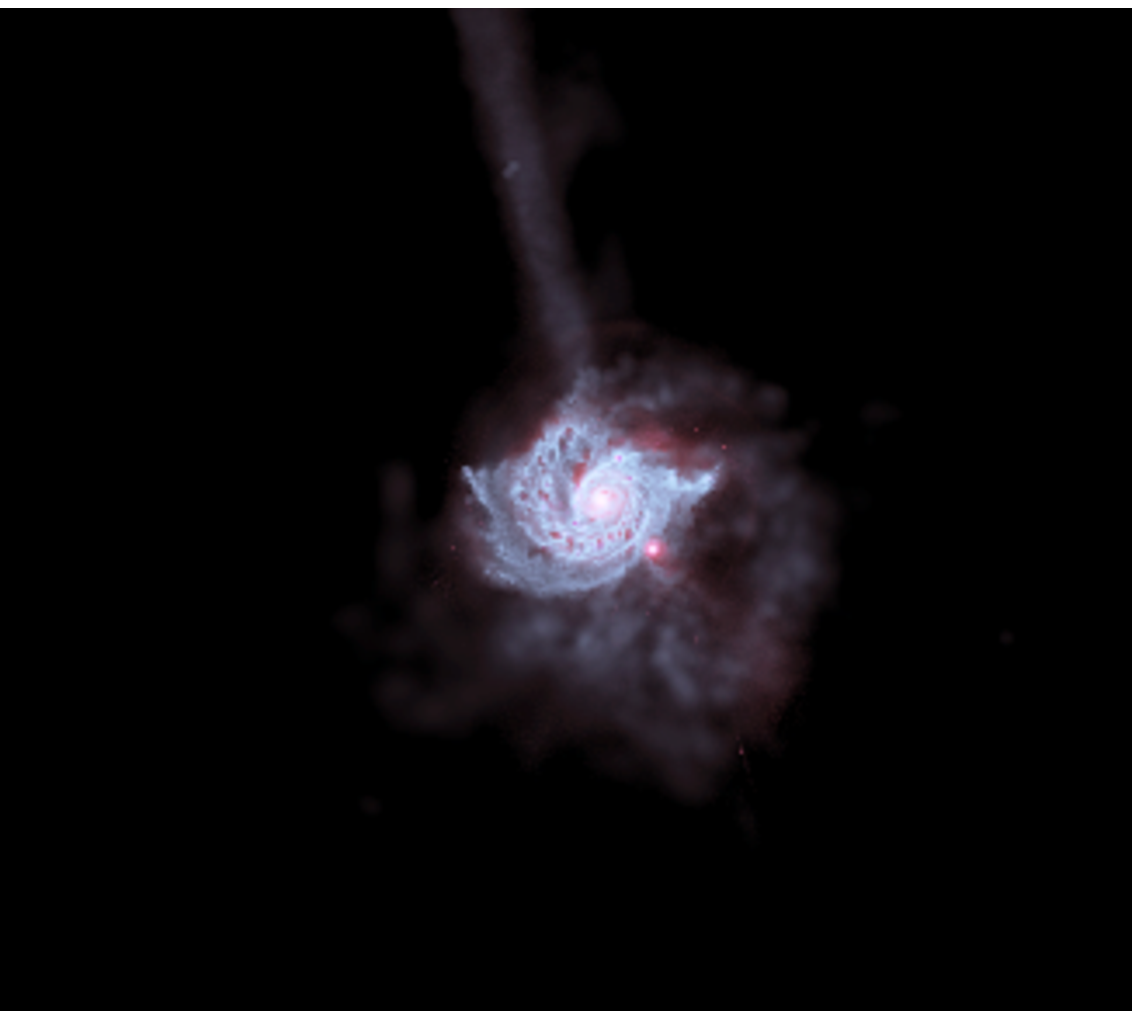}
\put (2,80) {\textcolor{white}{$7$}}
\put (26,80) {\textcolor{white}{Third apocentre}}
\end{overpic}
\hskip -1mm
\begin{overpic}[width=0.301\textwidth,angle=0]{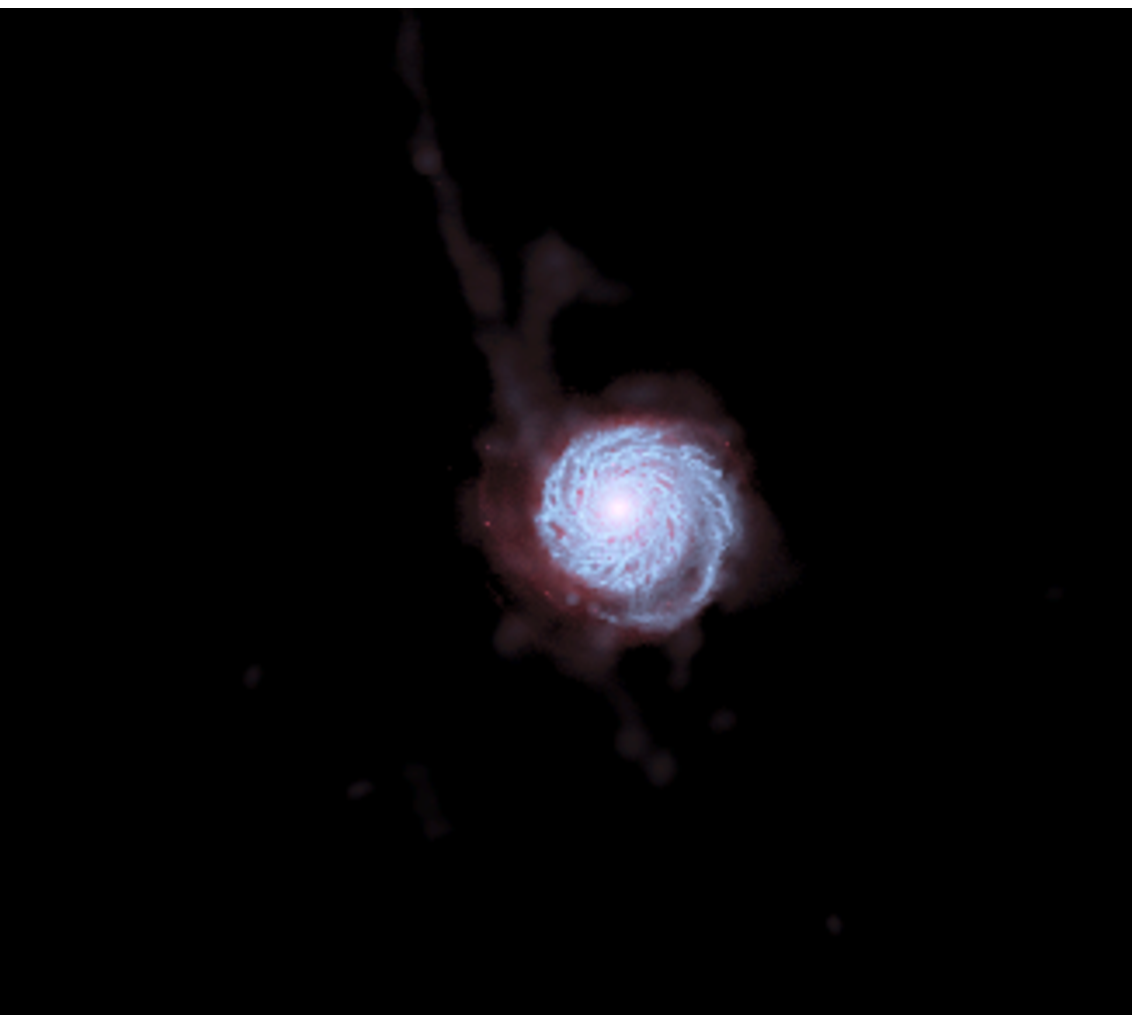}
\put (2,80) {\textcolor{white}{$8$}}
\put (8,80) {\textcolor{white}{End of the merger stage}}
\end{overpic}
\hskip -1mm
\begin{overpic}[width=0.299\textwidth,angle=0]{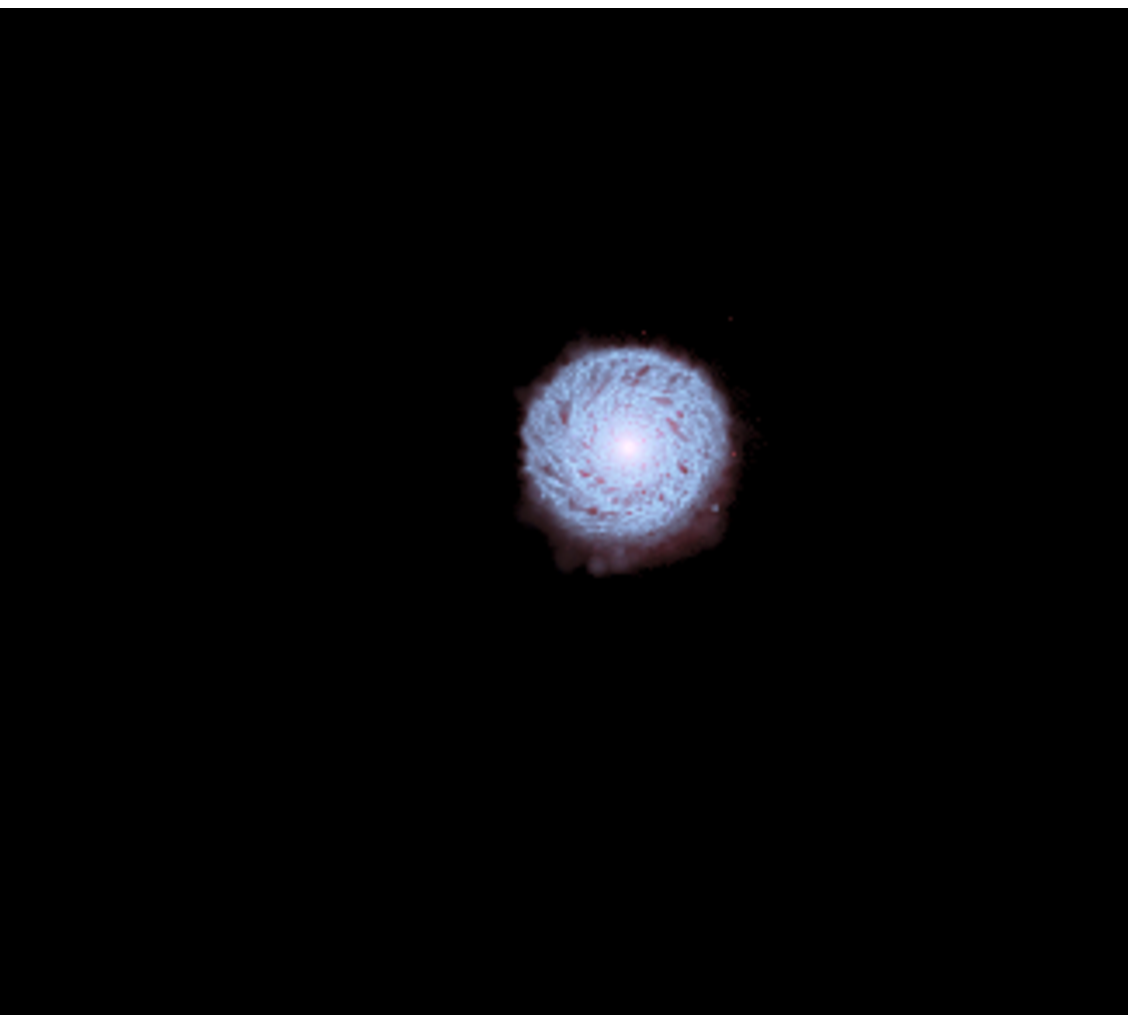}
\put (2,80) {\textcolor{white}{$9$}}
\put (10.5,80) {\textcolor{white}{End of the remnant stage}}
\end{overpic}
\vspace{5pt} 
\caption[Stellar and gas density snapshots]{Stellar (red) and gas (blue) density snapshots (viewed face-on) at representative times of the 1:6 coplanar, 
prograde--prograde merger. Times in Gyr are: (1) 0.20, (2) 0.30 (first pericentric passage), (3) 0.68 (first apocentric passage), (5) 1.10 (second pericentric passage -- end of the stochastic stage), (6) 1.25 (second apocentric passage), (7) 1.40 (third apocentric passage), (8) 1.60 (end of the merger stage), (9) 2.60 (end of the remnant stage),
The image size is 75x75 kpc. Gas density is over-emphasized with respect to stellar density to make it more visible.}
\label{fig:stellar_and_gas_density_snapshots_m6}
\end{figure*}

Broadly speaking, this behaviour is common to all simulations in our suite. However, as the mass ratio decreases,
$G_1$ becomes more and more insensitive to the dynamical presence of $G_2$.  Enhancements to BHAR and SFR in the merger phase are noticeable in $G_1$ for the 1:1, 1:2 and 1:4 mergers, but in the 1:6 and 1:10 mergers they become negligible. 
Conversely, $G_2$ is much more strongly affected by the merger dynamics as the mass ratio decreases.  Fig.~\ref{fig:4panelpro_m6} shows the properties of the 
1:6 merger, and Fig.~\ref{fig:stellar_and_gas_density_snapshots_m6} illustrates the morphology of the galaxies at different time-steps. 
In the 1:10 case, $G_2$ is almost completely disrupted at the third pericentre, and eventually its gas mixes completely with that in the centre of $G_1$, becoming fuel for the accretion and growth of $BH_1$. 
In general, we do not see any qualitative difference caused by the orbital inclination, by one of the galaxies being on a retrograde orbit, 
or by the different gas content.

\section{Temporal correlation between star formation rate and black hole accretion rate}\label{sec:timecorr}
In this section we discuss how SFR and BHAR vary temporally with respect to each other. 
We compare the BHAR to the SFR within shells of 100~pc ($\sfrn$) and 5 kpc ($\sfrg$) centred around each BH. 
These shells are our proxies for the nucleus and the entire galaxy. 

All merging galaxies calculated in this work qualify as star forming galaxies at least part of the time during the first two phases.  This means that they are on the ``main sequence'' \citep[or mass sequence,][]{elbaz2007,noeske2007,speagle}  in the SFR versus stellar mass plane. Discussing the time evolution of the merging galaxies in the SFR vs stellar mass plane is beyond the scope of the present paper.  We only use this property when trying to assess (or speculate on) the behaviour of larger merging galaxies that are not included in these calculations, but dominate current observational samples (see discussions in sections 6 and 7 below).

A delicate issue in this work is the comparison of the merger calculations with observations of samples of star forming galaxies and AGN. In doing so we assume that:

\begin{enumerate}

\item The collection of all calculations presented here represents a sample of galaxies with masses corresponding to the mass range covered by the calculations.

\item The comparison takes into account the relative duration of the three phases. For example, the remnant phase is the longest and, therefore, any comparison with real samples will include many more sources that are in this phase.

\item There is a way to extrapolate some of the general properties to larger systems that are not treated in the calculations. This is the most speculative part and we comment of it, more specifically, when addressing the properties in question.

\end{enumerate}

\begin{figure}
\centering
\includegraphics[width=\columnwidth,angle=0]{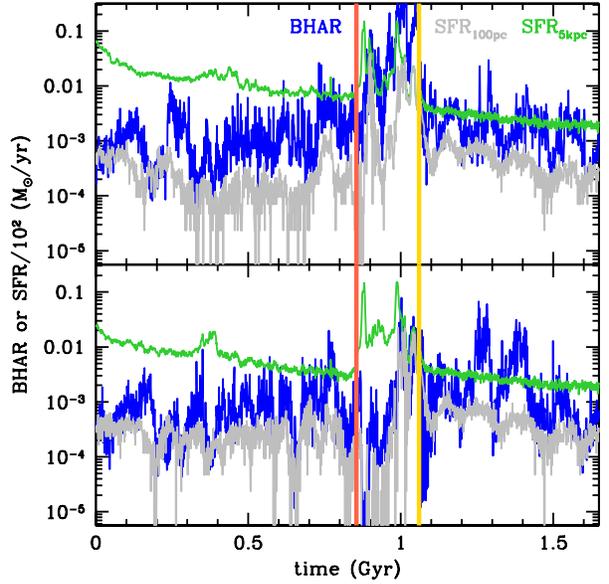}
\caption{BHAR and SFR in the stochastic phase (characterizing also quiescent galaxies in isolation), merger,
and remnant phase, for the 1:2 coplanar, prograde-prograde merger (note that the SFR is divided by 100 to fit in the same y-axis range as BHAR). 
Top panel: $G_1$. Bottom panel: $G_2$. We show the BHAR (blue, lower thin solid curve), SFR in the central 100~pc (gray, thick solid curve) 
and SFR inside 5~kpc (green, upper thin solid curve) all as a function of time. 
The vertical lines mark the transition between the three phases. In the stochastic phase, BHAR and nuclear SFR ($<$ 100~pc) show similar trends.  
The SFR in the outer region is less variable and less correlated with the BHAR compared with the nuclear SFR. }
\label{fig:sfravebharall}
\end{figure}

In Fig.~\ref{fig:sfravebharall} we show the evolution of $\sfrn$, $\sfrg$ and BHAR in the reference 1:2 merger, m2.gf0.3.pro, dividing the SFRs by 100 to fit in the same y-axis range as BHAR. The overall trend of decreasing global SFR is  related to the simulation being evolved in isolation.  The differences between phases derived from theses simulations can be summarised as follows.

In the first phase, the stochastic phase, BHAR and nuclear SFR ($\sfrn$) show similar patterns over  timescales of $\sim 0.05-0.1$ Gyr. We interpret this as caused by being the same gas that feeds the BH and fuels SF.  Both BHAR and $\sfrn$ show also shorter-term variations with large changes, sometime in excess of one order of magnitude over 0.01-0.1 Gyr. As a reference, the dynamical time at 100~pc is 0.01~Gyr. The latter is much less evident in $\sfrg$. SFR in the outer regions is  less variable, with changes typically less than a factor of 2. Furthermore, long time-patterns are almost absent.  As a reference, the dynamical time at 5~kpc is $\sim$0.2~Gyr.

 In the second, merger phase, the patterns for $\sfrn$ and BHAR continue to be similar, but  now $\sfrg$ also exhibits peaks and troughs over similar timescales as $\sfrn$ and BHAR.  The fluctuations in $\sfrn$, $\sfrg$ and BHAR increase with respect to the stochastic phase. The peaks in  $\sfrg$ occur during the second and third pericentres  (a small bump can be seen at the time of the first pericentre, t=0.32 Gyr, in Fig.~\ref{fig:sfravebharall}). After the fourth pericentre the BHs are separated by less than 5~kpc,  therefore $\sfrg$ is the same for both BHs. 

In the remnant phase, some of the temporal patterns are similar to what they were in the stochastic phase: BHAR and nuclear SFR ($\sfrn$) show similar behaviour, while $\sfrg$ does not show specific time patterns any longer.  Note that in the remnant phase the two BHs are separated by a few tens pc at most,  therefore  $\sfrg$ and $\sfrn$ are the same for both BHs.  In this phase the two BHs are also enclosed in the same gas density and temperature region, and BHAR is only modulated by the local dynamics, i.e., the relative velocity between each BH and the gas in the Bondi formulation.\\

\begin{figure}
\centering
\includegraphics[width=\columnwidth,angle=0]{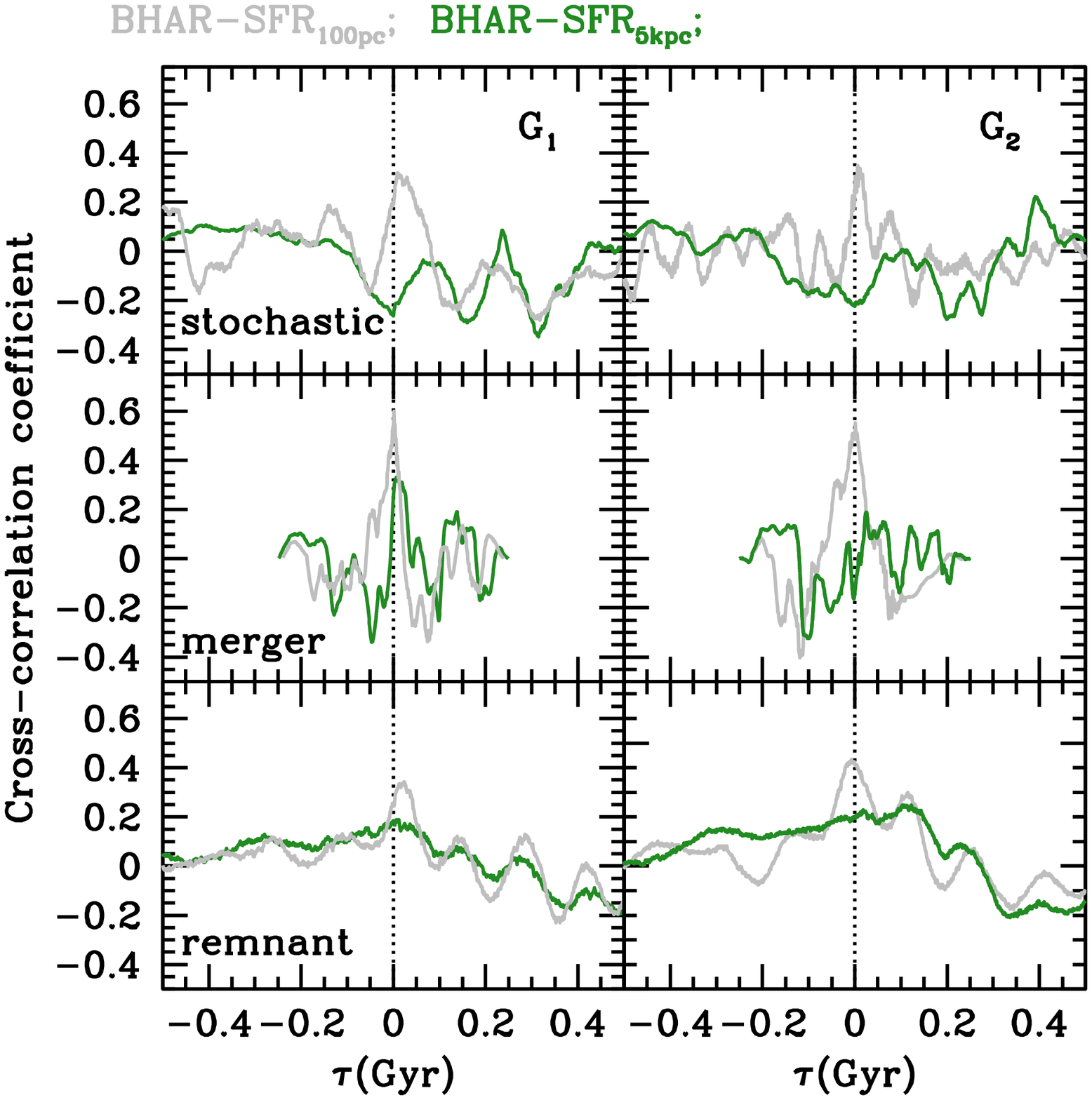}\vfill
\includegraphics[width=\columnwidth,angle=0]{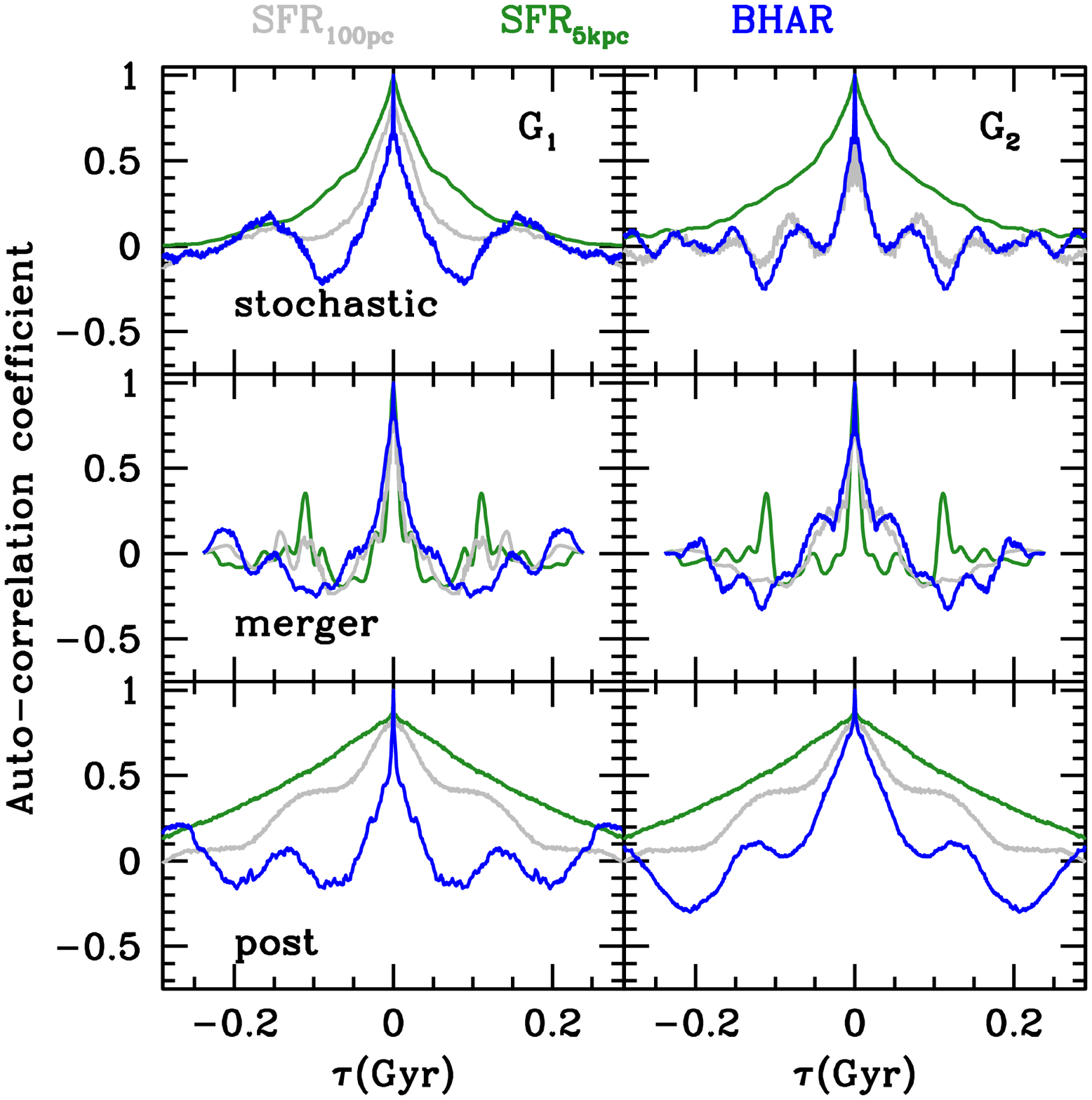}
\caption{Top: cross-correlation coefficient vs time lag, $\tau$,  for BHAR and SFR, based on the 1:2 coplanar, prograde-prograde merger. Left panel: $G_1$. Right panel: $G_2$.   Some degree of temporal correlation is present between BHAR and $\sfrn$ at all times. For $\sfrg$, the correlation is much weaker
and tends to be present only  during the merger proper. Qualitatively, this occurs in many mergers.  The behaviour in the remnant phase is initially erratic, 
and eventually becomes similar to the stochastic stage. Bottom: auto-correlation coefficient vs time lag for BHAR and SFR, based on the 1:2 coplanar, prograde-prograde merger.  Left panel: $G_1$. Right panel: $G_2$.  At early and late times, BHAR  varies on shorter timescales than the SFR. During the merger phase the typical timescales on all scales are similar.}
\label{fig:CCF_m2_pro}
\end{figure}

To disentangle the time correlation between BHAR and SFR on different scales, we calculated the cross-correlation function of BHAR with $\sfrn$ and with $\sfrg$  
 (Fig.~\ref{fig:CCF_m2_pro}; a positive $\tau$ means that BHAR lags the SFR).  While the general correlation at the peak of the cross-correlation function is not high,
it is clear that in all phases BHAR and $\sfrn$ are better correlated than BHAR and $\sfrg$.
The  peaks of the BHAR-$\sfrn$ cross-correlation occur close to $\tau=0$ Gyr underscoring that  they tend to 
occur simultaneously, and the correlation is strongest during the merger phase. The secondary peaks mark the longer-term patterns visible in 
Fig.~\ref{fig:sfravebharall} in the stochastic phase, or correlate one pericentre to another during the merger phase.  
The behaviour in the remnant phase is initially irregular, while at later times the conditions return to be similar to the stochastic phase.  
 The bottom panel shows the auto-correlation functions of BHAR, $\sfrn$ and  $\sfrg$, highlighting the differences in typical timescales characteristic of each process and scale.  The auto-correlation function is symmetrical around $\tau=0$, where is always peaks, as the function is identical to itself for no-lag. The presence of additional peaks mark the typical timescales over which the time-dependent quantity, BHAR, $\sfrn$ and  $\sfrg$ in our case, presents patterns or periodicities.  During the stochastic and remnant phase, BHAR and $\sfrn$ have characteristic timescales shorter than $\sfrg$, which does not show any peak other than that at $\tau=0$, out to more than 0.3~Gyr. During the merger phase the intrinsic timescales are similar for all BHAR, $\sfrn$ and  $\sfrg$. Recall that the dynamical times are  $\sim$0.01~Gyr at 100~pc and $\sim$0.2~Gyr at 5~kpc,  similar to the typical timescales for SF on these scales.

\begin{figure}
\centering
\includegraphics[width=\columnwidth,angle=0]{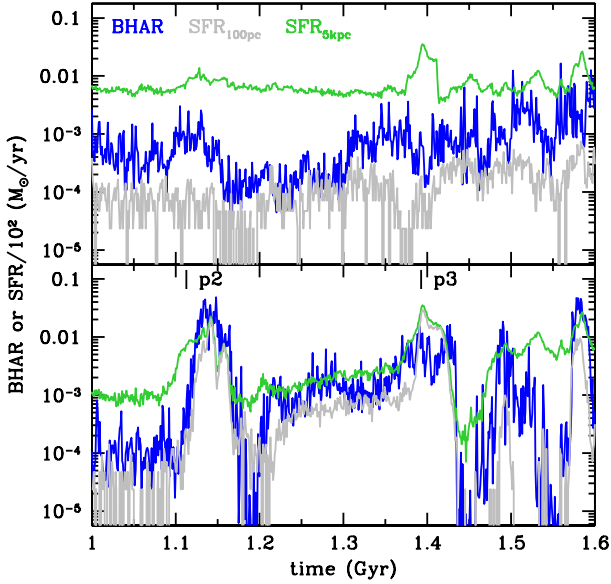}
\caption{BHAR and SFR in the merger phase for the 1:6 coplanar, prograde-prograde merger. 
Top panel: $G_1$. Bottom panel: $G_2$.  The times of the  second and third pericentres are marked as p$_2$ and p$_3$ respectively.}
\label{fig:sfravebharmergerm6}
\end{figure}

We can contrast the 1:2 merger used as reference to the 1:6 merger, where the dynamics is very different. As already noted, as the mass ratio decreases,  
$G_1$ is less and less affected by the merger.   Vice versa, $G_2$ feels much more strongly the dynamical effects. This is evident in Fig.~\ref{fig:sfravebharmergerm6}. 
At second pericentre, $G_1$ and $BH_1$ do not experience any burst of SF or BH activity. 
Instead, $BH_2$ increases its BHAR (and consequently luminosity) by more than two orders of magnitude. The SFR in $G_2$ also has a burst affecting the galaxy on all scales 
(i.e., $\sfrn$ and $\sfrg$). During and after the third pericentre, $\sfrg$ is enhanced, as $G_2$ is largely stripped of most of its gas 
After $t=1.5$ Gyr, the BHs are separated by less than 5~kpc,  hence the green curves  in both panels are identical.

 From the ensemble of our mergers (see the online-only material), we can identify several common patterns: 
(i) in the stochastic phase, which applies also to galaxies in isolation, $\sfrg$ and BHAR are temporally uncorrelated, while $\sfrn$ and BHAR show some  
degree of correlation: the cross-correlation function peaks with a correlation coefficient of $\sim 0.3$; 
(ii) in the merger phase, $\sfrn$ and BHAR become more strongly correlated, especially for $G_2$; $\sfrg$ and BHAR can have very different behaviours; e.g., be anti-correlated at $\tau=0$ in $G_1$ and positively correlated for $G_2$ (e.g., m4.gf0.3.pro).
(iv) in the remnant phase the behaviour returns similar to the stochastic phase.

We can only speculate on the expected behaviour in mergers of larger galaxies that we cannot simulate in this work.
We expect that the farther the gas is from the centre, the less likely SF will be correlated with BHAR, especially in the stochastic phase when $\sfrg$ is completely 
driven by local dynamics. At pericentre passages we expect that most of the galaxy will experience a strong perturbation, regardless of its size. Therefore, we expect a weakening of the correlation between $\sfrg$ and BHAR during the stochastic phase of larger systems, while $\sfrn$ and BHAR should behave 
similarly to the smaller systems treated here.

In summary, SFR and BHAR are both enhanced by the merger dynamics, and how similar temporal behaviour but only for a limited time;
 $\sim$ 0.2-0.3 Gyr for mass ratios of 1:1 to 1:4. 
At lower mass ratios only the smaller galaxy is significantly affected by the dynamics, with bursts of BHAR and SFR in coincidence with pericentric passages, 
while the larger galaxy is, for the most part, unaware of the merger taking place. 
A cross-correlation between BHAR and $\sfrn$ shows some level of correlation even during the stochastic and remnant phases, while BHAR and $\sfrg$ are 
uncorrelated in these two phases. In the merger phase, $\sfrn$ and BHAR tend to become more correlated, while $\sfrg$ and BHAR can be either correlated or
anti-correlated at $\tau=0$.

\section{Time variability of star formation rate and black hole accretion rate}\label{sec:timevar}
As discussed in the previous section, the cross-correlation function of BHAR and SFR shows some temporal correlation between BHAR and SFR on small scales ($<100$ pc) at all times, 
and between the BHAR and the SFR on larger scales (5~kpc) during the merger phase. Our goal in this section 
is to test whether the BHAR variability differs between the stochastic and merger phases, what is the connection with the SFR variability,
 and whether there is a dependence on the scales over which SFR is measured.

\begin{figure}
\centering
\includegraphics[width=\columnwidth,angle=0]{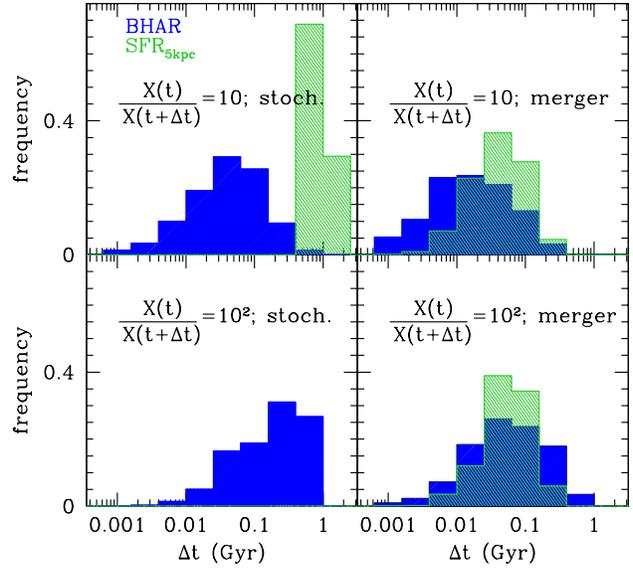}
\includegraphics[width=\columnwidth,angle=0]{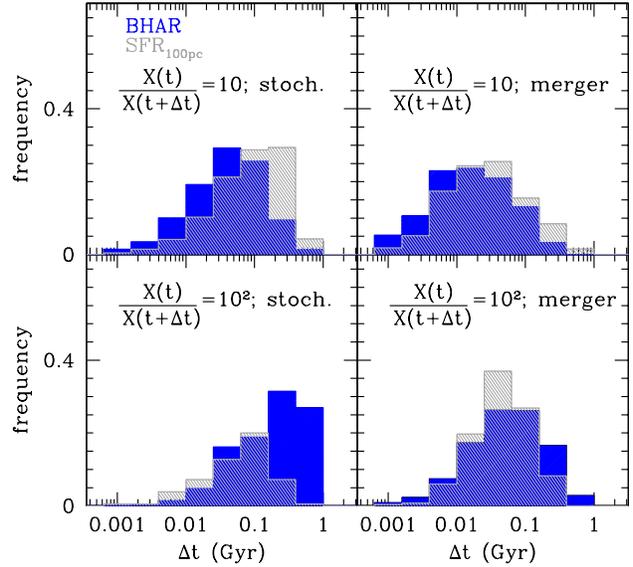}
\caption{Distribution of the time, $\Delta t$,  needed for a quantity X, where X=BHAR (blue),  X=$\sfrn$ (gray) or  X=$\sfrg$ (green),
 to vary by a factor of 10 or 100. 
In the stochastic phase the BHAR varies on much shorter timescales than $\sfrg$ and the distinction between the 
two distributions is clear.  In the merger phase, large variations of BHAR and SFR occur over similar timescales. BHAR and $\sfrn$ vary over  similar timescales in all cases. 
The behaviour in the  remnant phase (not shown here for clarity) is intermediate between stochastic and merger.}
\label{fig:variability}
\end{figure}

As discussed in several recent publications, the different variability time scales of BHAR and SFR may be responsible for the lack of correlation  
between SFR and BHAR found in galaxy samples \citep[see also Fig. 9 in Netzer 2009]{Aird2012,Mullaney2012,2013ApJ...773....3C,Hickox2014}. 
To extract information on the time variability of BHAR and SFR, we estimate the time, $\Delta t$,  needed for each of these quantities to vary by a factor of 10 
or a factor of 100. At each time-step we search forward in time for the first subsequent time-step, within the same phase, where a given quantity  
(BHAR, $\sfrn$ and $\sfrg$) varies by one of these factors. Note that 1~Myr is the minimum timescale for SF, i.e., the SF outputs are separated by $10^6$ years.
The distributions we show in Fig.~\ref{fig:variability} include all the $\Delta t$ for all the runs. We collect  $BH_1$ and $BH_2$, and $G_1$ and $G_2$, 
for all the simulations listed in Table~\ref{tab:params}, as we find no statistical difference if we separate the primary and secondary galaxy and BH.

The $\Delta t$ distributions show that in the stochastic phase BHAR varies more rapidly than $\sfrg$, and by a larger factor. For instance, we do not see variations in $\sfrg$ of a factor of 100 in the stochastic phase, which lasts about one Gyr in our runs.  BHAR and $\sfrn$ vary instead over comparable timescales. In the merger phase, the distributions of $\sfrn$ and BHAR remain very similar, and $\sfrg$ shows higher variability,  approaching that of  BHAR. When we compare runs with 30\% and 60\% gas fractions, the only difference we find is that $\sfrn$ varies more rapidly in  the stochastic phase in the high gas fraction case (the distribution peaks at $\Delta t=$0.04~Gyr instead of $\Delta t=$0.1~Gyr for the factor of 100 case).

We speculate, again,  on what would happen in larger galaxies. In the stochastic phase, the peaks of SFR variability seem to occur over a few dynamical times 
(1~Gyr corresponds to 5 dynamical times at 5~kpc,  and 0.1~Gyr corresponds to 10 dynamical times at 10~kpc), therefore more massive and extended galaxies would present longer time-variability when considering the galaxy-wide SFR. The nuclear SFR will be instead slower (faster) for galaxies poorer (richer) in gas.

In summary, our simulations, owing to their very high temporal and spatial resolution,  confirm the hypothesis that the variability of BHAR is higher 
than that of  SFR measured on large scales (several kpc). However, if the nuclear SFR can be resolved, and SFR measured over short timescales ($<100$~Myr), we 
predict that SFR and BHAR will show similar time variability.

\section{Relative growth of stellar and black hole mass}\label{sec:growth}
\begin{figure}
\centering
\includegraphics[width=\columnwidth,angle=0]{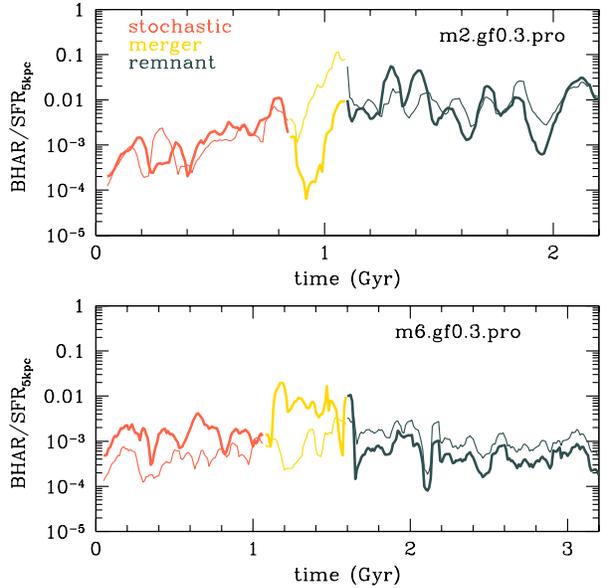}\vfill
\caption{Ratio of BHAR to $\sfrg$, averaging both quantities in bins of 50~Myr for a 1:2 and a 1:6 mass ratio mergers. 
We distinguish stochastic (red), merger (gold) and remnant (dark gray) phases. Thin curve:  $G_1$; thick curve: $G_2$. 
Recall that from half-way through the merger phase, the two BHs are separated by less than 5~kpc, meaning that $\sfrg$ is the same for both BHs.
In the minor merger case (mass ratio of 1:6), the larger galaxy is almost unaffected by the merger while the secondary is strongly perturbed.}
\label{fig:sfr_bhar_ratio}
\end{figure}

We now turn to examine the stellar and BH mass growth. For this we use the information in Fig.~\ref{fig:sfr_bhar_ratio} where we show, explicitly, the ratio of BHAR 
versus SFR within 5~kpc, for the 1:2 and a 1:6 mass ratio mergers, averaging both quantities in bins of 50~Myr (figures for all other simulations are available as online-only material). 
In the stochastic phase  $\sfrg$ is $\sim 10^{3}\times$BHAR.   If the newly formed stars end up in the bulge, the stochastic phase leads to a scaling between BH mass 
and bulge mass close to the ``canonical'' value in the local universe \citep[$10^{-3}$,][]{MarconiHunt2003,Haring2004}.
This feature, which  is common to all our simulations, is not surprising, as we have adjusted the AGN feedback efficiency to agree with the observed 
BH to bulge mass ratio. Specifically, with a feedback efficiency $\epsilon_f=$0.001 $\sfrg$ is $\sim 10^{3}\times$BHAR in the stochastic and remnant phases, 
and with a feedback efficiency $\epsilon_f=$0.005 $\sfrg$ is $\sim 10^{3}\times$BHAR in the merger phase (see the Appendix for additional tests of feedback efficiency).

The merger phase leads, typically, to a higher ratio of BHAR to SFR, and therefore of BH mass to stellar mass. However, 
there are short episodes where the BHAR drops significantly with respect to the SFR. For instance, for $BH_2$ at time=0.87 Gyr in m2.gf0.3.pro 
(top panel of Fig.~\ref{fig:sfr_bhar_ratio}). This is caused by a strong burst of SF triggered at the second pericentre, when $\sfrg$ increases by a factor of 25. The ensuing supernova feedback depletes the nucleus of gas: the gas mass within the central 100~pc decreases by 2 orders of magnitude, more than the mass consumed in forming stars within the same time.  Once additional gas replenishes the BH environs at the third pericentre, accretion restarts at high levels, with the fiducial ratio between BHAR and SFR, on all scales,  of $\sim$ few $\times 10^{-3}$. 

While globally the BHAR to SFR ratio is enhanced during the merger phase, periods where the BHAR  
to SFR ratio is suppressed exist, and they typically follow either bursts in SF or in BH accretion. As noted above, for low mass ratios (the 1:6 and 1:10 mergers), both $BH_1$ and $G_1$ do not ``notice'' that they are involved in a merger, as the secondary galaxy 
is only a negligible perturbation.

\begin{figure}
\centering
\includegraphics[width=\columnwidth,angle=0]{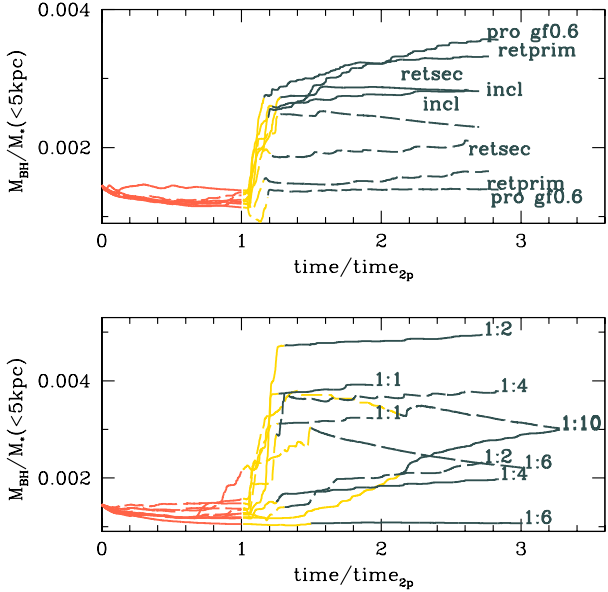}
\caption{Ratio of BH mass to stellar mass within 5~kpc as a function of time, normalized to the time of the second pericentre. Solid curves: $BH_1$ and $G_1$. 
Dashed curves: $BH_2$ and $G_2$. Top panel: different orbital orientations and gas fractions for the same mass ratio of 1:2. Bottom panel: different mass ratios, from 1:1 to 1:10 for the same coplanar, prograde-prograde orbital configuration. Note how $G_1$ in the 1:6 merger is almost completely unaffected by the merger. In the remnant phase it would make more sense to sum the masses of $BH_1$ and $BH_2$, as well as the stellar mass in both galaxies, but this would create a discontinuity in the curves.}
\label{fig:MBH_mstar5kpc}
\end{figure}

Since the bulge mass cannot be reliably measured while galaxies are disturbed, we quantify the relative growth of BH and galaxy by estimating the ratio of  BH mass to stellar mass within 5~kpc as a function of time (Fig.~\ref{fig:MBH_mstar5kpc}). During the remnant phase, BH accretion remains at levels that are
slightly higher than before the merger, keeping  the ratio of  BH mass to stellar mass roughly 
constant. At the end of the remnant phase, when the galaxy is relaxed and the bulge mass can be measured, the BH to bulge mass ratio is between 0.0025 and 0.004,  i.e. about a factor 1.25 to 2 higher than in the initial conditions.

\begin{figure}
\centering
\includegraphics[width=\columnwidth,angle=0]{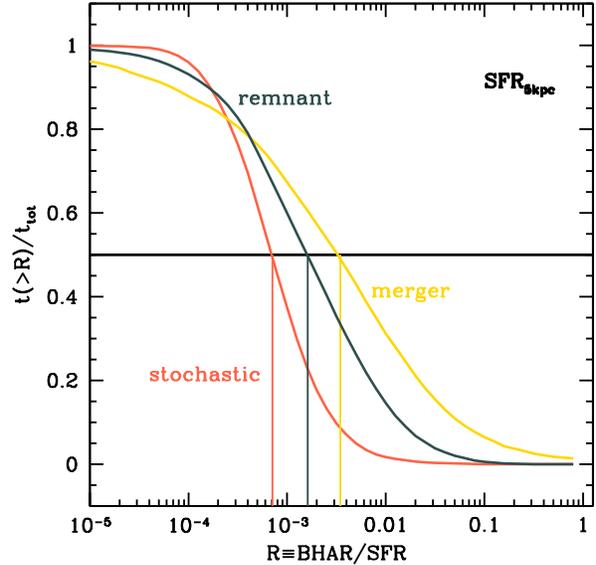}
\caption{Cumulative time fraction above a given ratio of BHAR/SFR within 5~kpc. The numbers apply to all simulations together and include both $G_1$ and $G_2$ 
but we distinguish for each of them the stochastic, merger  and remnant phases. 
The horizontal line marks the 50\% level and the vertical lines the BHAR/SFR ratios above which the system spends 50\% of its time.}
\label{fig:cumtime}
\end{figure}

A second diagnostic of the enhancement of BHAR relative to the SFR during the merger is the cumulative time fraction spent by the BH and galaxy above 
a given ratio of BHAR/SFR.  
This is shown in figure~\ref{fig:cumtime}.  For this figure we combined all our simulations together, and include both $G_1$ and $G_2$, since the difference between the two galaxies is negligible. The horizontal line marks  the 50\% level, and the vertical lines the BHAR/SFR ratio above which the system spends 50\% of its time.  This ratio increases by a factor $\sim$ 5  during the (transitory) merger phase, meaning that during this phase the BH grows more efficiently than its host's stellar mass, skewing the BH to stellar  mass ratio to higher values.  We find a very weak dependence of this ratio on orbital configurations, and gas fraction, and a somewhat 
stronger dependence on the mass ratio. For example, the enhancement in the merger phase in the cases of mass ratios 1:6 and 1:10 is completely 
dominated by $BH_2$ and $G_2$. The Appendix gives more details about the dependence on feedback efficiency and resolution. 

Regarding more massive galaxies that are included in observational samples, we speculate that their behaviour can be inferred from the present calculations using some well know properties of star forming galaxies. In such systems the stellar mass and SFR are coupled to form the ``main sequence'', and in general SFR$\propto M_*^{\alpha}$ where $\alpha \approx 0.7-1$ with some hints for changes with redshift. Therefore, for galaxies on the main sequence, SFR would increase approximately linearly with stellar mass. The galaxies we simulate start on the main sequence, but eventually they move away from it as they consume gas. As a consequence, adopting the linear evolution of SFR with mass characteristic of the main sequence is not rigorously correct at all times, but we can use this approach to infer some trends. 

\begin{figure}
\centering
\includegraphics[width=\columnwidth,angle=0]{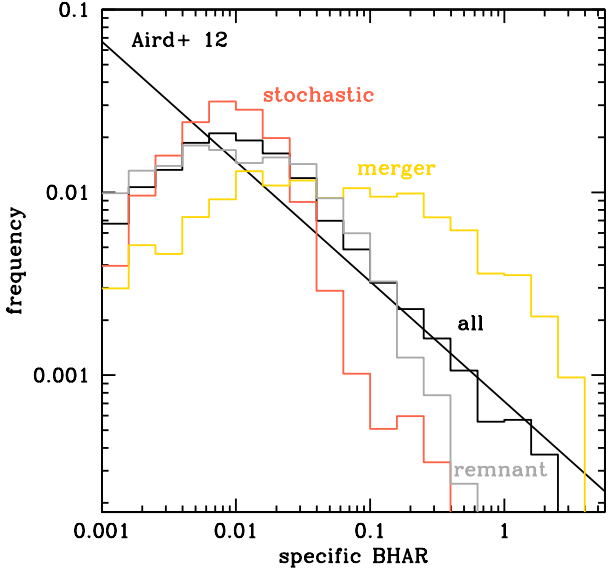}
\caption{Distribution of specific BHAR, i.e. BHAR relative to the stellar mass of the host galaxy, for our simulated galaxies. We distinguish the  stochastic 
(red), merger (gold)  and remnant (dark gray) phases. We also combine all phases together (`all', black). As a reference we show 
the distribution proposed by Aird et al. 2012 (solid black line).}
\label{fig:aird}
\end{figure}

We can conjecture how the BHAR would scale with the galaxy mass taking as a starting point the work by  \cite{Aird2012}. They find that in a sample of AGN, at $0.2<z<1.0$,  
the probability of finding an AGN with a specific accretion rate, i.e. BHAR relative to the stellar mass of the host galaxy, can be described by a power-law 
distribution, with slope $-0.65$,  independent of stellar mass \citep[see also][where they find a slope closer to unity]{2012MNRAS.427.3103B}. 
Calculating  the same quantity for the collection of all our simulated 
galaxies (Fig.~\ref{fig:aird}) we find a power-law with a slope of $-0.8$ over three orders of magnitude, in reasonable agreement with Aird et al 
(2012). Perhaps more important are the slopes of the individual phases. We find that the slope is steeper for the quiescent and remnant phases, and shallower for the merger phase. 
If the specific accretion rate is self-similar, we expect that in a larger galaxy we would have similar specific accretion rates,
 therefore the BHAR would increase linearly with the stellar mass.

Based on these conjectures, therefore, both BHAR and SFR would increase approximately linearly with stellar mass. The results in Figs.~\ref{fig:sfr_bhar_ratio}, showing the ratio of BHAR/SFR, and~\ref{fig:MBH_mstar5kpc}, showing the ratio of BH mass to stellar mass, would therefore be in first approximation similar in larger galaxies.

\section{Relative magnitude of star formation rate and black hole accretion rate}\label{sec:magcorr}
\begin{figure}
\centering
\includegraphics[width=\columnwidth,angle=0]{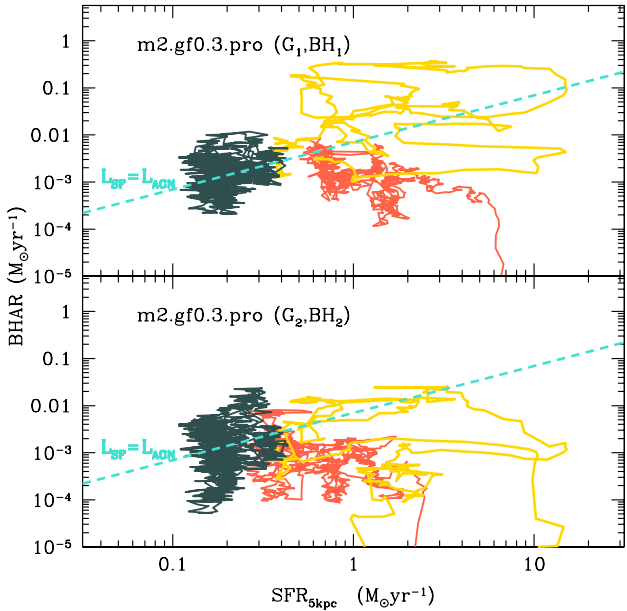}
\caption{Time evolution tracks of  BHAR versus SFR for a 1:2 merger with prograde, coplanar configuration and gas fraction 0.3. Individual measurement are spaced by 1~Myr. 
We distinguish the stochastic (red), merger (gold) and remnant (dark gray) phases. 
Tracks for all other mergers are available as online-only material.}
\label{fig:tracks}
\end{figure}

Fig.~\ref{fig:tracks} shows the time evolution of BHAR versus SFR for the reference merger. This figure provides a visual representation of the trajectory of 
BHAR and $\sfrg$ over time. It complements the analysis of \cite{2014MNRAS.443.1125T}, who present the time evolution of BHAR versus SFR in a series of 
1:1 mergers where they vary the BH accretion and AGN feedback.  The dashed line marks the boundary of the AGN versus SF dominated regions, 
i.e the regions where $L_{AGN}$ is respectively higher or lower than $L_{SF}$. We calculated the AGN luminosity assuming a fixed radiative efficiency 
$\epsilon_{\rm r}=0.1$, and the far-infrared luminosity by assuming $10^{10}$ solar luminosities per SFR of one solar mass per year.  

In all our simulations, galaxies inhabit the SF dominated region during the stochastic phase and move between the AGN and the SF dominated regions during the remnant phase.
 However, during the merger phase the evolution is complex and chaotic (see tracks for all simulations in the online-only material). 
In general, not all mergers lead to an appreciable enhancement of AGN activity and not all mergers lead to an appreciable enhancement of SF. For instance, the bottom panel of Fig.~\ref{fig:tracks} is a clear example of a case where BH accretion is not enhanced.

\begin{figure*}
\centering
\includegraphics[width=\columnwidth,angle=0]{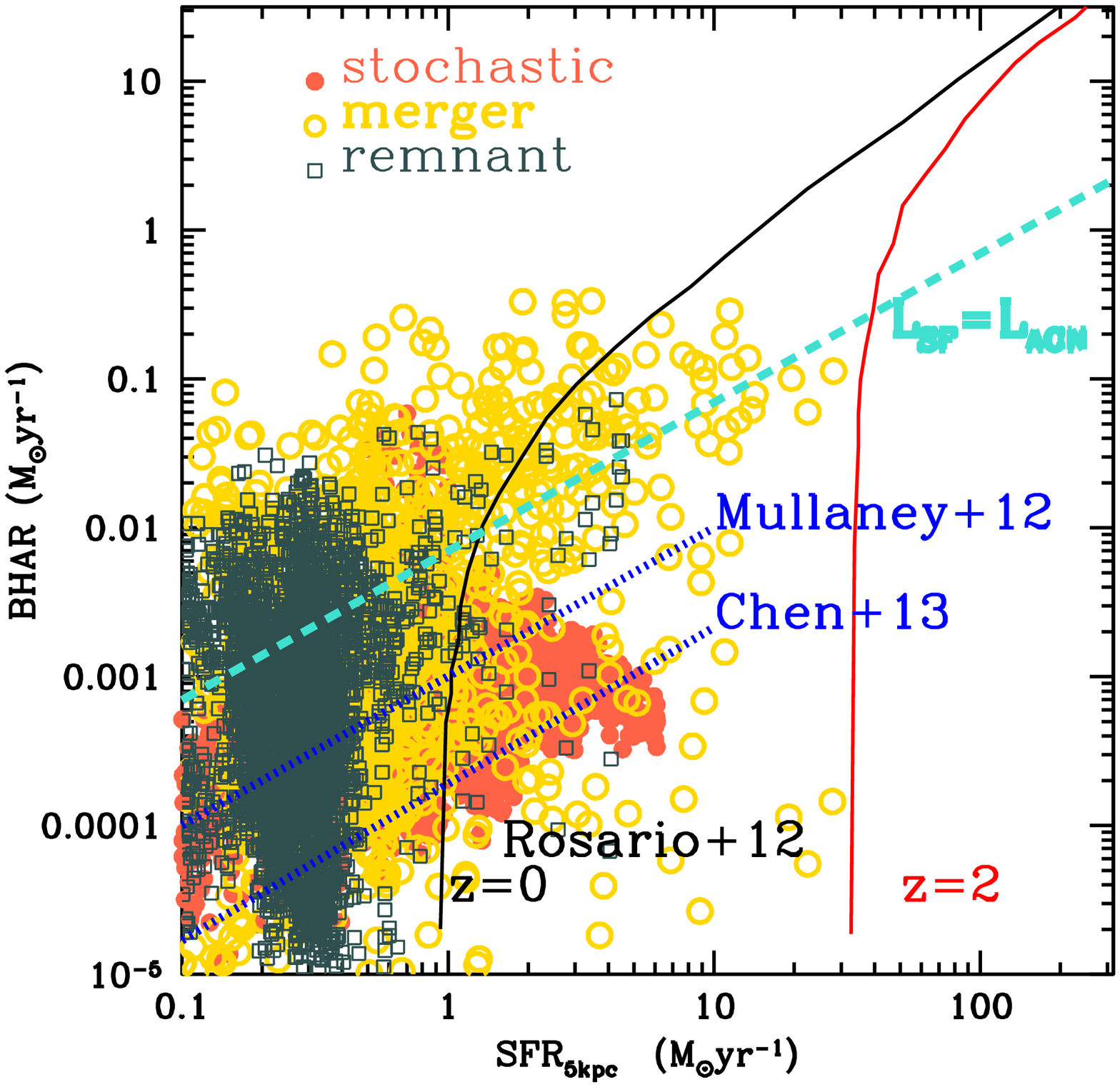}
\includegraphics[width=\columnwidth,angle=0]{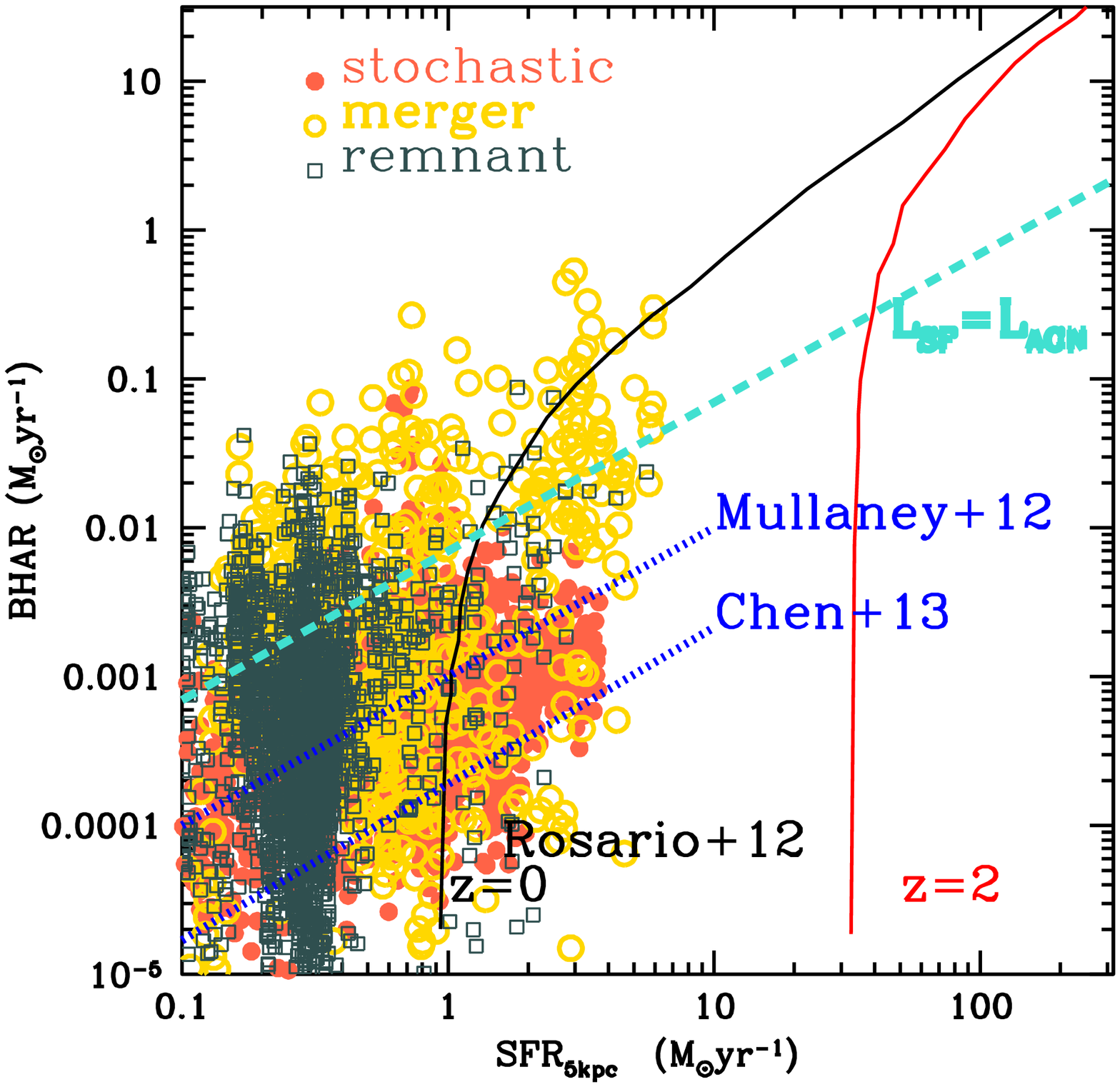}
\includegraphics[width=\columnwidth,angle=0]{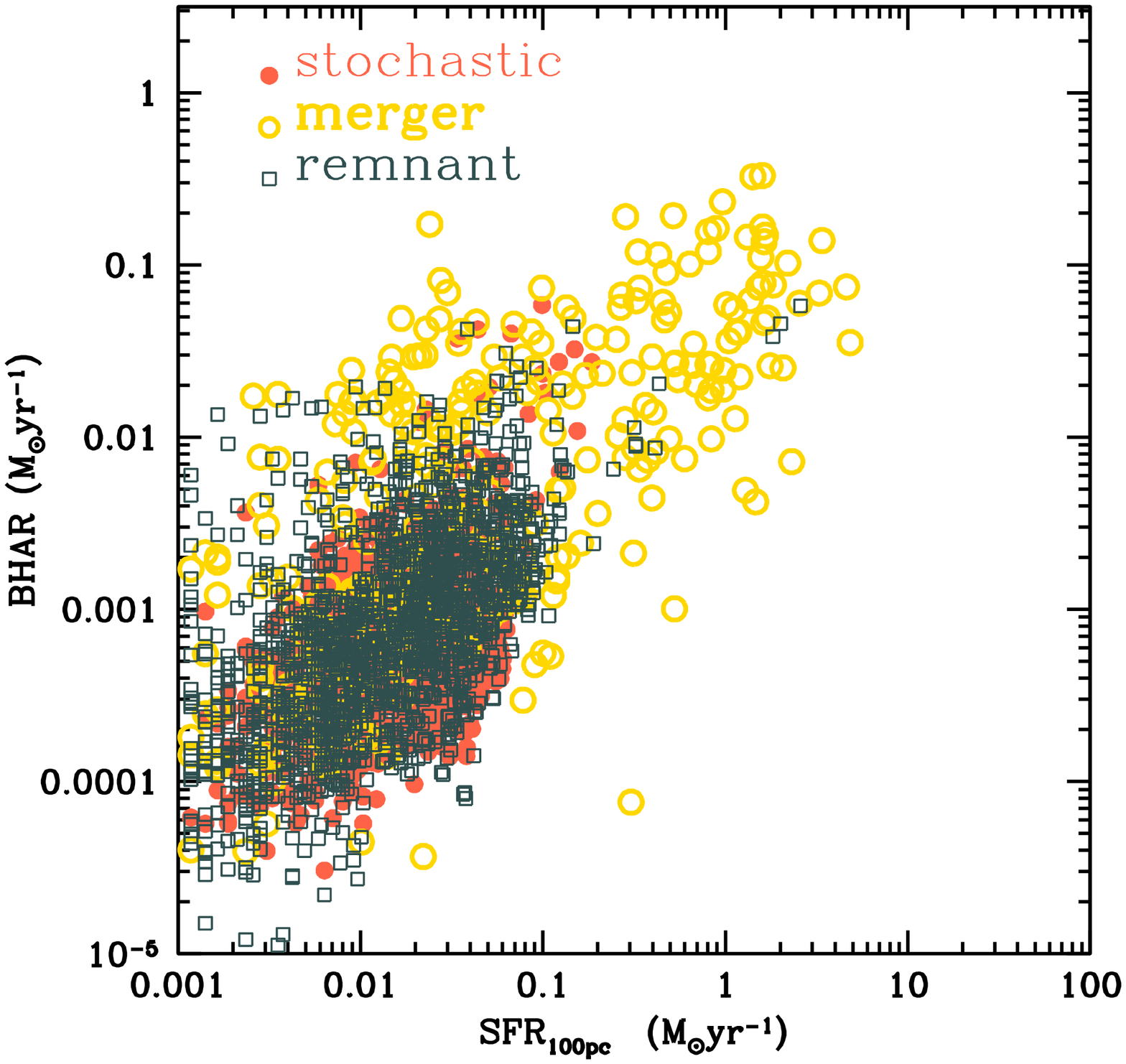}
\includegraphics[width=\columnwidth,angle=0]{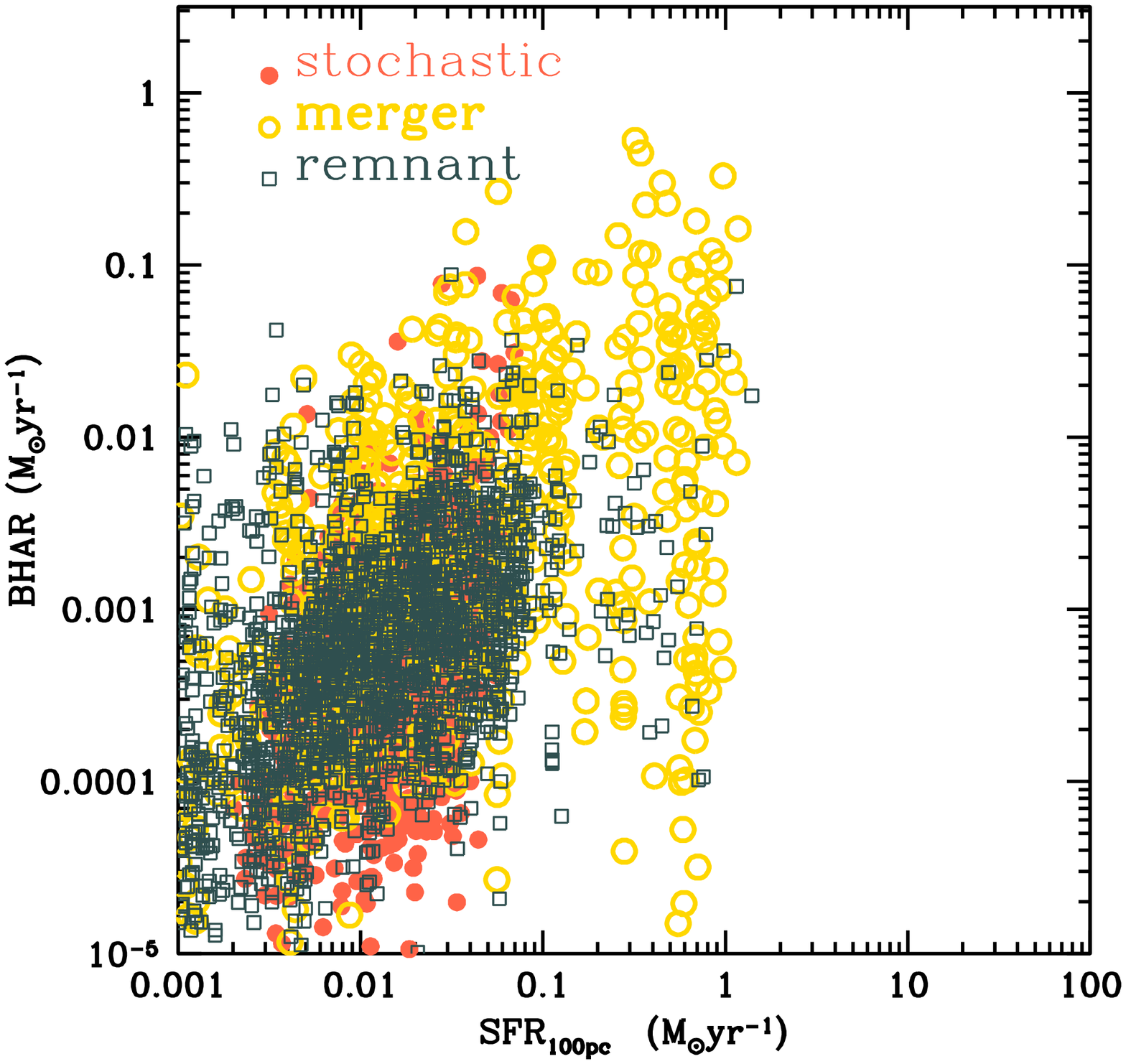}
\caption{BHAR versus SFR within 5~kpc (top), and within 100~pc (bottom). 
In the two left panels we average both quantities in bins of 10~Myr. 
In the two right panels we average the SFR over the last 100 Myr before the time-step used for the BHAR calculation, to mimic observational SFR indicators. 
  In the top panels we include curves from Rosario et al. (2012, AGN, red and black), and from Chen et al. (2013, star-forming galaxies, blue). 
We also mark the AGN and SF dominated regions, respectively above and below the light blue line. }
\label{fig:sfr_bhar_2w_app}
\end{figure*}

We can translate the BHAR-SFR tracks into a different representation by averaging SFR and BHAR in all simulations over the same time span, e.g., 
10~Myr (Fig.~\ref{fig:sfr_bhar_2w_app} left panels). BHAR and $\sfrn$ show some degree of correlation, although the scatter is large (2 dex in BHAR at fixed $\sfrn$). 
If we were able to measure the BHAR and the SFR that occurred within the same time span, we would find that BHAR and $\sfrn$ correlate for both quiescent and merging hosts.  BHAR and $\sfrg$, instead,  do not correlate.

So far we have addressed the question of whether a correlation between SFR and BHAR exists a priori. In the following, we discuss our results 
taking the observers' view and asking whether a correlation between SFR and BHAR can be inferred from the observations, 
by considering the different timescales probed by measurements of the AGN luminosity and SFR.   
Most SFR diagnostics measure the ongoing, rather than instantaneous SFR. 
To mimic this we average the SFR  at each time-step over the previous 100 Myr.   BHAR is averaged instead over 1~Myr time-steps.  Only a random subsample of the points in each phase matching the number of points in the left panels is shown, to avoid the figure being overcrowded.  We bundled all our simulations together, but we distinguished for each of them the stochastic (red), merger (gold) and remnant (dark gray) phases. Fig.~\ref{fig:sfr_bhar_2w_app}, right panels, shows the results of this analysis.  To guide the eye, we include curves from observational studies focusing on AGN \citep{2012A&A...545A..45R} or galaxies \citep{2012ApJ...753L..30M,2013ApJ...773....3C}. We postpone the detailed comparison with observations to a forthcoming dedicated paper. 

In the stochastic phase, the timescale over which SFR is estimated has a limited effect,  as the SFR varies over relatively long timescales (cf. Fig.~\ref{fig:variability}) and the magnitude of the changes in SFR and BHAR is small.  The main effect is in removing the low- and high-SFR peaks, and as a consequence slightly tilting the BHAR-$\sfrg$ distribution. In the merger phase several effects should be taken into account. In the first few 100~Myr of the merger phase,
 the average SFR includes part of the stochastic phase. Since in this phase the SFR  is better correlated with the BHAR, 
the averaging washes out this underlying correlation. This is particularly evident for $\sfrn$ (bottom panel of Fig.~\ref{fig:sfr_bhar_2w_app}). 
A very similar average SFR is associated to a wide range of BHAR, as during the merger phase both BHAR and SFR vary by a large factor over short timescales 
(cf. Fig.~\ref{fig:variability}). The main effect can be seen at the high-SFR end of the BHAR-$\sfrn$ distribution, as the average $\sfrn$ in the merger phase is high, 
while the instantaneous BHAR varies by several orders of magnitude. The remnant phase is different again. Here there is no difference between measuring 
the ongoing or the instantaneous SFR, except in the first few 100~Myr that include part of the merger phase. 
 In summary, for merging and post-merger galaxies, a time-averaged SFR worsens the BHAR-$\sfrn$ correlation, without affecting much the behaviour of BHAR versus $\sfrg$.

\section{Conclusions}\label{sec:Conclusions}

We present a suite of simulations devoted to the detailed study of BH and galaxy properties during various type mergers.
This is the first suite that reaches consistent high time (data on BHAR is extracted every 0.1~Myr, data on SFR is extracted every 1~Myr), 
mass ($3.3 \times 10^{3}$ and $4.6 \times 10^{3}\; {\rm M}_{\odot}$, for stars and gas respectively) and spatial (10~pc and 20~pc, for stars and gas respectively) resolution 
for a range of mass ratios that includes minor mergers, down to 1:6 and 1:10. 
We set  the initial pericentre distance near 20 percent of the virial radius of $G_1$, and start from a separation equal to the sum of the virial radii of the two galaxies, 
in order to be consistent with cosmological orbits.  We run the simulation for a time which is long enough to include the return of the final remnant phase to
a quiescent state. This allows us to capture the post-merger phase, which may represent the state where most AGN are observed, as they show no hint of a companion galaxy. 
The set-up allows us to resolve nuclear gas inflows in the innermost region ($<$100~pc) where BHAR takes place, as well as to resolve the SF within the galaxy. 
We can also meaningfully compare BHAR and SFR timescales, both in an abstract way (i.e. instantaneous SFR and BHAR occurring within the same time-step) and in a realistic 
way more appropriate for a comparison with the observations (ongoing SFR). We analyze three phases that we dub `stochastic' (corresponding to a galaxy in isolation or in the early 
phases of an encounter), `merger' proper (when the merger dynamics dominates), and `remnant' (from the end of the merger to the return to quiescence). 
Our findings can be summarized as follows:

\begin{enumerate}
\item The temporal patterns of BHAR and $\sfrg$ are generally uncorrelated, except in some cases during the strongest pericentre passages. 
$\sfrn$ and BHAR show some level of correlation during all the stochastic, merger and remnant phases.
\item For galaxies in quiescence or post-merger, BHAR varies more rapidly than $\sfrg$, but on timescales similar to $\sfrn$. 
For AGN in the merger phase BHAR and SFR are expected to show similar time variability even on galactic scales. 
\item The merger phase leads, in most cases, to a higher BHAR/SFR, by a factor of a few, when averaged over time, and therefore to high BH mass to stellar mass. 
However, there are short  episodes when BHAR drops with respect to the SFR, because of supernova and AGN feedback triggered by a previous burst of SF or BH activity.
\item When measured over the same timescales, BHAR and $\sfrn$ are proportional to each other. The correlation lessens if the ongoing rather than the
 instantaneous SFR is measured, since the average SFR is associated to a wide range of BHAR.
\item The timescale over which SFR is measured affects less strongly the interpretation of BHAR versus $\sfrg$.
BHAR and $\sfrg$ show different behaviour during the three stages. From a rough proportionality in the stochastic phase, 
with BHAR $\sim 10^{-3}\sfrg$  (and a large scatter), to  BHAR$ \sim 10^{-2} \sfrg$  for the most  luminous AGN in the merger phase. 
In the remnant phase, galaxies occupy a  region where a limited range in  $\sfrg$ corresponds to a large range of BHARs.
\end{enumerate}

A major conclusion of our study is that any comparison between BH activity and SFR must take into account the different stages
of the merger process since those properties can change dramatically. While in the stochastic phase galaxies would, for the most part, not be 
considered AGN, the merger phase is when AGN and SF activity is close to their peak. The remnant phase is characterised by a large range of BHARs, 
moving, at times, the galaxy into the AGN-dominated region.  An AGN can be caught sometime (up to 1.5~Gyr) after the merger and starburst actually 
took place.  Study of BH activity and SFR in large AGN and galaxy samples, must take into account the different durations of the various phases. Finally,  we have shown that SFR diagnostics that provide a measure of the recent, rather than present one, affect the recovery of the underlying population properties.

\section*{Acknowledgements}
MV thanks D. Alexander, E. Daddi and V. Wild for valuable suggestions and discussions. MV acknowledges funding support from NASA, through award ATP NNX10AC84G, from SAO,  through award TM1-12007X, from NSF, through award AST 1107675, and from a Marie Curie FP7-Reintegration-Grant within the 7th European Community Framework Programme (PCIG10-GA-2011-303609). HN acknowledges support by the Israel Science Foundation grant 284/13 
This work was granted access to the HPC resources of TGCC under the allocations 2013-t2013046955 and 2014-x2014046955 made by GENCI. This research was supported in part by the National Science Foundation under grant no. NSF PHY11-25915, through the Kavli Institute for Theoretical Physics and its program `A Universe of Black Holes'. PRC thanks the Institut d'Astrophysique de Paris for hosting him during his visits.

\scalefont{0.94}
\setlength{\bibhang}{1.6em} 
\setlength\labelwidth{0.0em}
\bibliographystyle{mn2e}

\appendix
\section{Parameter study: resolution and feedback efficiency}
\label{appendix}

In this section we discuss the robustness of our results against resolution and AGN feedback efficiency 
\citep[see also][]{2014MNRAS.443.1125T}. We have performed three lower-resolution simulations of the 1:2 coplanar, prograde-prograde merger with 30\% gas fraction (m2.gf0.3.pro), where we degraded the mass resolution by a factor of 4, and the softening by a factor $4^{1/3}$, one with the same feedback efficiency of $\epsilon_f=0.001$ as in the high-resolution suite, one with $\epsilon_f=0.002$  and one with $\epsilon_f=0.005$. We have also performed an additional high-resolution simulation with $\epsilon_f=0.005$.

\begin{figure}
\centering
\includegraphics[width=\columnwidth,angle=0]{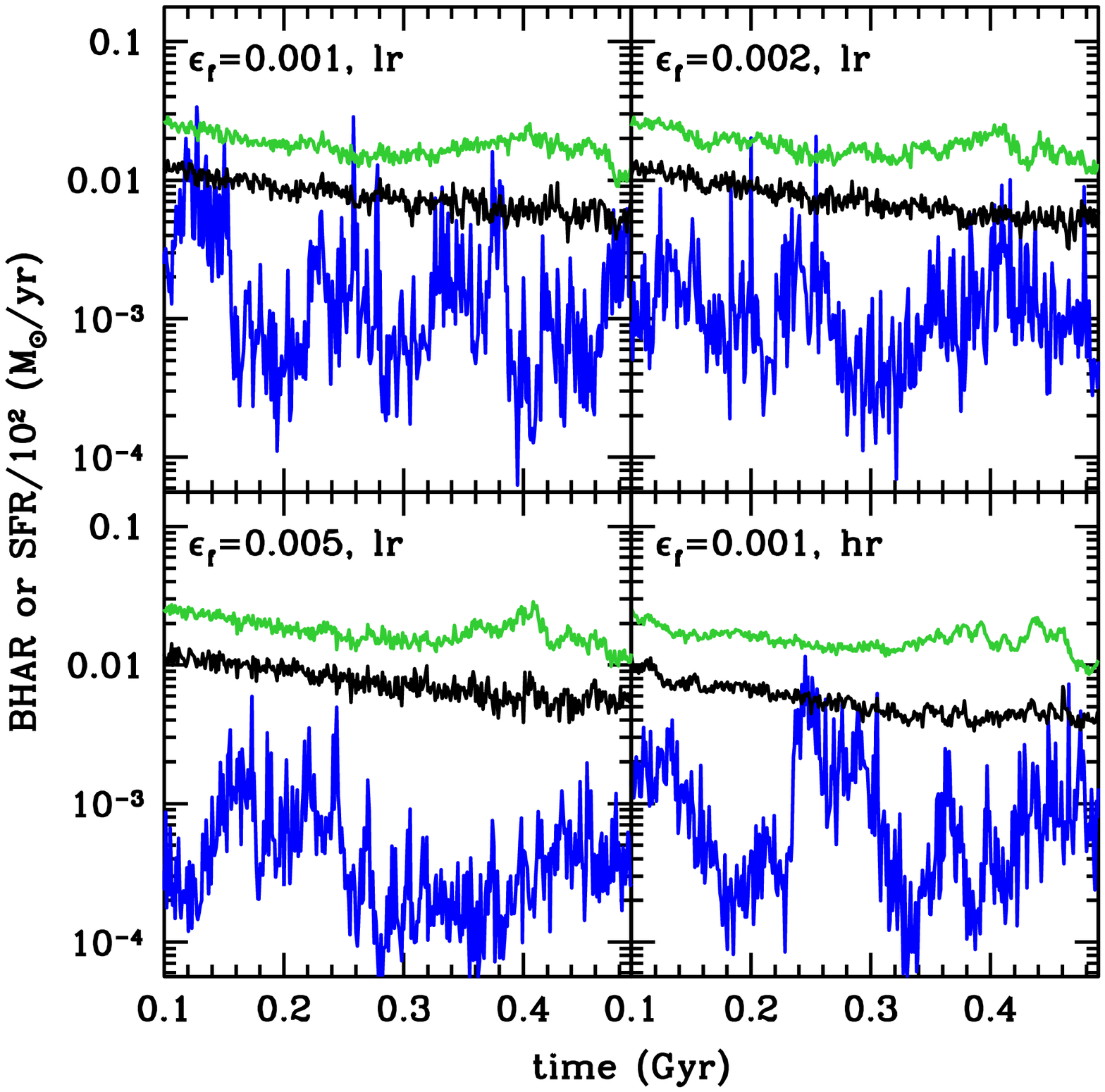}
\includegraphics[width=\columnwidth,angle=0]{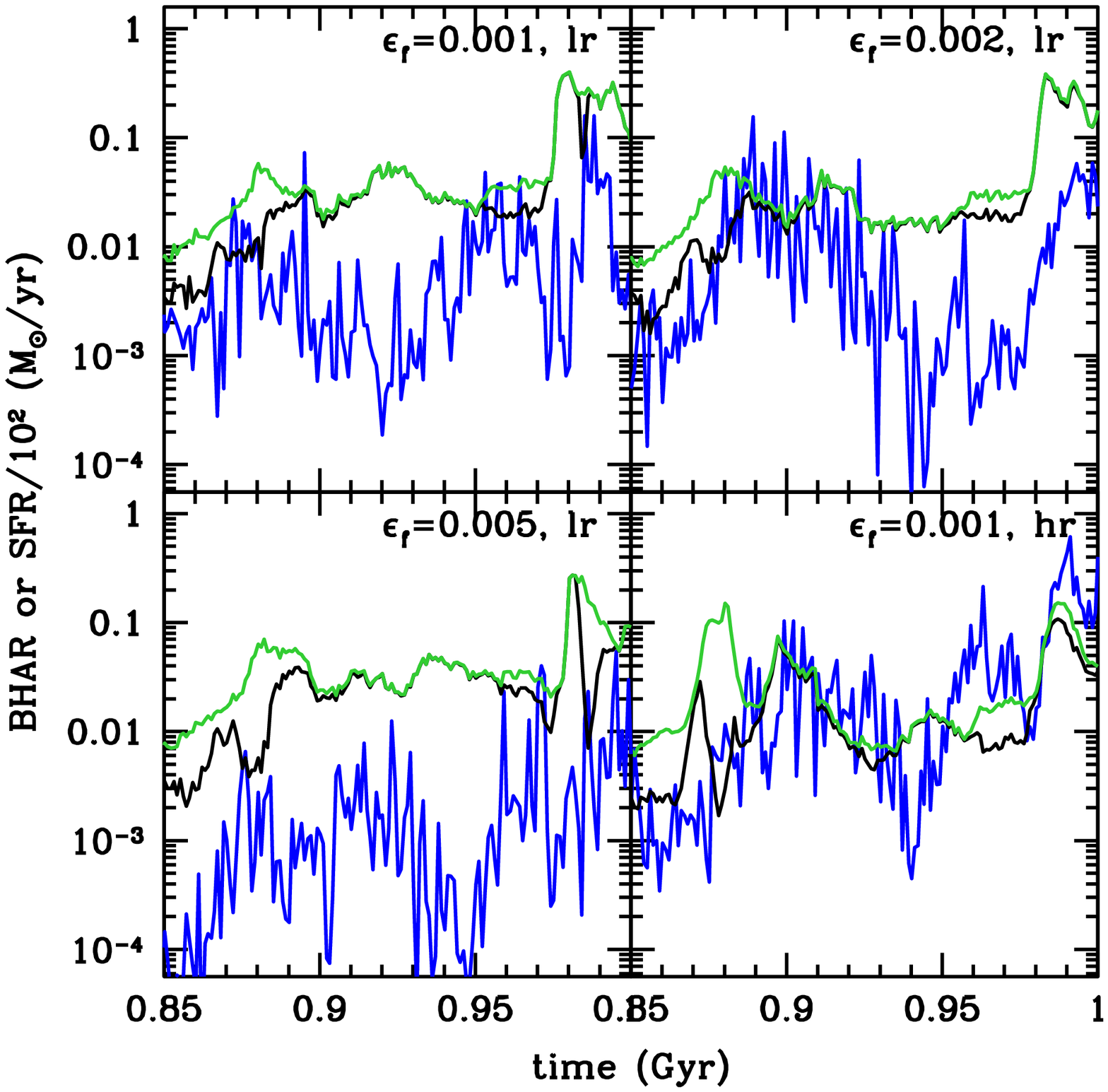}
\caption{Effect of changing resolution (lower resolution by a factor of four) and feedback efficiency ($\epsilon_f=0.002$ or 0.005) for $G_1$ in the 1:2 coplanar, 
prograde-prograde merger with a gas fraction of 30\%. 
Feedback efficiency and resolution are labelled in each panel.
For each simulation we select the same time-steps (top four panels: stochastic phase; bottom four panels: merger phase). We show the BHAR (blue, lower thin solid curve) 
and the SFR in the central 1~kpc (black, thick solid curve) and 5~kpc (green, upper thin solid curve) all as a function of time.}
\label{fig:sfr_bhar_res}
\end{figure}

\begin{figure}
\centering
\includegraphics[width=\columnwidth,angle=0]{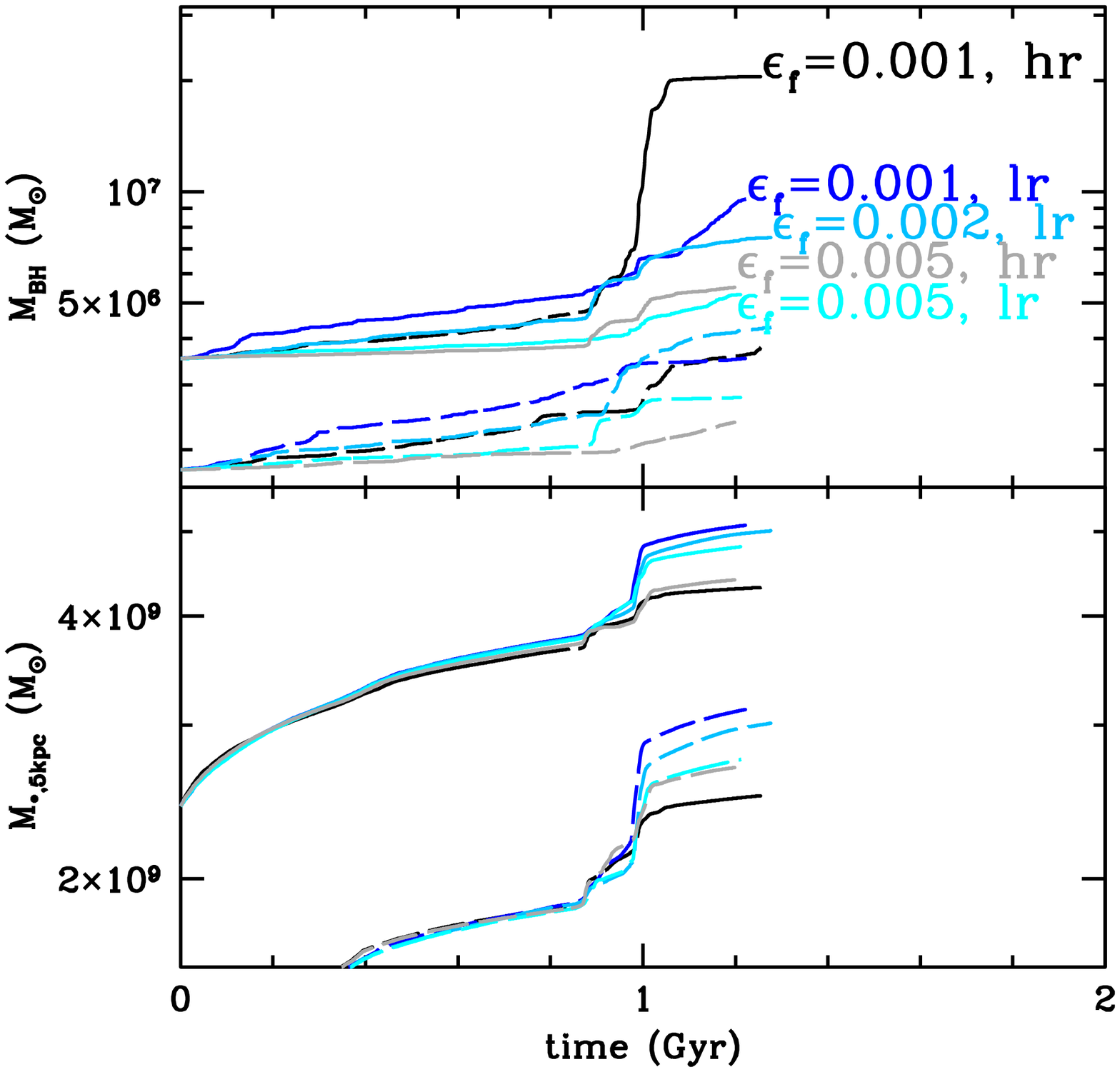}
\caption{Effect of changing resolution (lower resolution by a factor of four) and feedback efficiency ($\epsilon_f=0.002$ or 0.005) on the BH (top, $BH_1$: solid; $BH_2: dashed$) and stellar growth (bottom, $G_1$: solid; $G_2: dashed$)  in the 1:2 coplanar, prograde-prograde merger with gas fraction 30\%. The color coding is labelled for $BH_1$ and applies to all curves.}
\label{fig:bhar_res}
\end{figure}

In Fig.~\ref{fig:sfr_bhar_res} we compare the BHAR and SFR for these four runs of the same merger 
(`lr' and `hr' in the figure captions stands for low-resolution and high-resolution respectively), excluding the nuclear SFR, $\sfrn$, as we do not resolve
 adequately that  region in the low-resolution runs. 
We supplement this figure with the SFR within 1~kpc as a proxy for the central regions, however.  It is clear that $\sfrg$ is robust against changes in 
resolution or AGN feedback efficiency during the stochastic phase, while the BHAR decreases as AGN feedback strength increases.  
In the merger phase local dynamics becomes more important than feedback efficiency for BHAR, and the low-resolution run with $\epsilon_f=0.001$ has a BHAR similar 
to the low-resolution runs with $\epsilon_f=0.002$ and $\epsilon_f=0.005$. The new stellar mass formed changes by at most 30\% among runs with different $\epsilon_f$. On the other hand, resolution has an important effect: small-scale gravitational torques and small-scale 
over-densities cannot be resolved at low-resolution, leading to an `average' BH growth that may be higher than (e.g., in the case of $BH_1$) or similar to  (in the case of $BH_2$)  the high-resolution run, where the region near the BHs is well resolved (see Fig.~\ref{fig:bhar_res}, $0.85<t<1.1$~Gyr). The stellar mass formed changes  by up to 50\% in runs with different resolution.

\begin{figure}
\centering
\includegraphics[width=\columnwidth,angle=0]{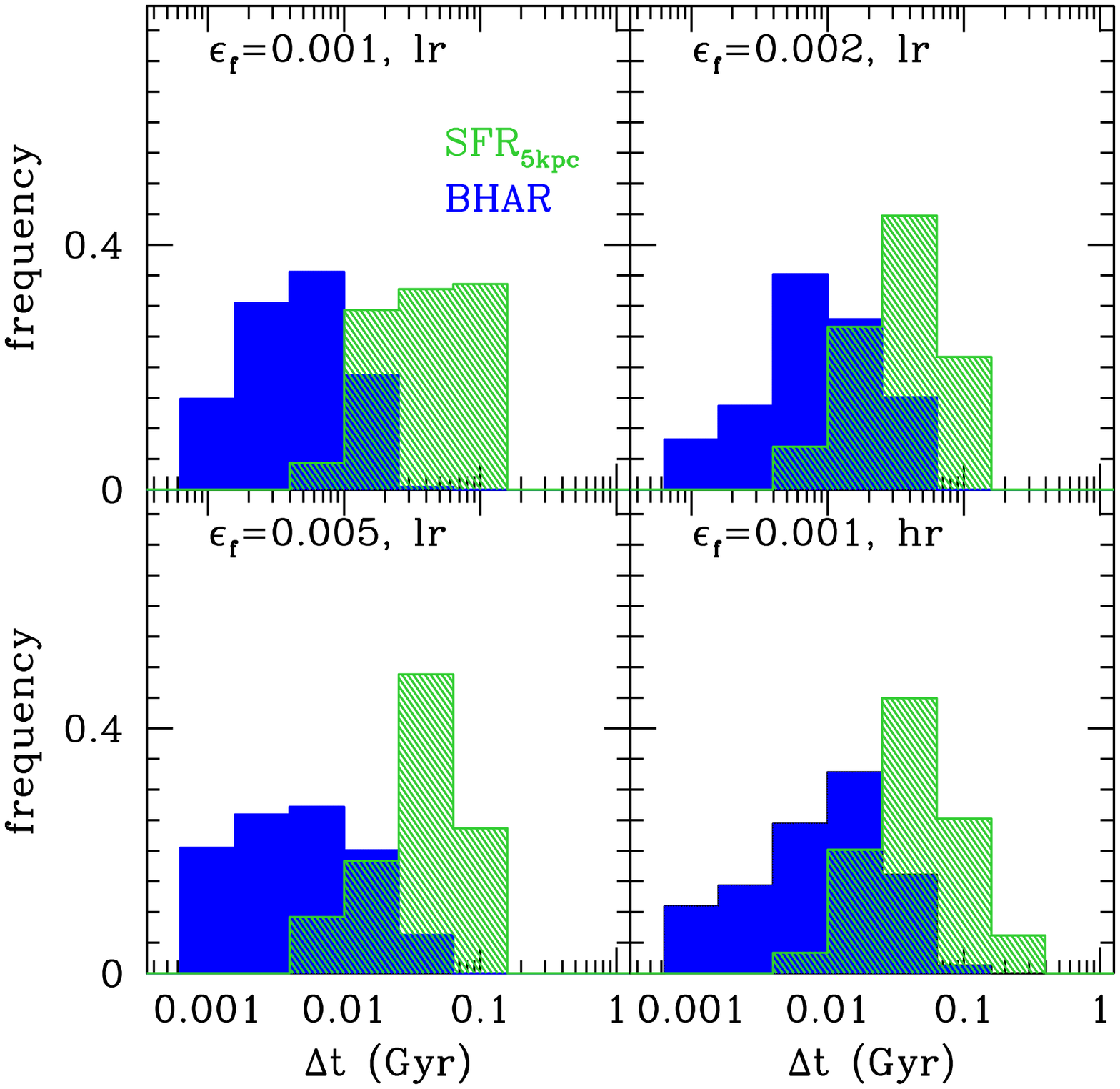}
\caption{Distribution of the time, $\Delta t$,  needed for a quantity X, where X=BHAR (blue) and X=$\sfrg$ (green),  to vary by a factor of 10 during the merger phase.  The results on temporal variability are robust against different resolution or feedback efficiency choices.}
\label{fig:variabilityapp}
\end{figure}

 In general, different AGN feedback strength or resolution modify the normalization of the SFR or BHAR, but not their temporal trends (Fig.~\ref{fig:variabilityapp}). 
Therefore, the results discussed in Sections~\ref{sec:General} and~\ref{sec:timecorr} are robust against the choice of parameters and resolution, 
keeping in mind the caveat that $\sfrn$ cannot be measured for the low-resolution runs as the 100~pc region is not sufficiently resolved. We also find that the results on variability presented in Section~\ref{sec:timevar} are also robust: the time over which SFR and BHAR vary is not much affected by the AGN feedback efficiency or resolution.

The results discussed in sections~\ref{sec:growth} and~\ref{sec:magcorr} are more sensitive to  changes in the normalization of SFR and BHAR, and 
we warn the reader to keep this caveat in mind. In particular, in all but the low-resolution, $\epsilon_f=0.001$ run,
 we see consistently an increase in the ratio between BHAR and SFR in the merger phase compared to the stochastic phase. 
In the low-resolution, $\epsilon_f=0.001$ run shown in Fig.~\ref{fig:bhar_res} the stellar growth is the largest, while the BH growth is limited. 
If we take the ratio of BH mass to stellar mass,  in the stochastic phase, this ratio tends to decrease, and the decrease is stronger the higher is $\epsilon_f$. 
Therefore, this trend is caused primarily by the effect of AGN feedback on the BHAR (SFR is almost unaffected as discussed above). 
The ratio shows a high value plateau during the merger phase for all mergers, and returns to roughly constant or decreasing values.

\begin{figure}
\centering
\includegraphics[width=\columnwidth,angle=0]{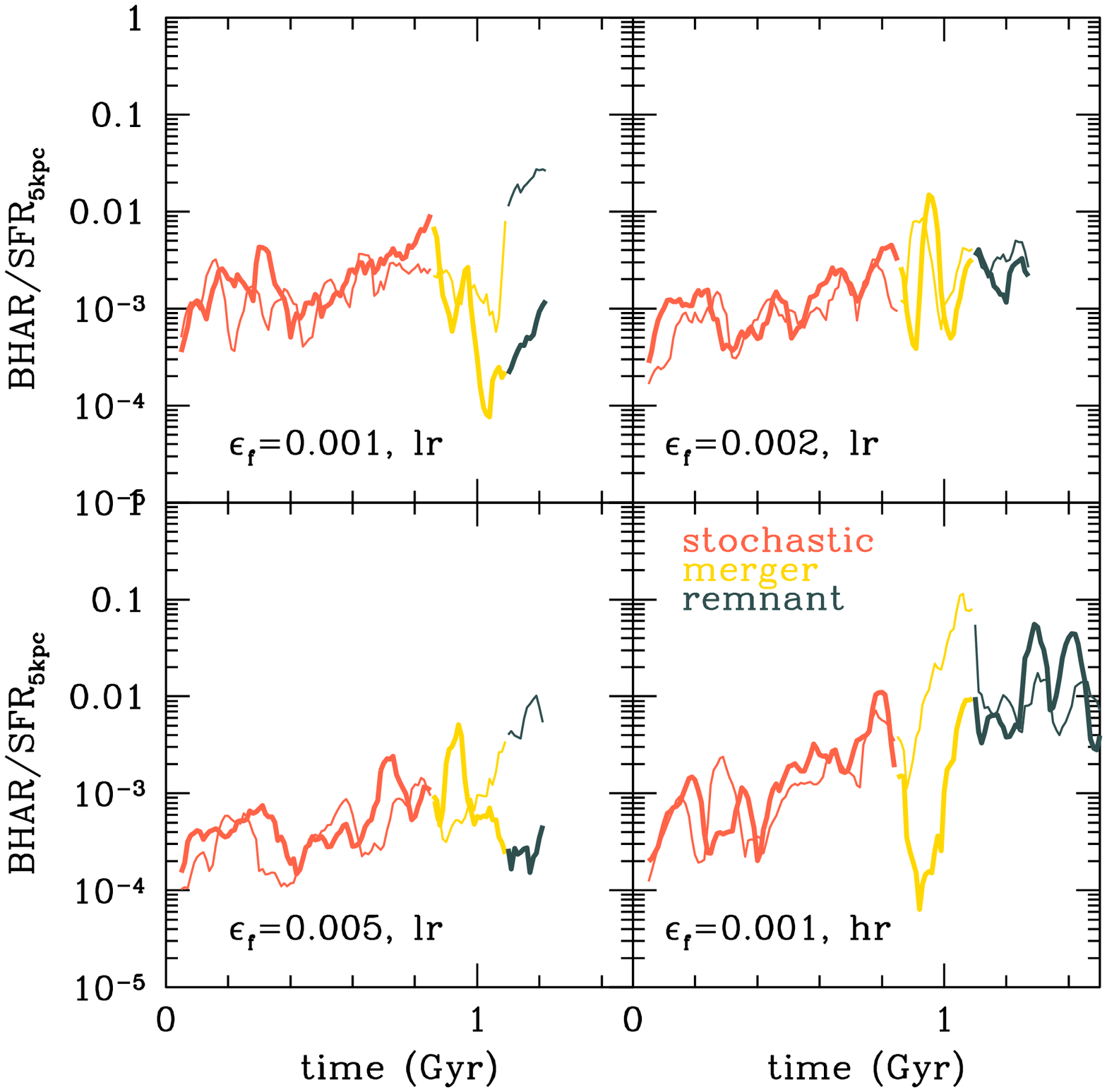}\vfill
\caption{Ratio of BHAR versus SFR within 5~kpc, averaging both quantities in bins of 10~Myr illustrating the effect
of changing resolution (lower resolution by a factor of four) and feedback efficiency ($\epsilon_f=0.002$ or 0.005).}
\label{fig:sfr_bhar_ratio_res}
\end{figure}

Fig.~\ref{fig:sfr_bhar_ratio_res} explicitly shows how the ratio BHAR/SFR within 5~kpc, averaging both quantities in bins of 50~Myr, 
depends on resolution and feedback efficiency. By increasing the feedback efficiency the BHAR is affected more than $\sfrg$, 
i.e., the ratio decreases as $\epsilon_f$ increases.  
As a consequence,  in the run with $\epsilon_f=0.005$, 
longer time is spent at a lower ratio between BHAR and $\sfrg$: 50\% of the time in the stochastic phase is spent at a ratio $<2\times 10^{-4}$, 
and 50\% of the merger phase has ratio $<10^{-3}$. We nevertheless find a relative enhancement between the stochastic and merger phase. 
We note that if AGN feedback were as strong as $\epsilon_f=0.005$  BHs would require growth boosts driven by mergers in order to attain, 
over cosmic history, a mass compatible with the BH-bulge correlation. The feedback strength we chose for the reference runs allows the BHs to grow towards the 
BH-bulge correlation through stochastic   low-level activity, rather than through merger-driven events \citep{2013ApJ...779..136B}.

\begin{figure}
\centering
\includegraphics[width=\columnwidth,angle=0]{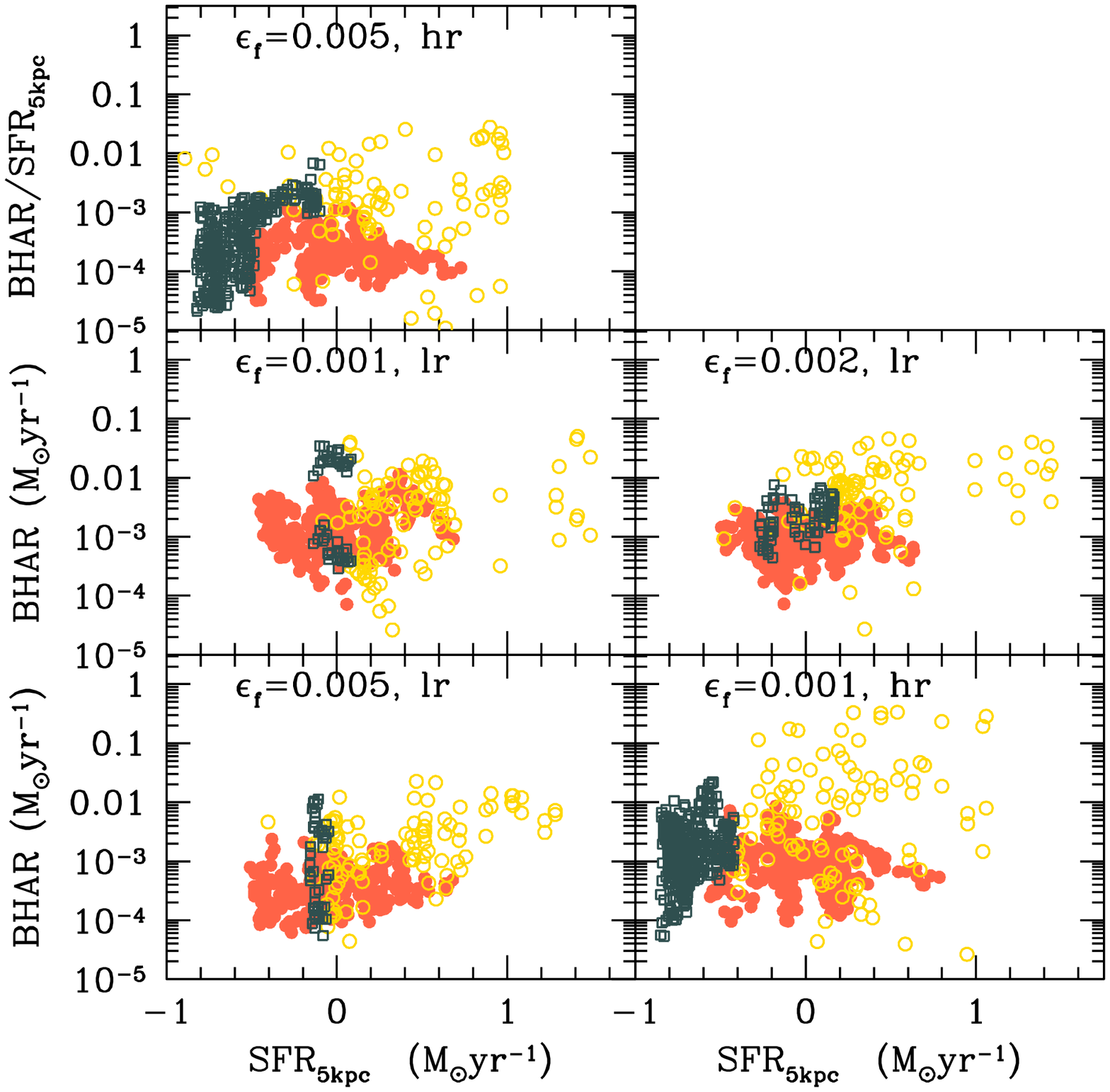}
\caption{BHAR versus SFR within 5~kpc (top) and within 100~pc (bottom), averaging both quantities in bins of 10~Myr illustrating the effect
of changing resolution (lower resolution by a factor of four) and feedback efficiency ($\epsilon_f=0.002$ or 0.005).}
\label{fig:sfr_vs_bhar_res}
\end{figure}

Finally, if we plot BHAR versus $\sfrg$ (Fig.~\ref{fig:sfr_vs_bhar_res}; cf. Fig.~\ref{fig:sfr_bhar_2w_app}), we find consistent trends regardless of resolution 
and AGN feedback efficiency. In the stochastic phase we obtain a roughly spherical blob spanning the same BHAR and SFR as in Fig.~\ref{fig:sfr_bhar_2w_app} (red-orange points), while in the merger phase the blob expands to higher BHAR and SFR, as in the reference run (golden points).  Low resolution runs show, visually, a  somewhat better SFR-BHAR trend for all feedback efficiencies.

\clearpage
\section{Supplementary figures}
\clearpage

\begin{figure}
\vspace{0.5cm}
\centering
\includegraphics[width=\columnwidth,angle=0]{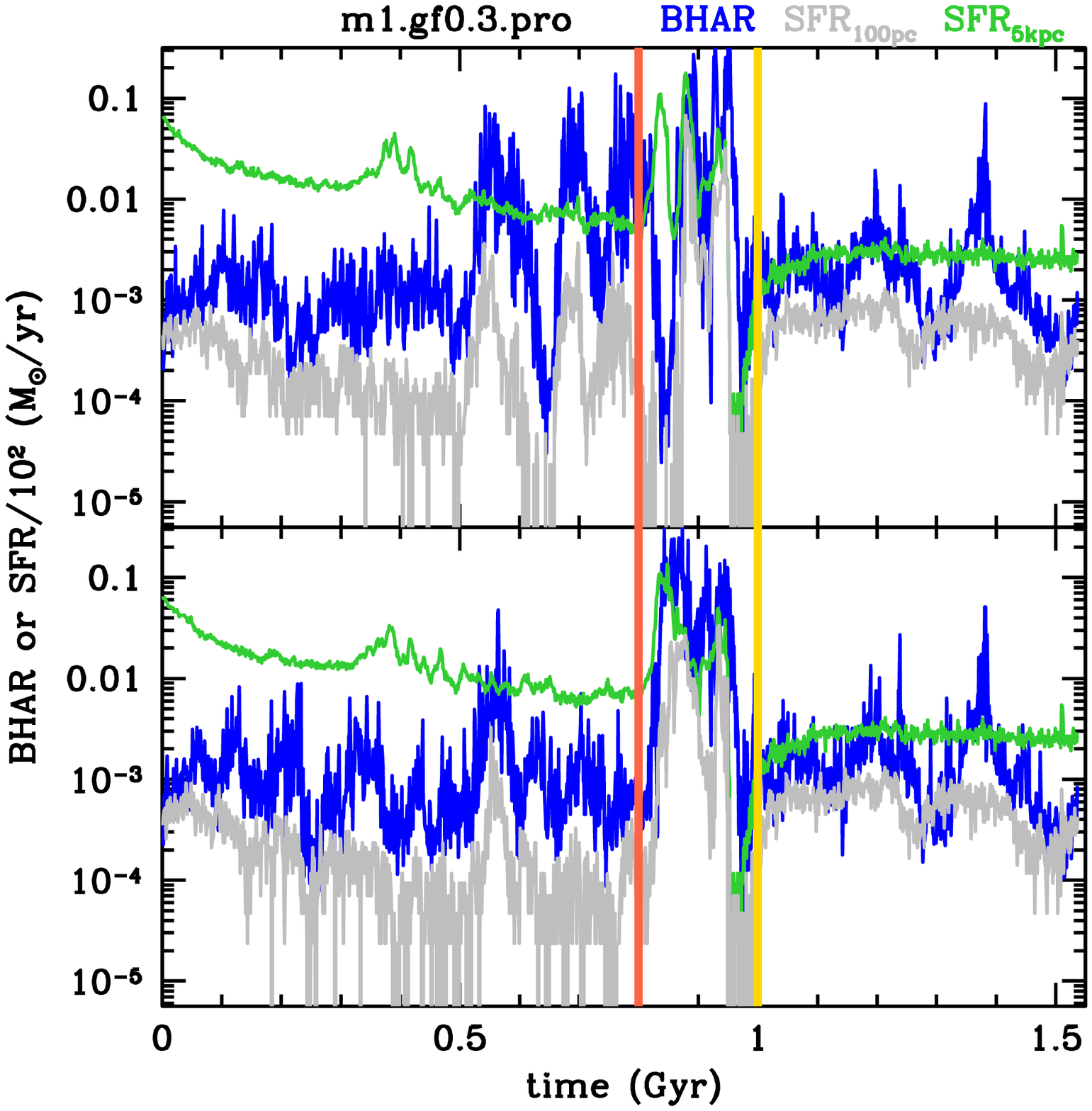}
\caption{BHAR and SFR  for the 1:1 coplanar, prograde-prograde merger. Top panel: $G_1$. Bottom panel: $G_2$.}
\end{figure}

\begin{figure}
\centering
\includegraphics[width=\columnwidth,angle=0]{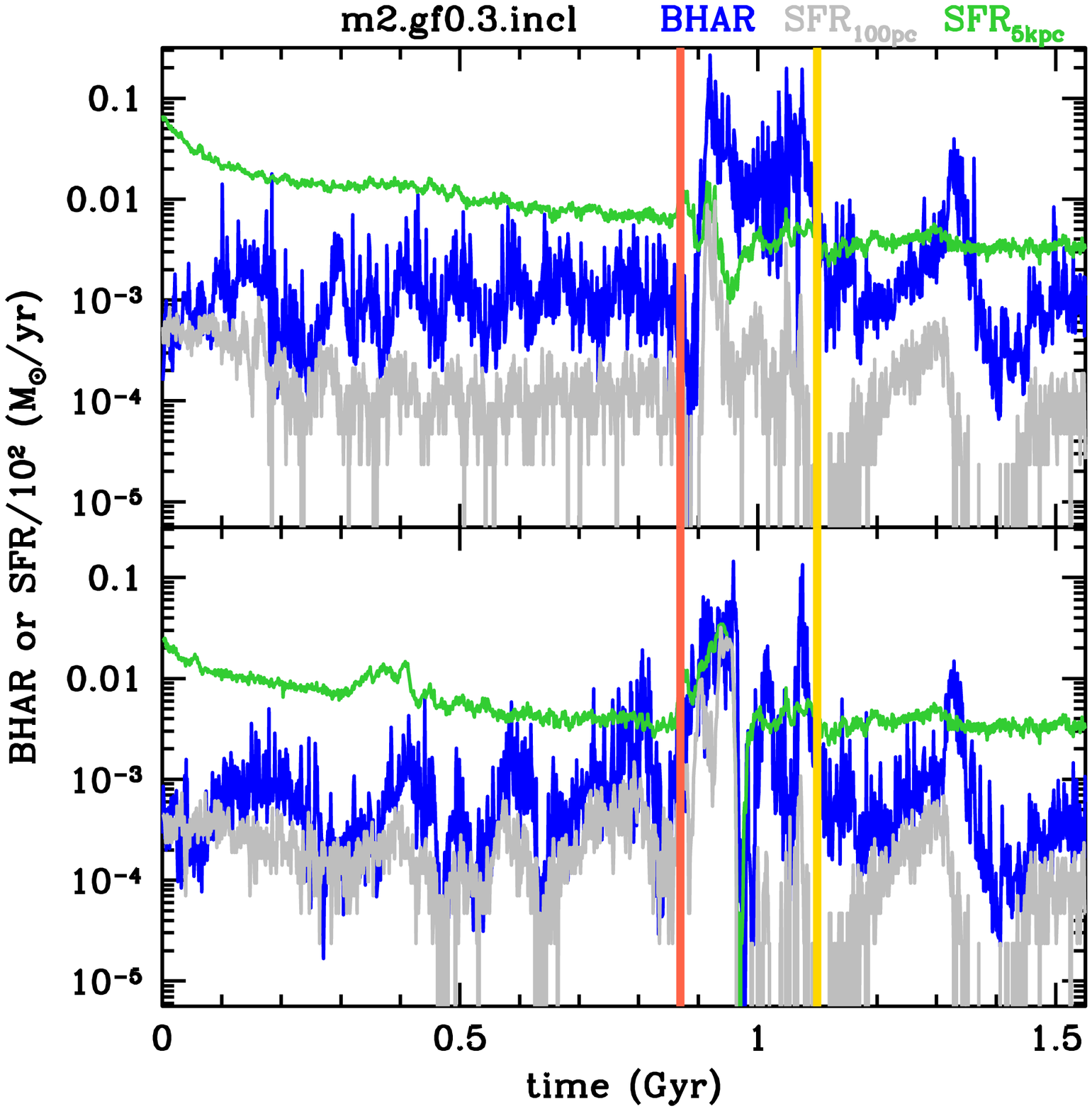}
\caption{BHAR and SFR  for the 1:2 inclined, prograde-prograde merger. Top panel: $G_1$. Bottom panel: $G_2$.}
\end{figure}

\begin{figure}
\vspace{0.5cm}
\centering
\includegraphics[width=\columnwidth,angle=0]{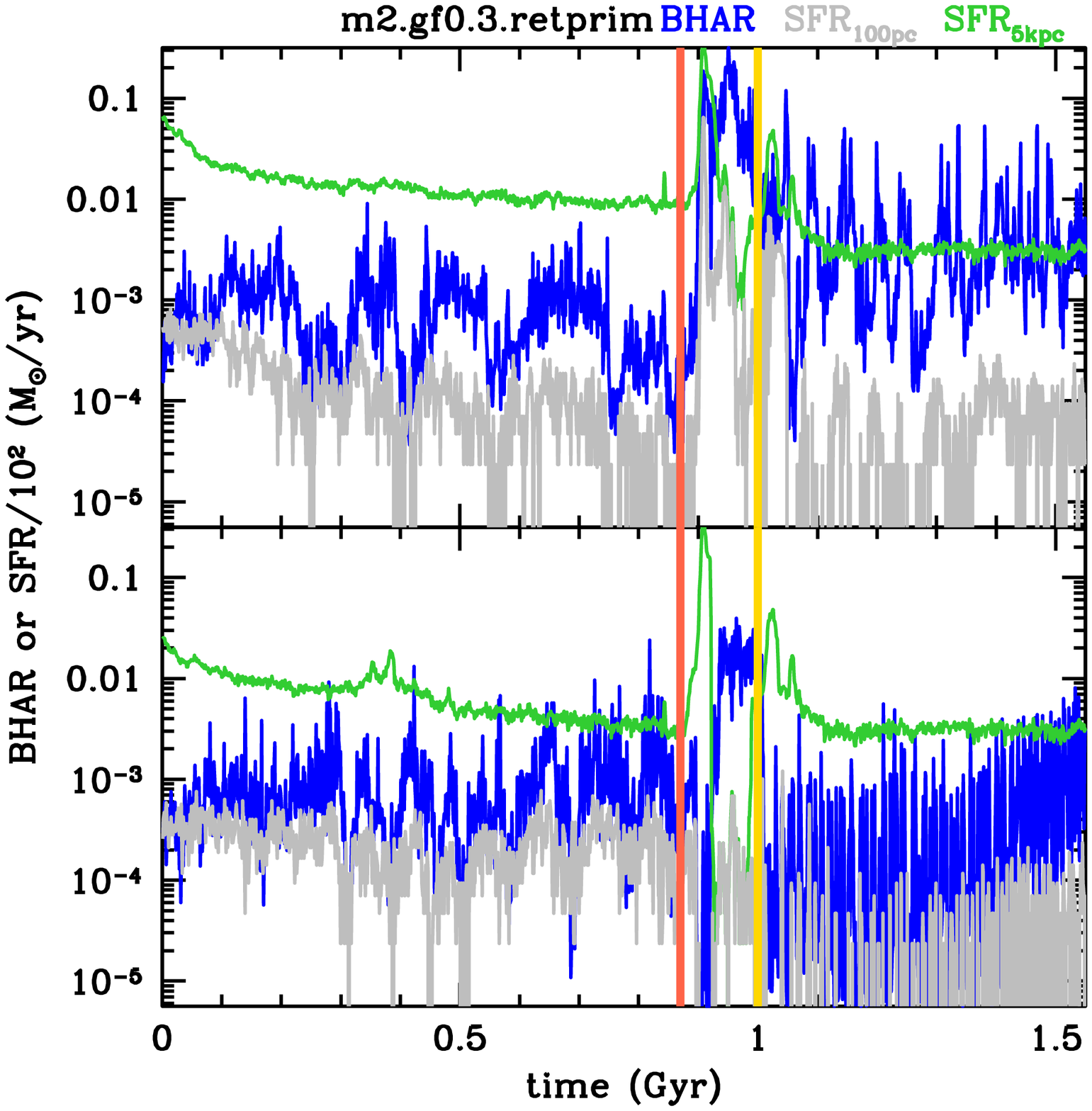}
\caption{BHAR and SFR  for the 1:2 coplanar, retrograde-prograde merger. Top panel: $G_1$. Bottom panel: $G_2$.}
\end{figure}

\begin{figure}
\centering
\includegraphics[width=\columnwidth,angle=0]{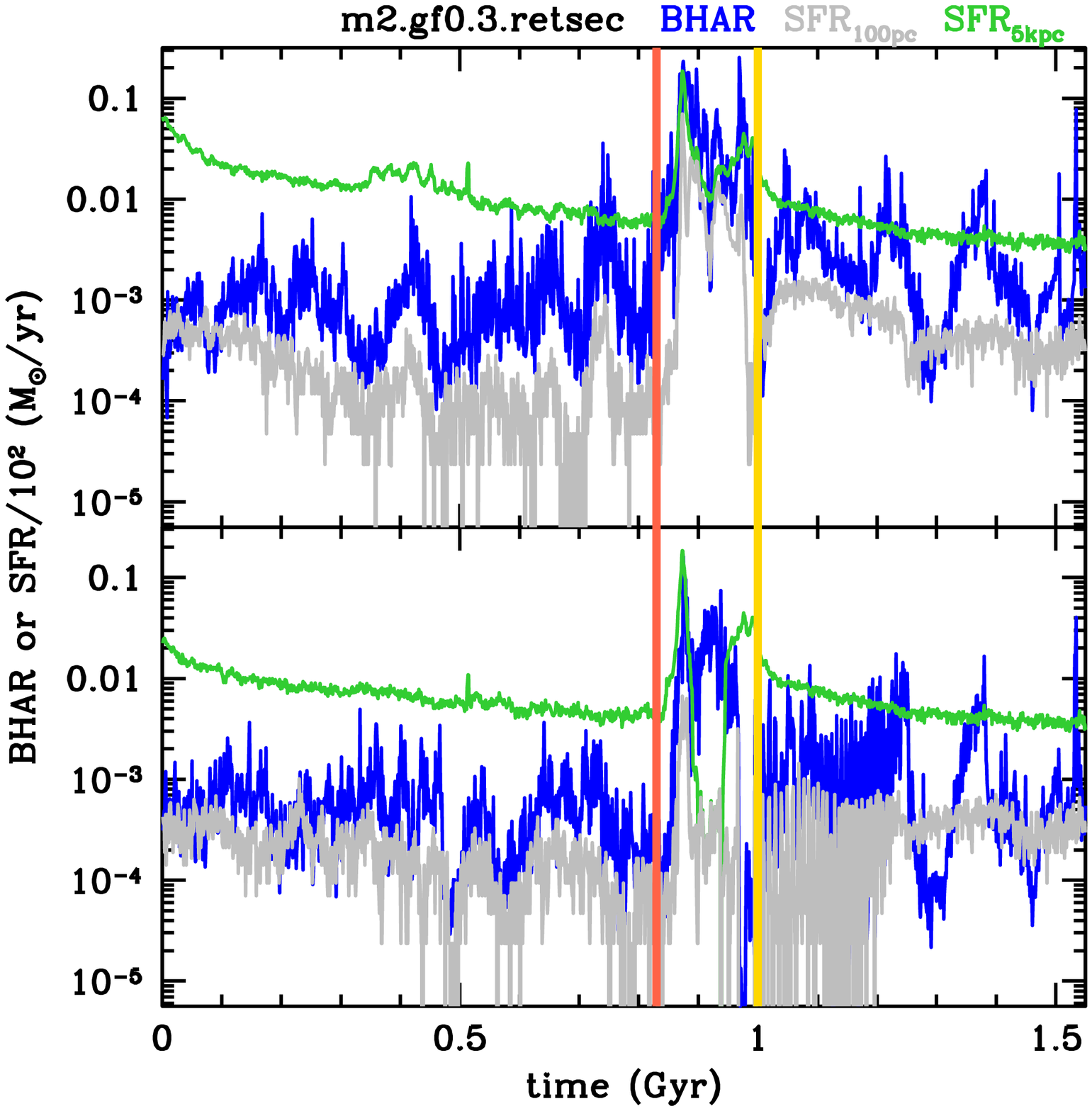}
\caption{BHAR and SFR  for the 1:2 coplanar, prograde-retrograde merger. Top panel: $G_1$. Bottom panel: $G_2$.}
\end{figure}

\begin{figure}
\centering
\includegraphics[width=\columnwidth,angle=0]{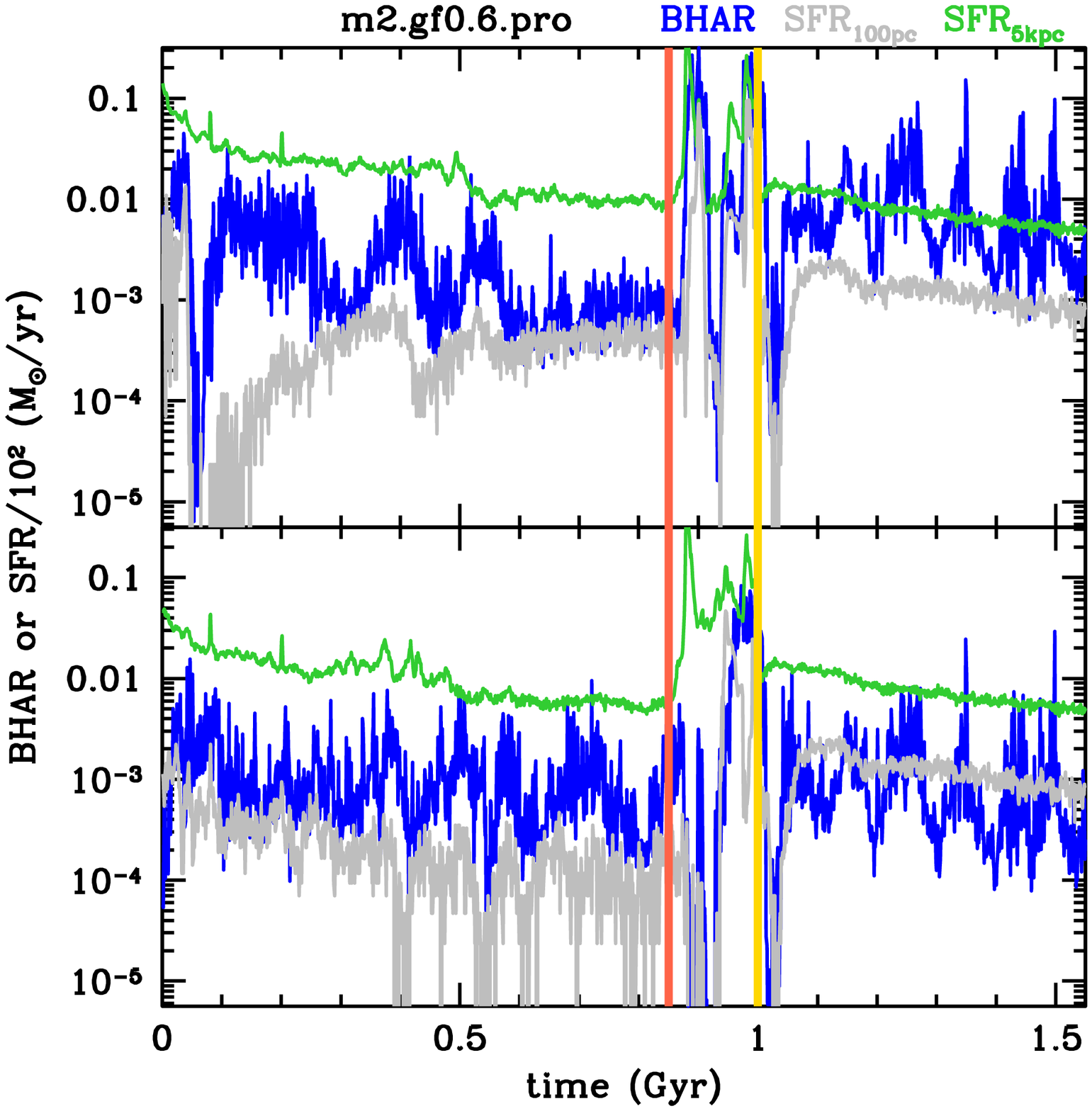}
\caption{BHAR and SFR  for the 1:2 coplanar, prograde-prograde merger, 60\% gas fraction. Top panel: $G_1$. Bottom panel: $G_2$.}
\end{figure}

\begin{figure}
\centering
\includegraphics[width=\columnwidth,angle=0]{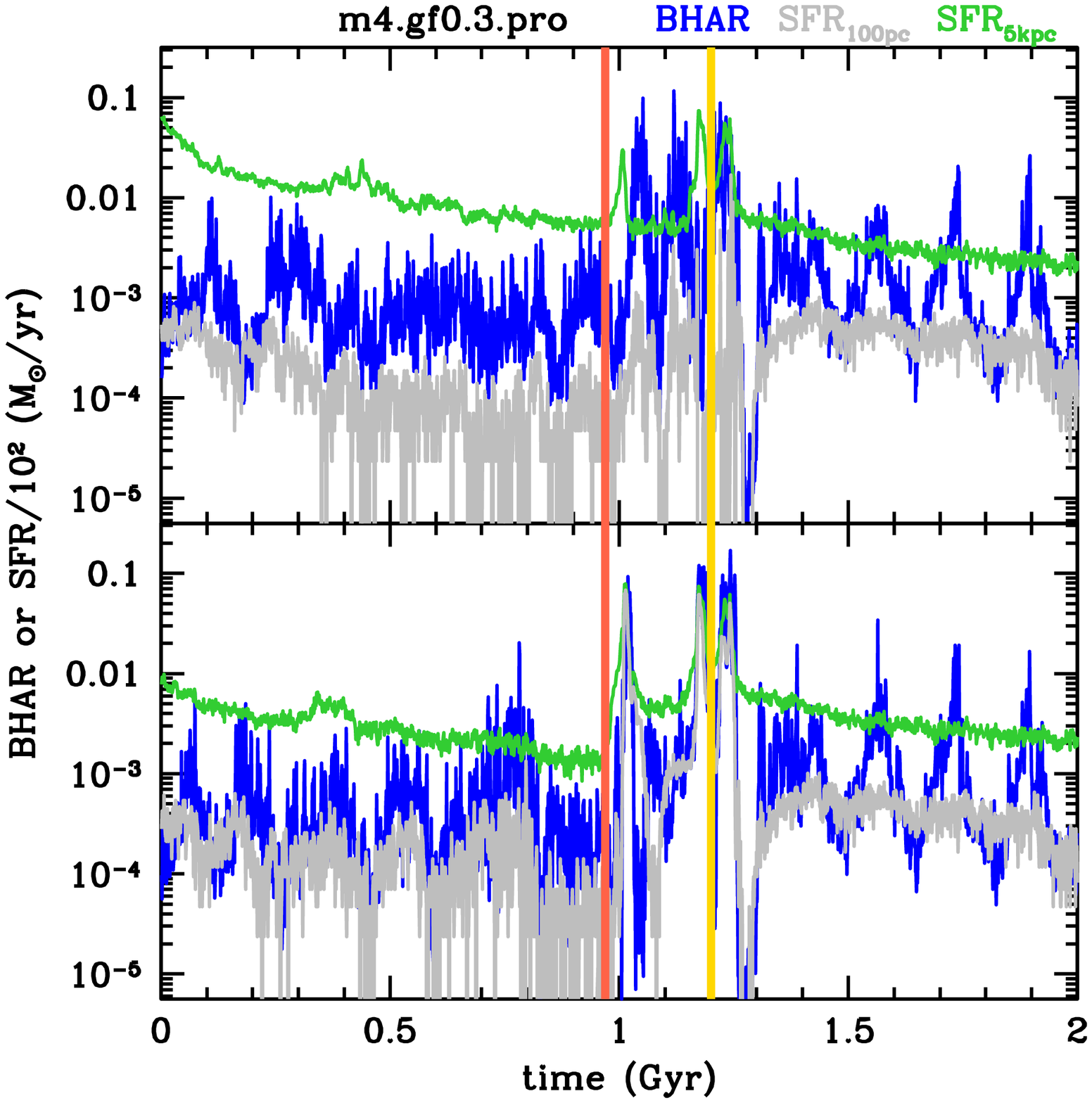}
\caption{BHAR and SFR  for the 1:4 coplanar, prograde-prograde merger. Top panel: $G_1$. Bottom panel: $G_2$.}
\end{figure}

\begin{figure}
\centering
\includegraphics[width=\columnwidth,angle=0]{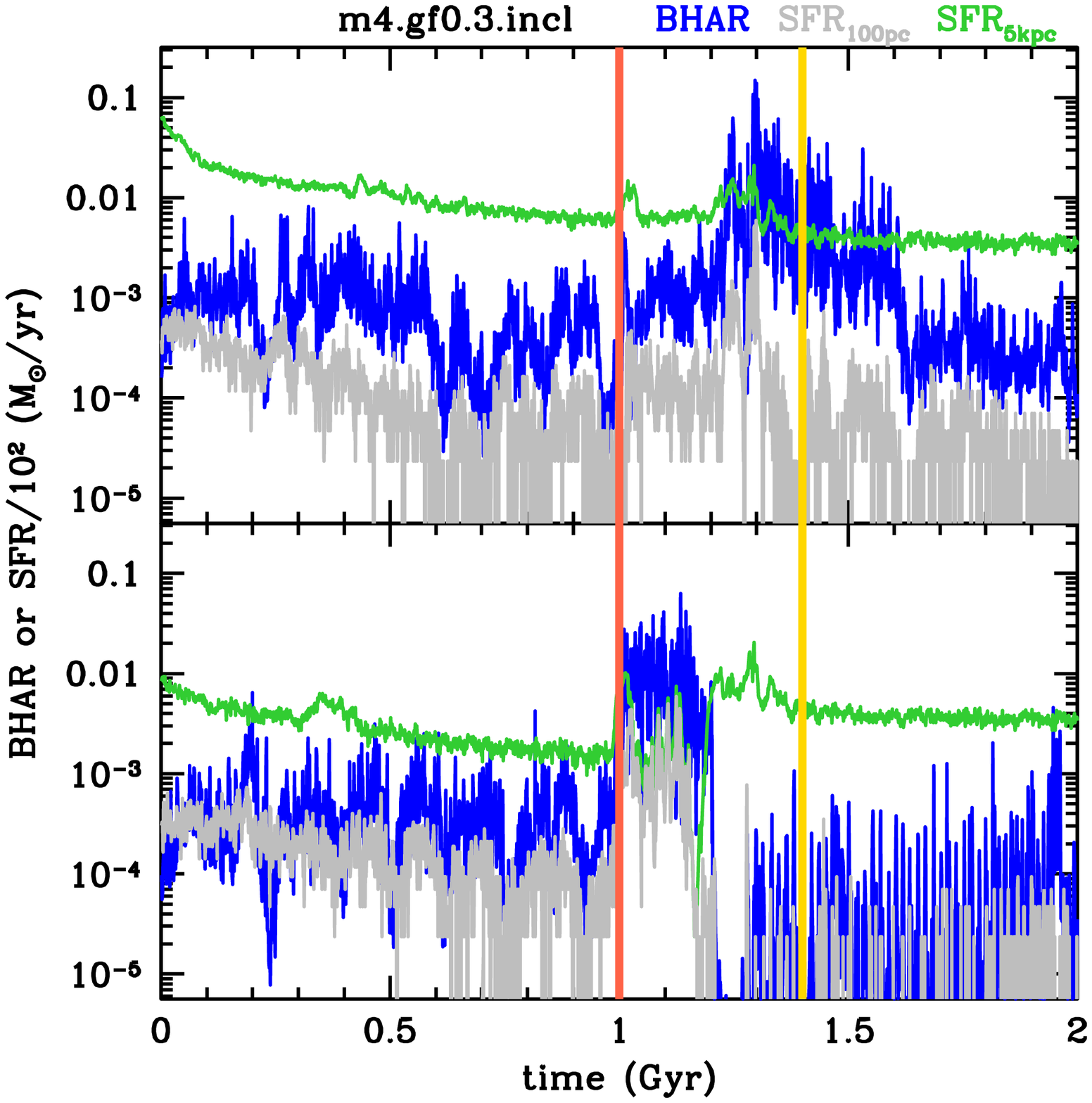}
\caption{BHAR and SFR  for the 1:4 inclined, prograde-prograde merger. Top panel: $G_1$. Bottom panel: $G_2$.}
\end{figure}

\begin{figure}
\centering
\includegraphics[width=\columnwidth,angle=0]{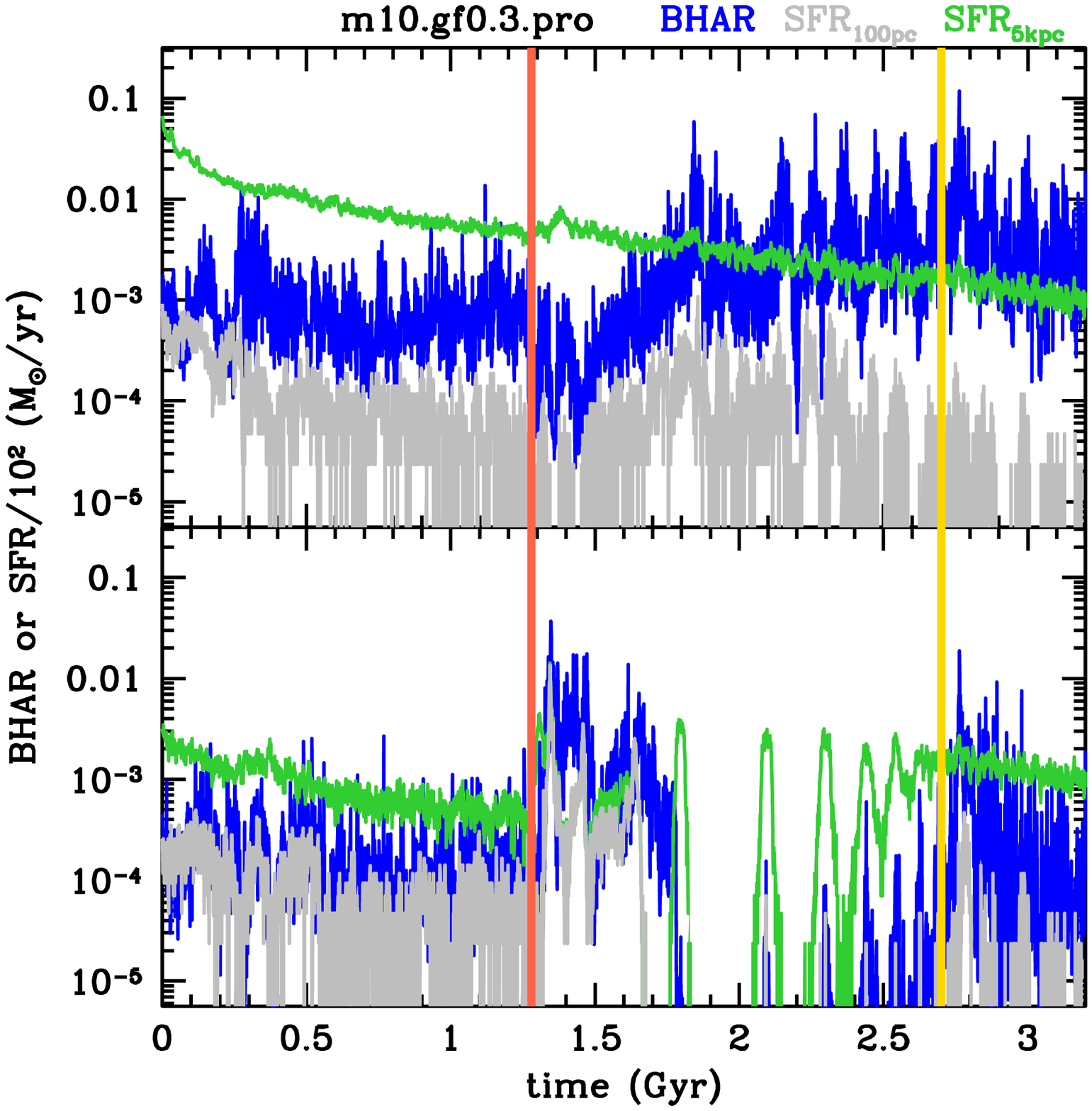}
\caption{BHAR and SFR  for the 1:10 coplanar, prograde-prograde merger. Top panel: $G_1$. Bottom panel: $G_2$.}
\end{figure}

\begin{figure}
\centering
\includegraphics[width=\columnwidth,angle=0]{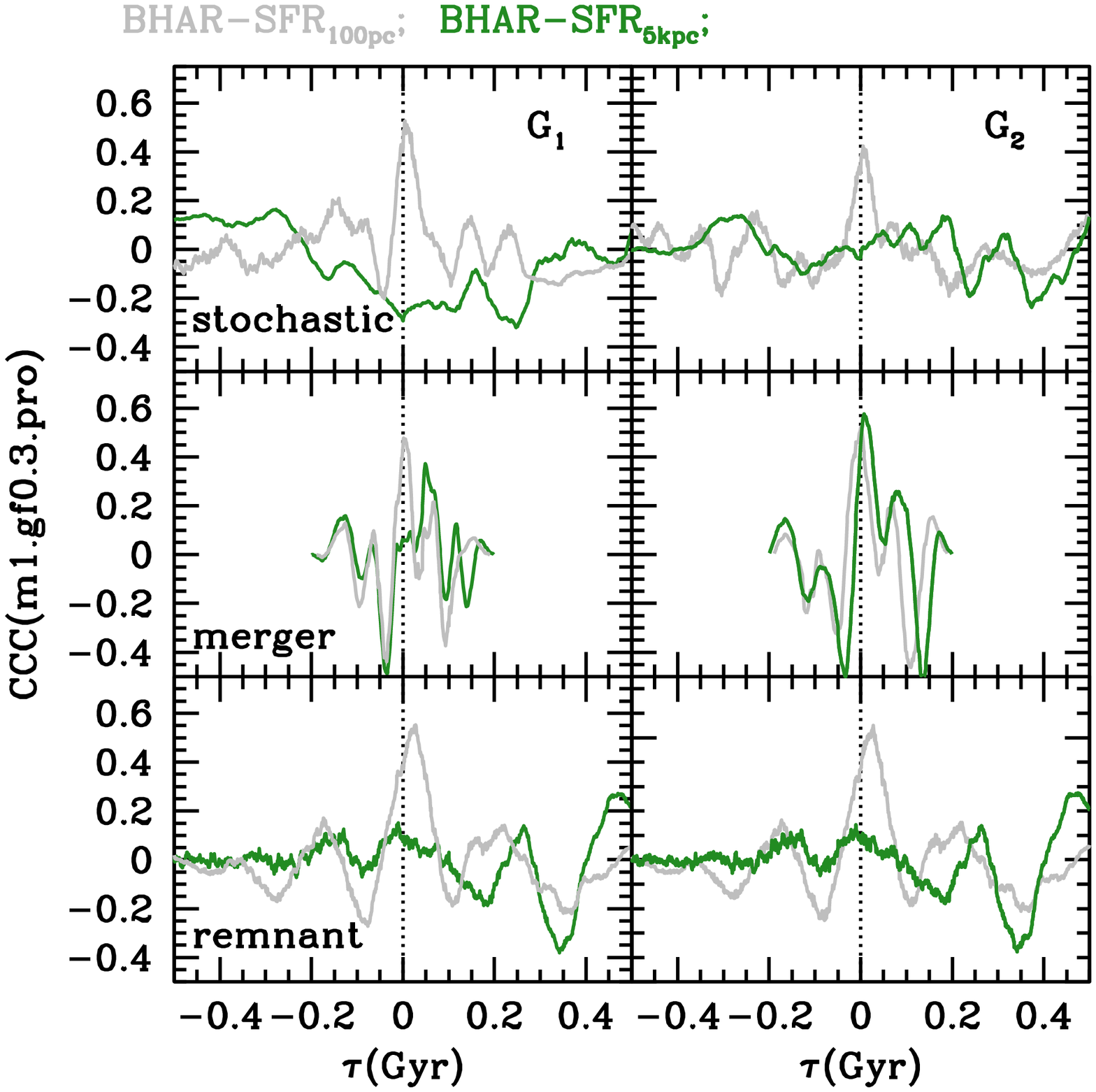}
\caption{Cross-correlation function for BHAR and SFR  for the 1:1 coplanar, prograde-prograde merger.}
\end{figure}

\begin{figure}
\vspace{-0.5cm}
\centering
\includegraphics[width=\columnwidth,angle=0]{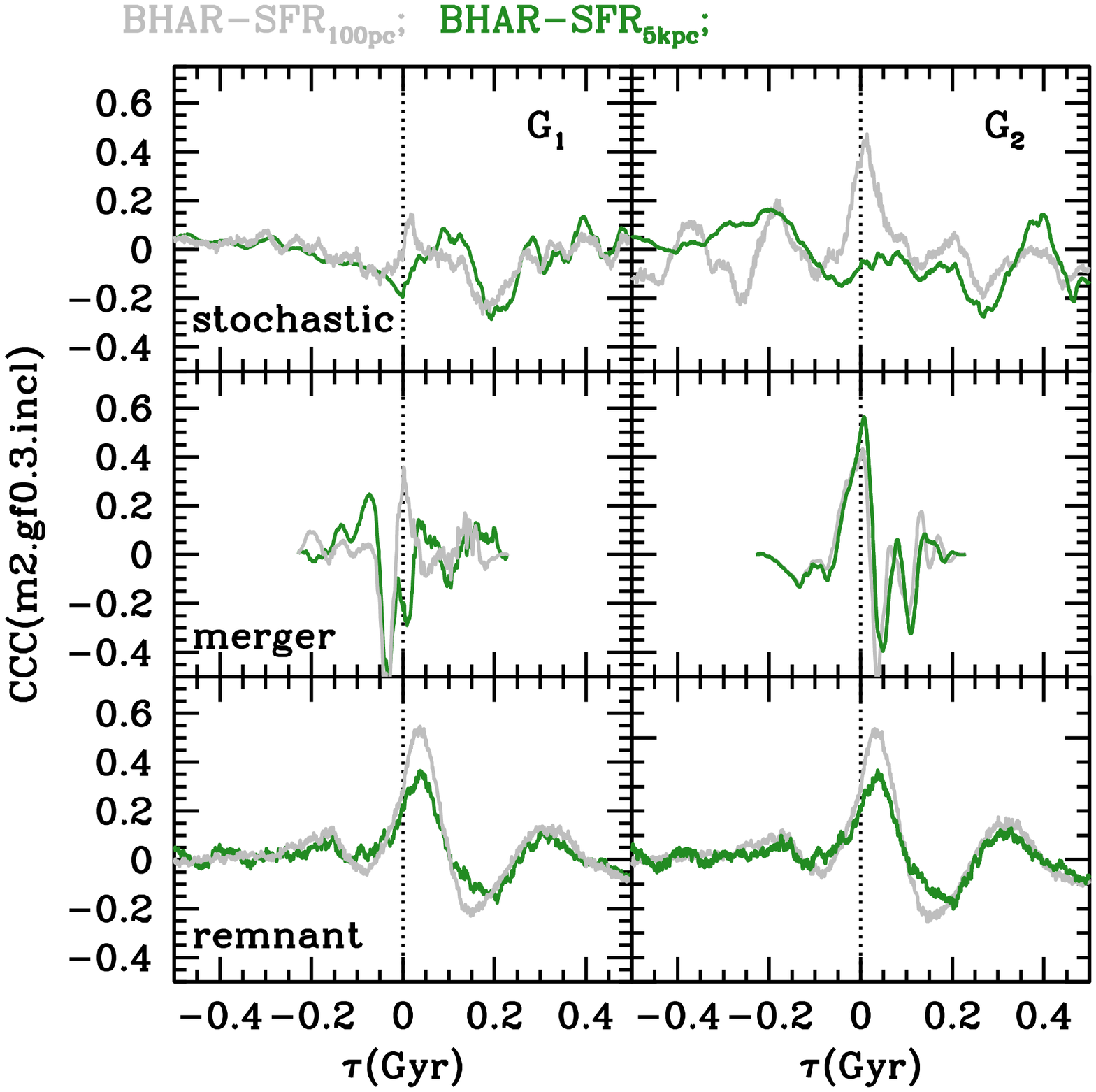}
\caption{Cross-correlation function for BHAR and SFR  for the 1:2 inclined, prograde-prograde merger.}
\end{figure}

\begin{figure}
\centering
\includegraphics[width=\columnwidth,angle=0]{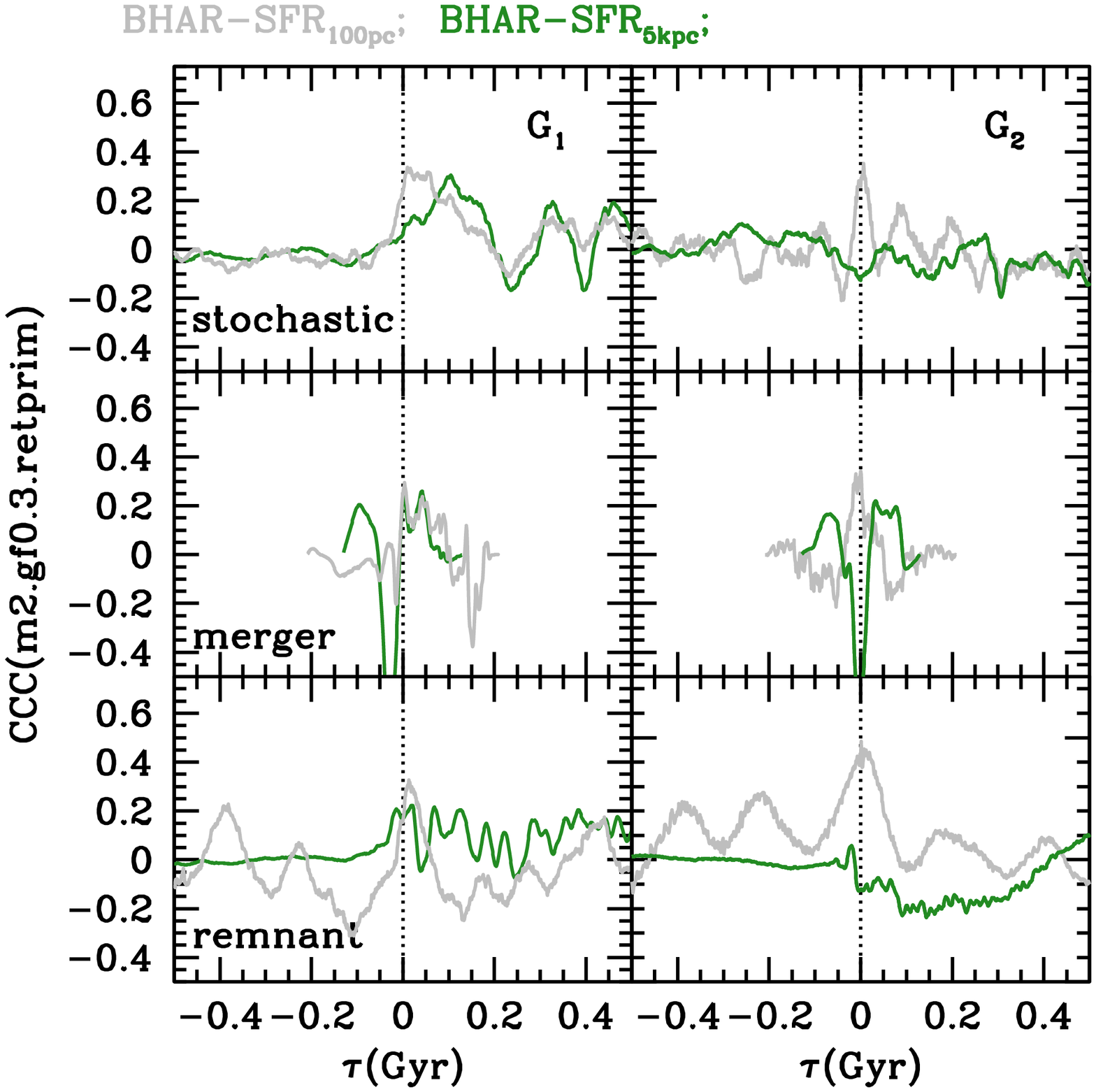}
\caption{Cross-correlation function for BHAR and SFR  for the 1:2 coplanar, retrograde-prograde merger.}
\end{figure}

\begin{figure}
\vspace{-0.5cm}
\centering
\includegraphics[width=\columnwidth,angle=0]{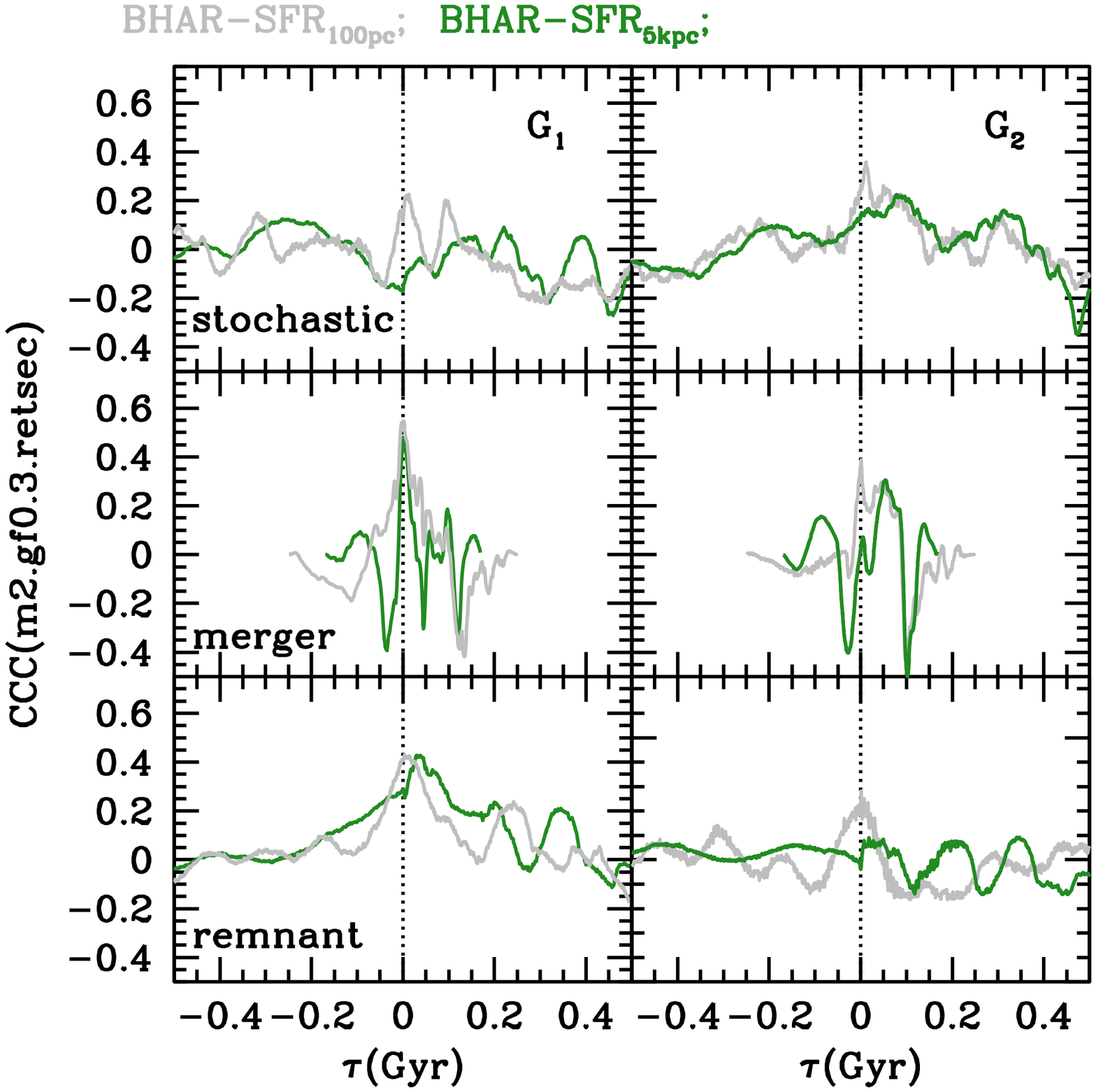}
\caption{Cross-correlation function for BHAR and SFR  for the 1:2 coplanar, prograde-retrograde merger.}
\end{figure}

\begin{figure}
\centering
\includegraphics[width=\columnwidth,angle=0]{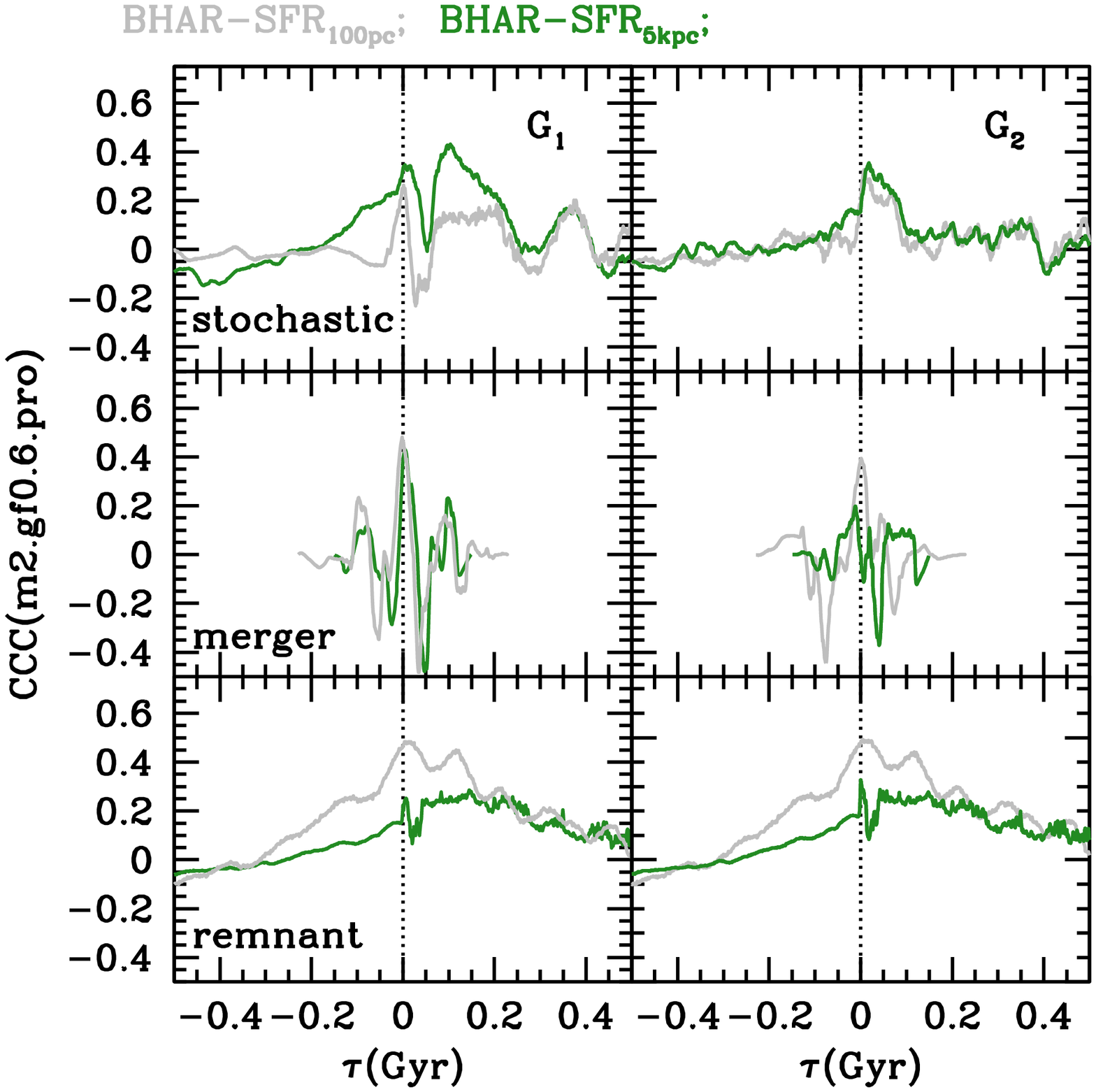}
\caption{Cross-correlation function for BHAR and SFR  for the 1:2 coplanar, prograde-prograde merger, 60\% gas fraction.}
\end{figure}

\begin{figure}
\centering
\includegraphics[width=\columnwidth,angle=0]{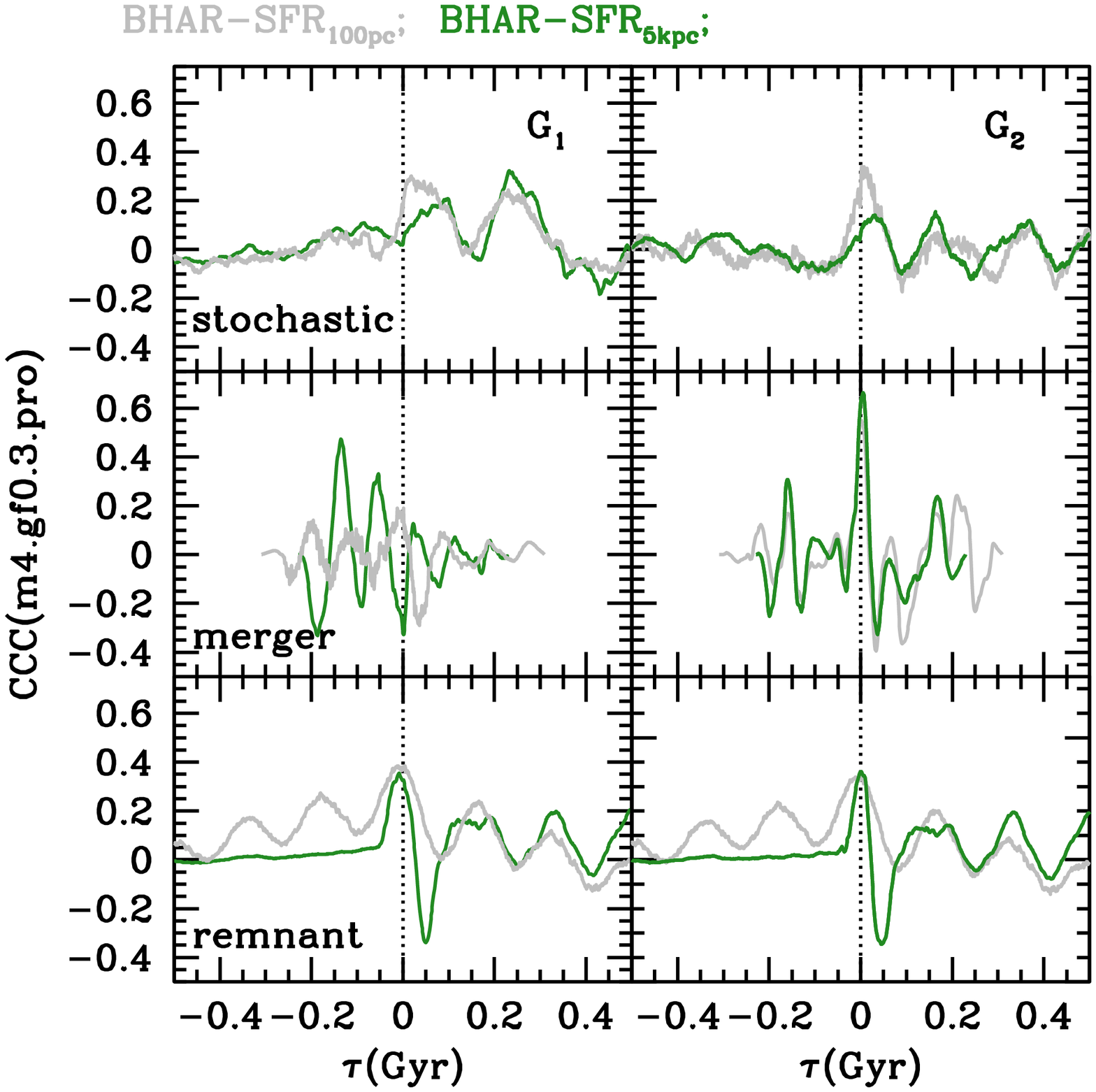}
\caption{Cross-correlation function for BHAR and SFR  for the 1:4 coplanar, prograde-prograde merger.}
\end{figure}

\begin{figure}
\centering
\includegraphics[width=\columnwidth,angle=0]{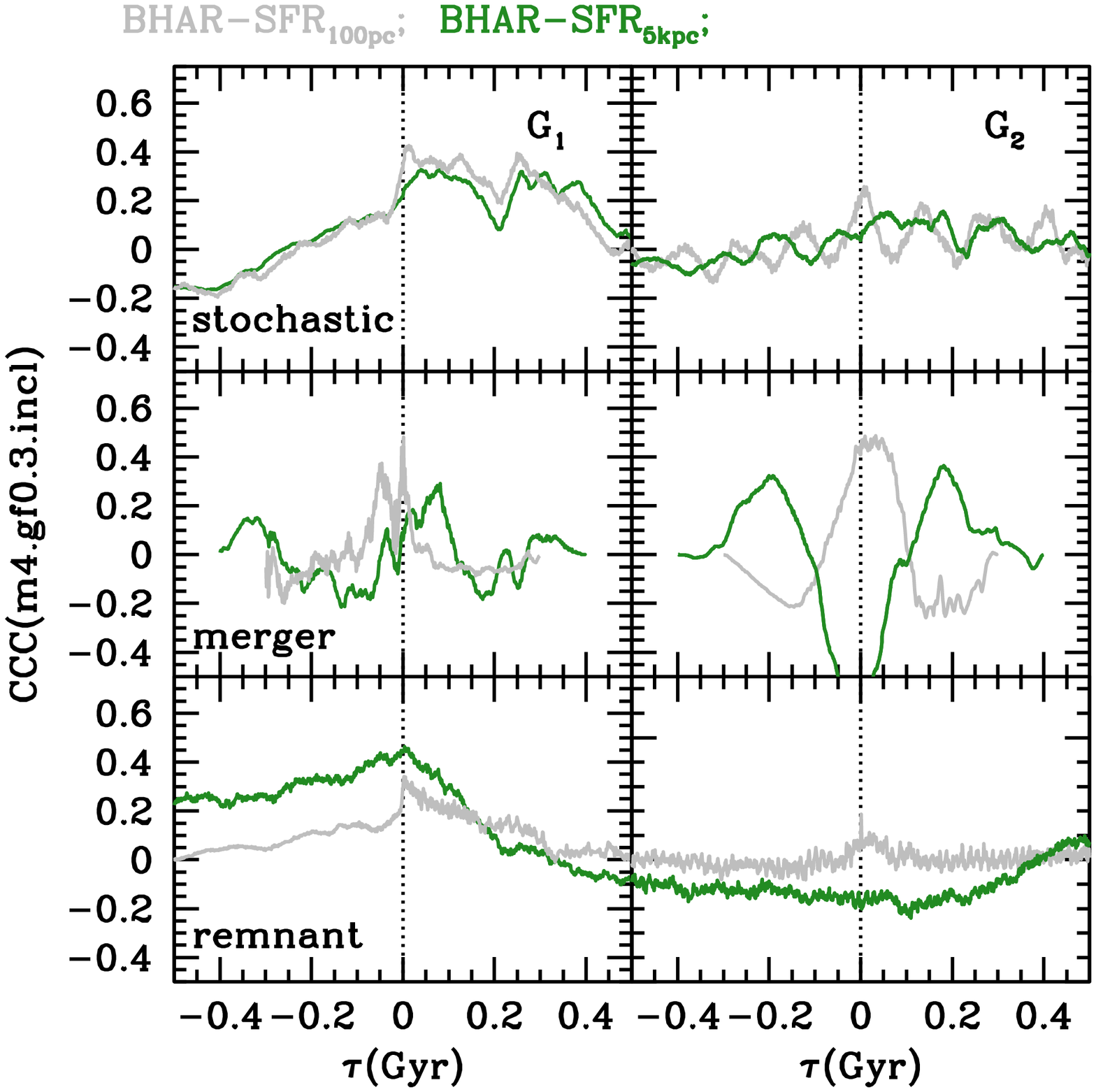}
\caption{Cross-correlation function for BHAR and SFR  for the 1:4 inclined, prograde-prograde merger.}
\end{figure}

\begin{figure}
\centering
\includegraphics[width=\columnwidth,angle=0]{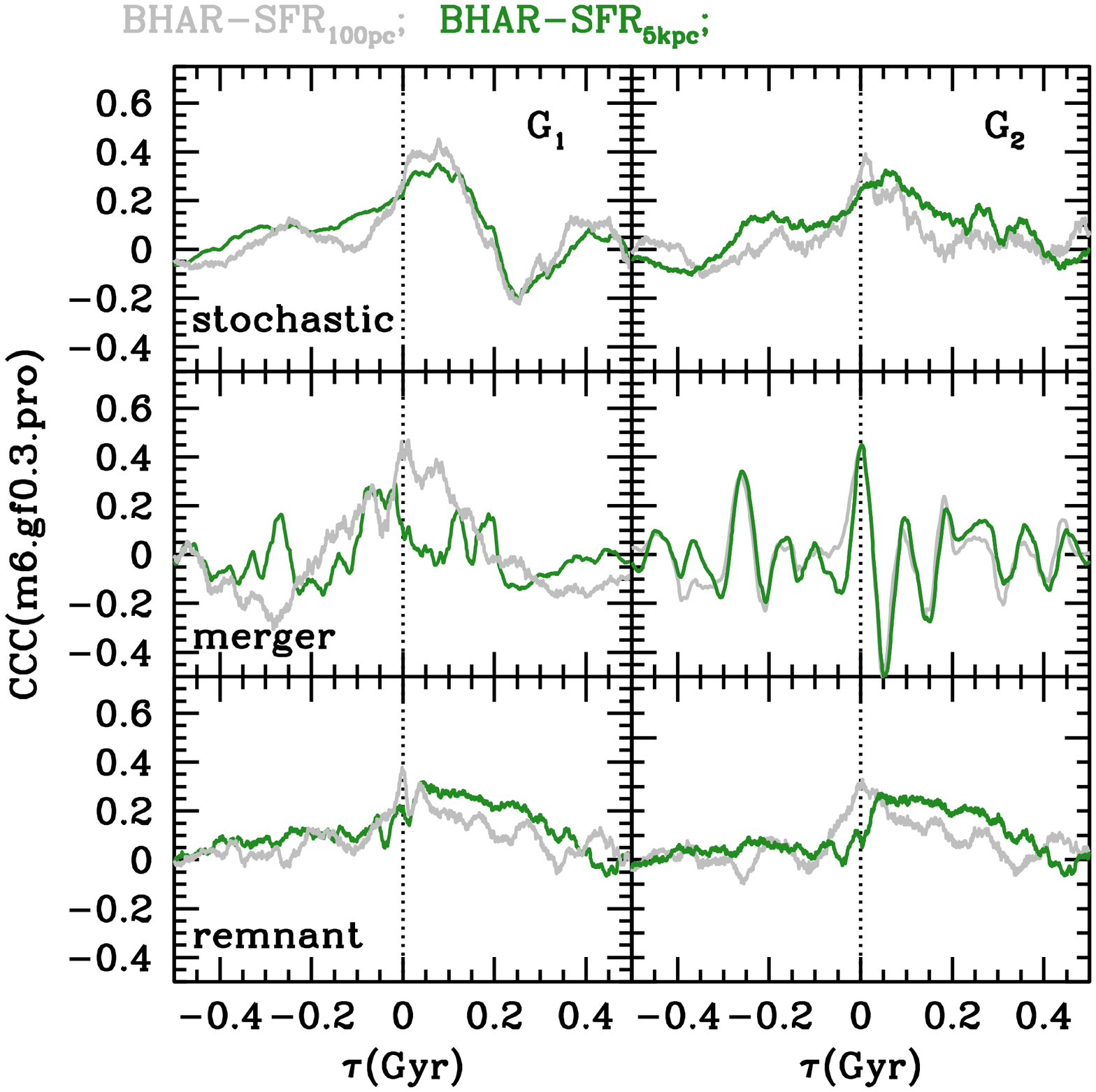}
\caption{Cross-correlation function for BHAR and SFR  for the 1:6 coplanar, prograde-prograde merger.}
\end{figure}

\begin{figure}
\centering
\includegraphics[width=\columnwidth,angle=0]{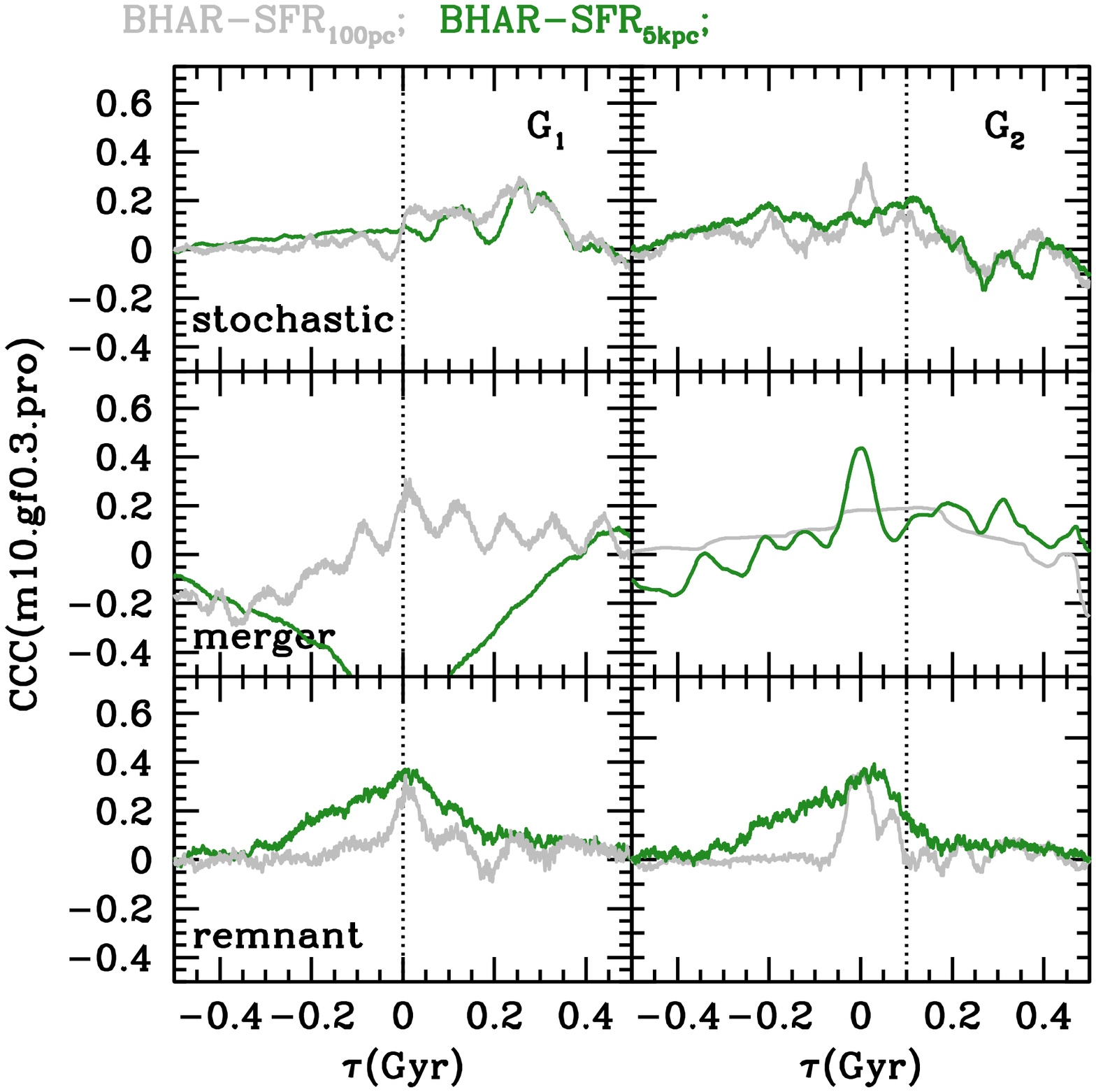}
\caption{Cross-correlation function for BHAR and SFR  for the 1:10 coplanar, prograde-prograde merger.}
\end{figure}

\clearpage

\begin{figure}
\centering
\includegraphics[width=\columnwidth,angle=0]{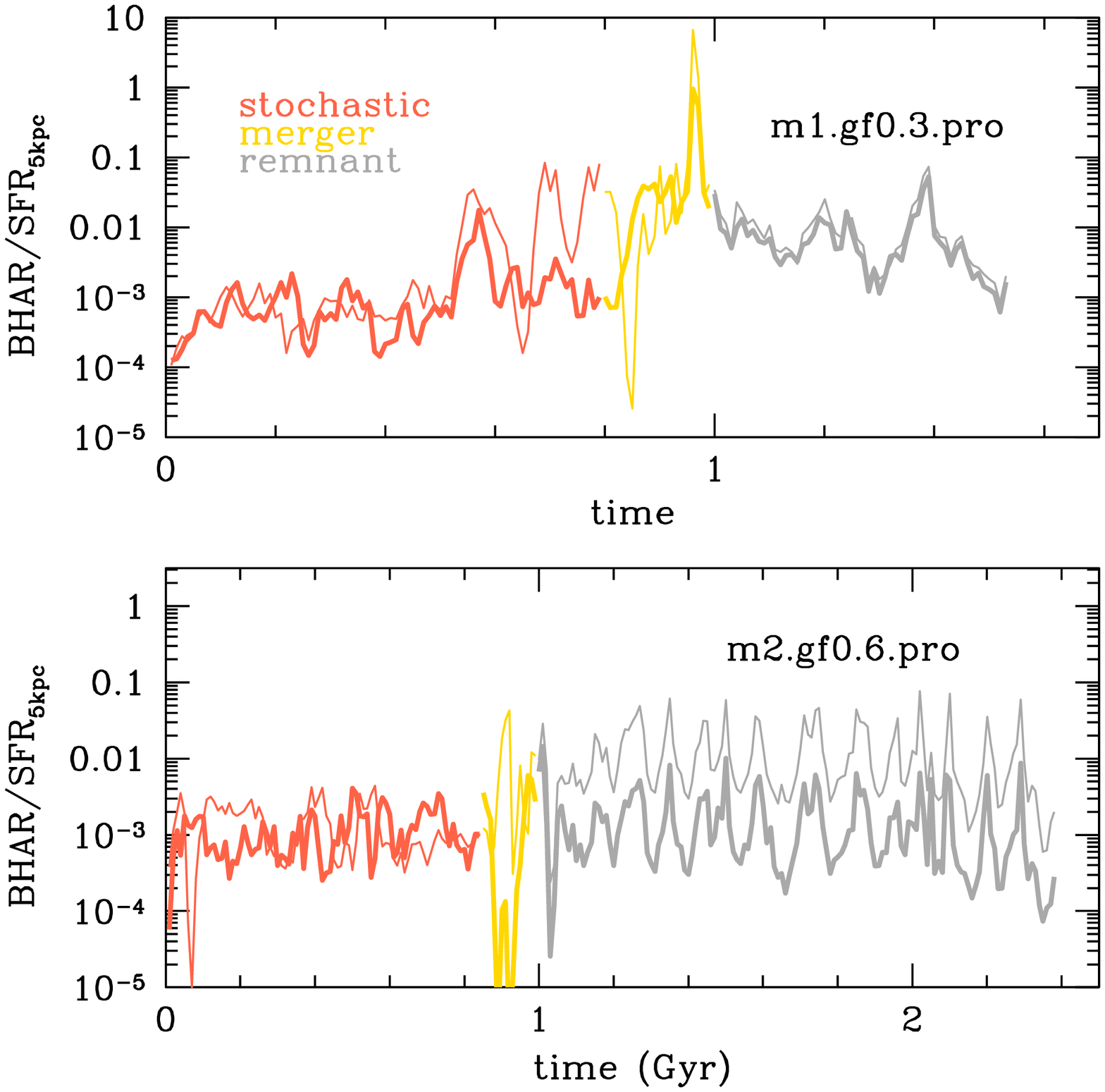}
\caption{Ratio of BHAR to $\sfrg$, averaging both quantities in bins of 50~Myr for the 1:1 coplanar, prograde-prograde and the 1:2 coplanar, prograde-prograde 60\% gas fraction mergers. Thin curve: $G_1$; thick curve: $G_2$. }
\end{figure}

\begin{figure}
\vspace{-0.85cm}
\centering
\includegraphics[width=\columnwidth,angle=0]{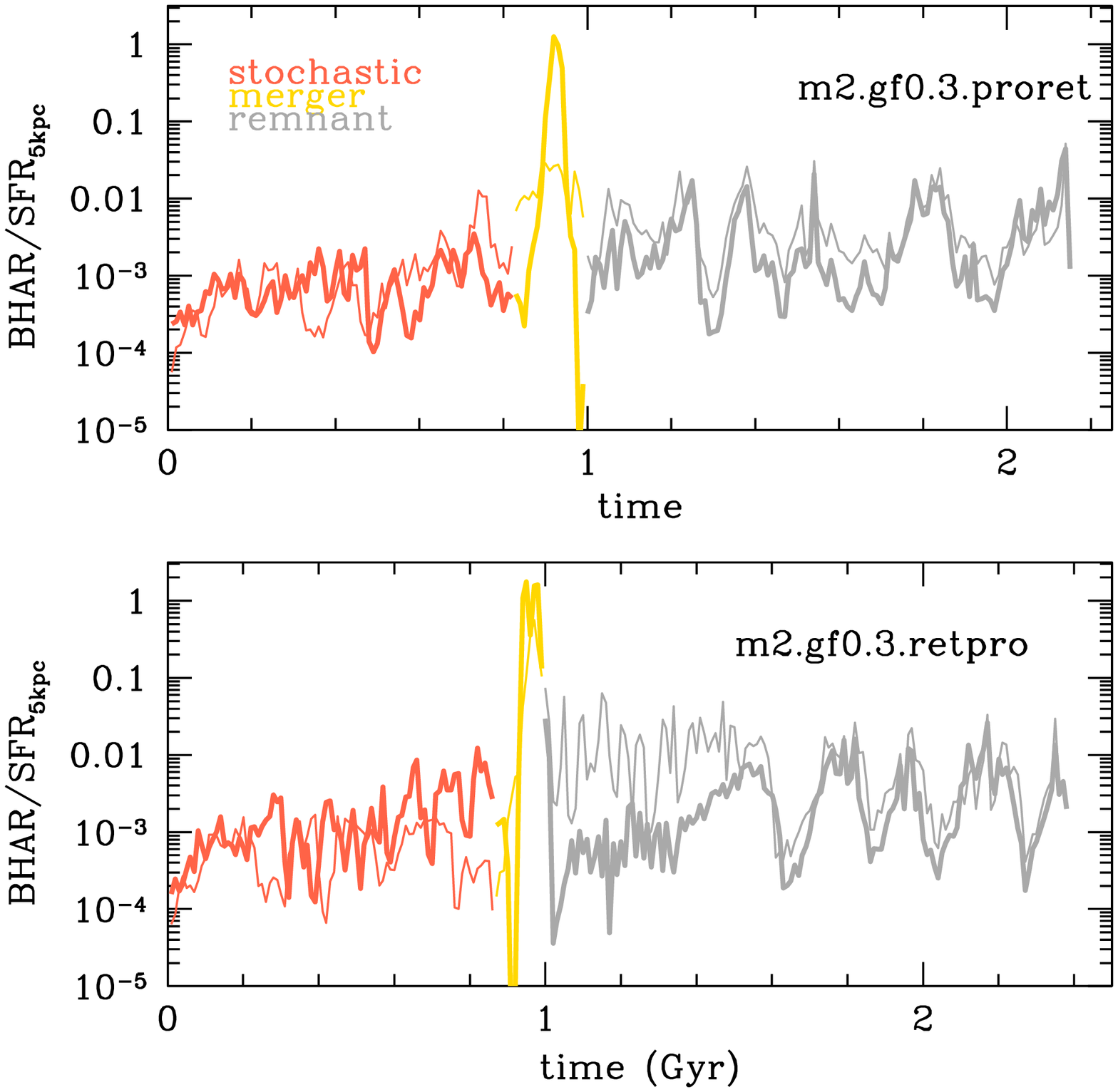}
\caption{Ratio of BHAR to $\sfrg$, averaging both quantities in bins of 50~Myr for the 1:2 coplanar, retrograde-prograde and the 1:2 coplanar, prograde-retrograde mergers. Thin curve: $G_1$; thick curve: $G_2$. }
\end{figure}

\begin{figure}
\centering
\includegraphics[width=\columnwidth,angle=0]{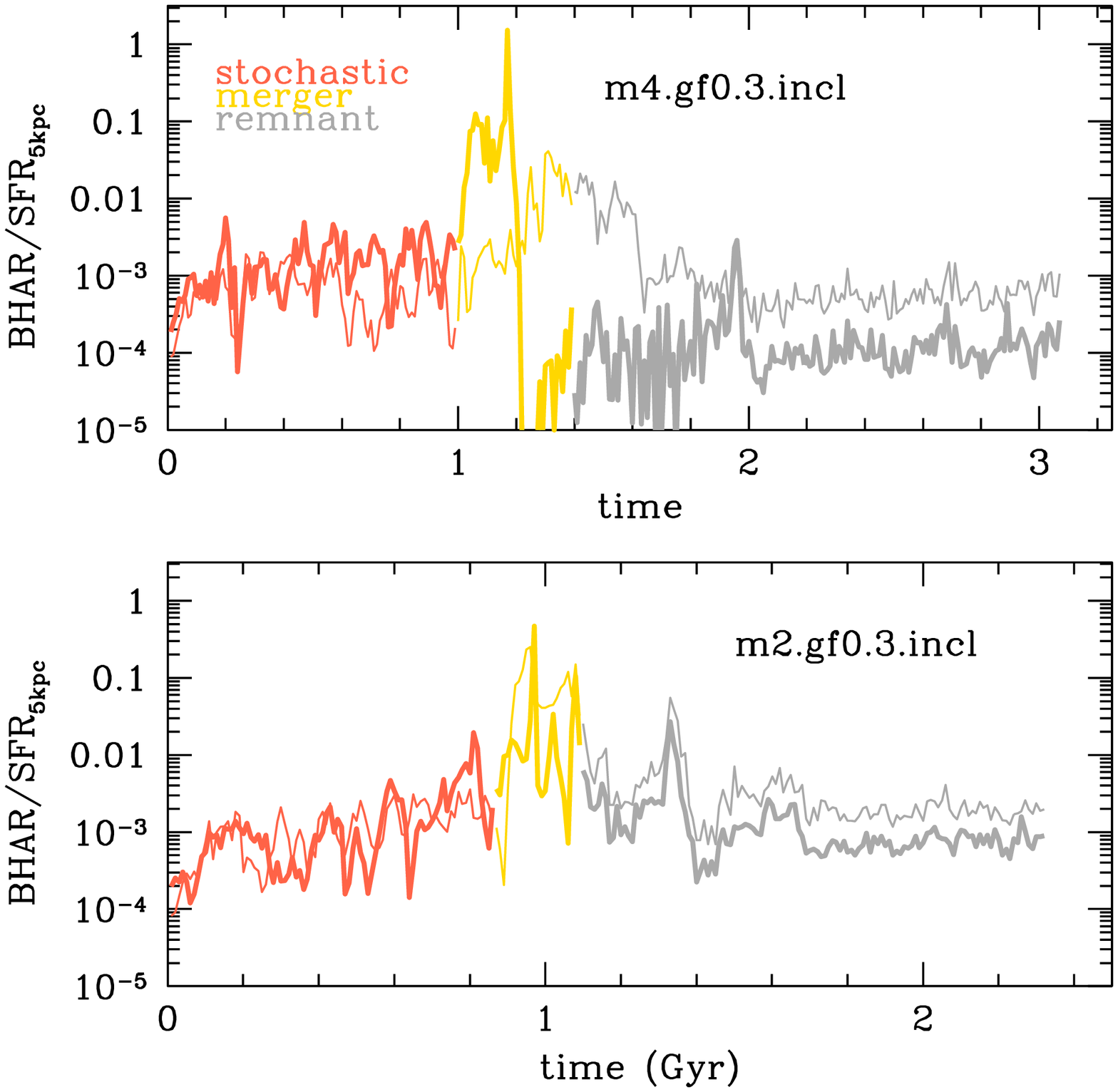}
\caption{Ratio of BHAR to $\sfrg$, averaging both quantities in bins of 50~Myr for the 1:2 inclined, prograde-prograde  and the 1:4 inclined, prograde-prograde mergers. Thin curve: $G_1$; thick curve: $G_2$. }
\end{figure}

\begin{figure}
\vspace{-0.85cm}
\centering
\includegraphics[width=\columnwidth,angle=0]{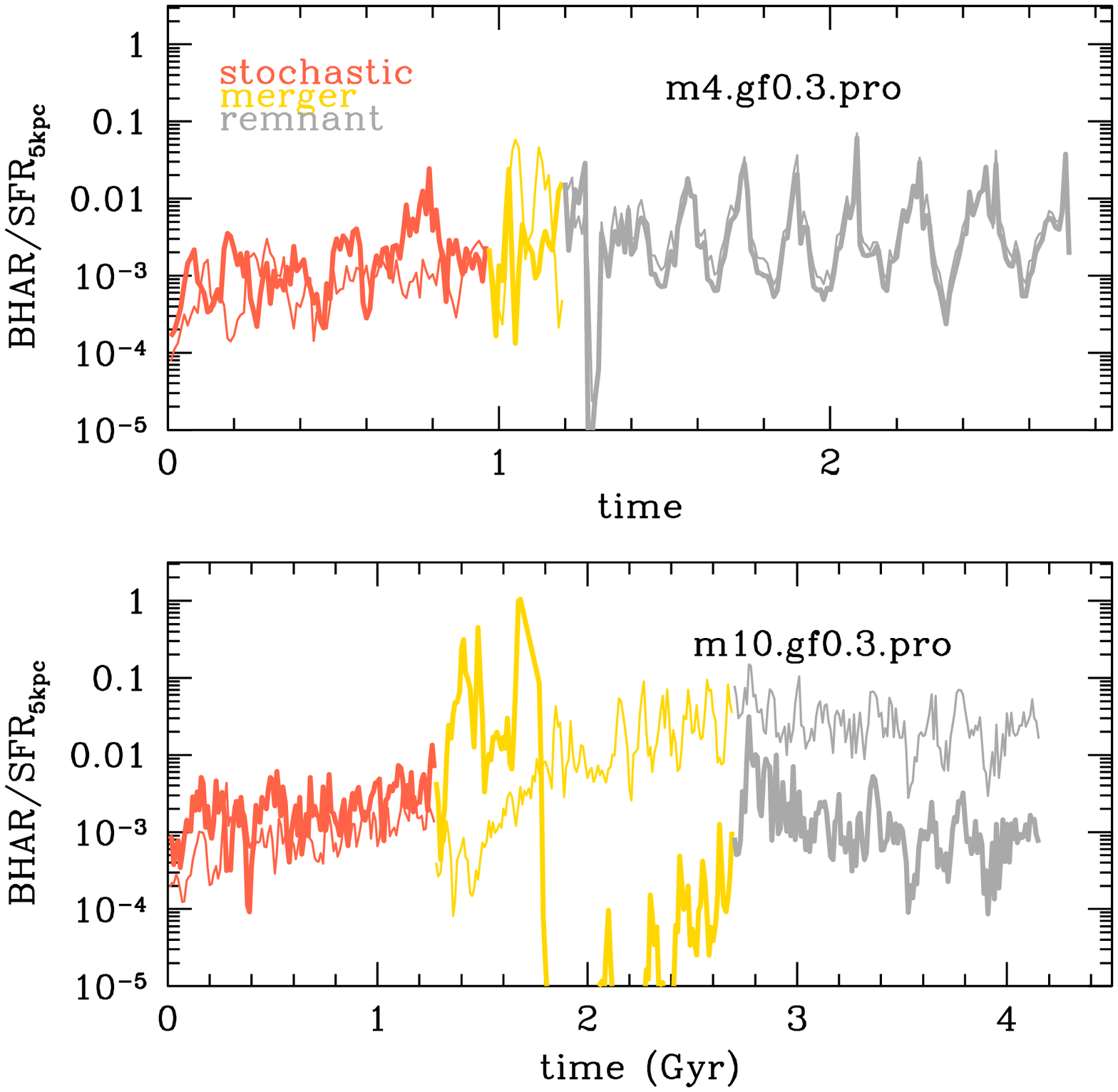}
\caption{Ratio of BHAR to $\sfrg$, averaging both quantities in bins of 50~Myr for the 1:4 and 1:10 coplanar, prograde-prograde mergers. Thin curve: $G_1$; thick curve: $G_2$. }
\end{figure}

\begin{figure*}
\centering
\includegraphics[width=\textwidth,angle=0]{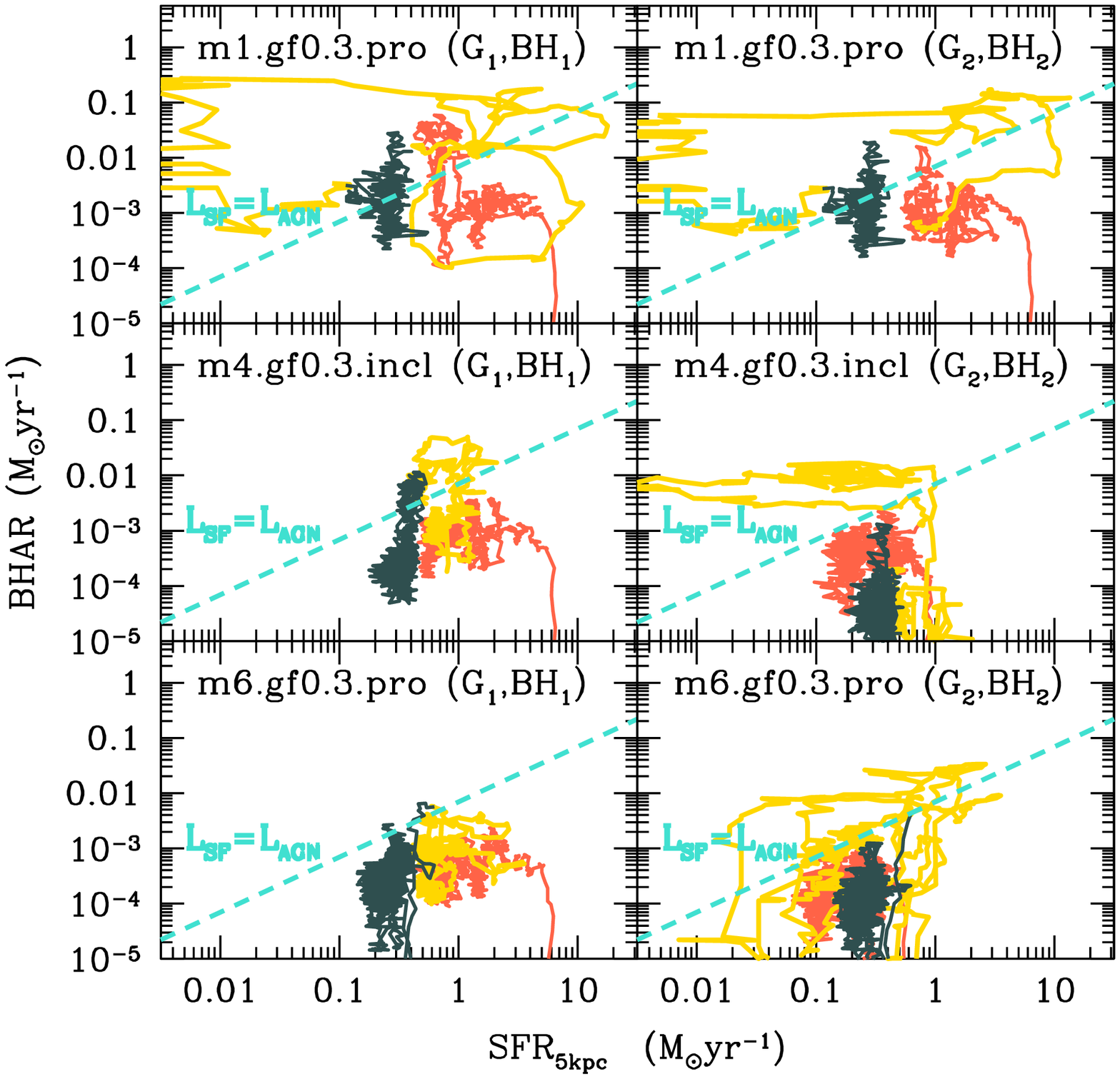}
\caption{Tracks in the BHAR and SFR for the 1:1, 1:4 and 1:6 coplanar, prograde-prograde mergers.}
\end{figure*}

\begin{figure*}
\centering
\includegraphics[width=\textwidth,angle=0]{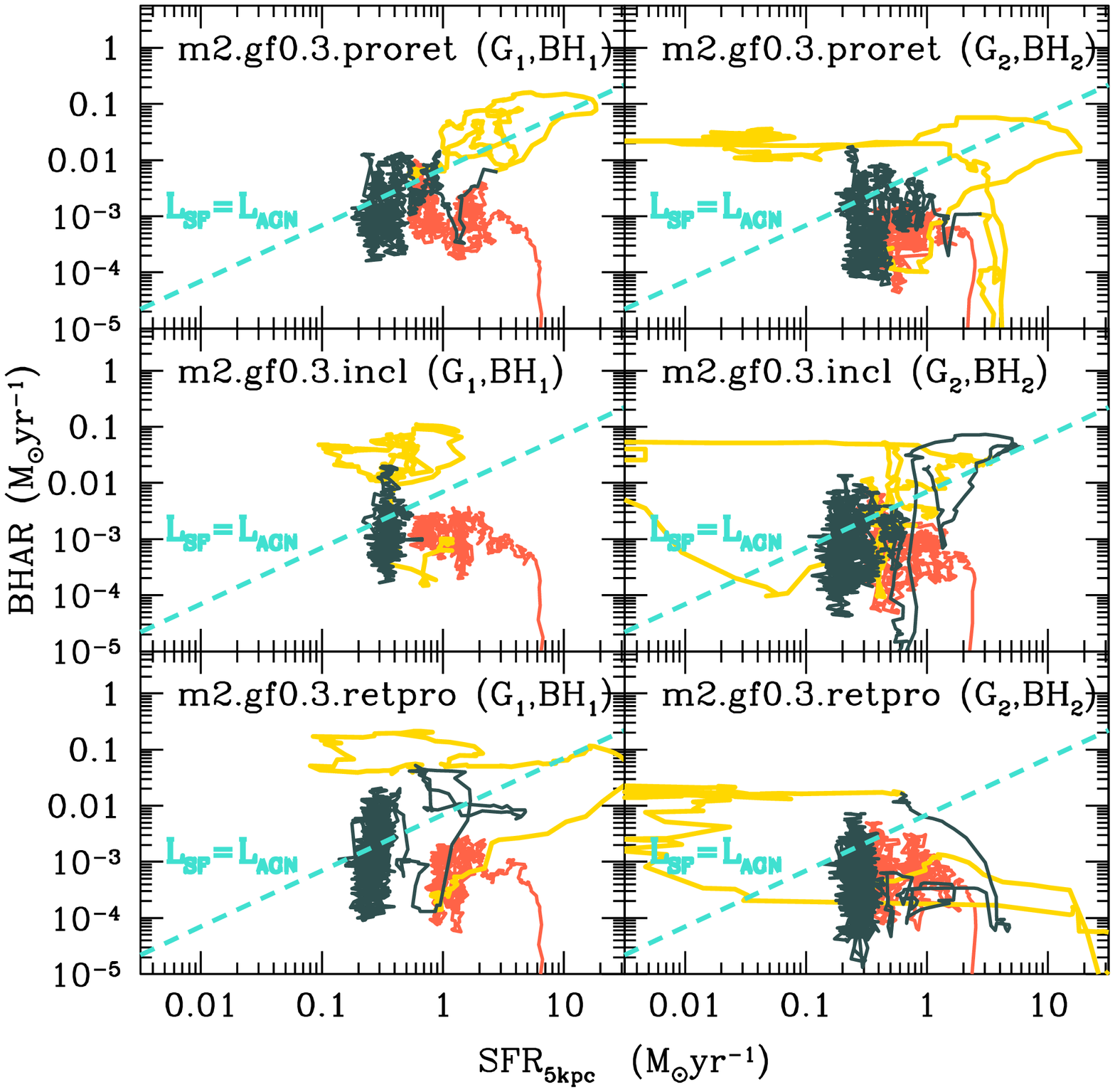}
\caption{Tracks in the BHAR and SFR for the 1:2 mergers with inclined or retrograde configurations.}
\end{figure*}

\normalsize

\end{document}